\pdfoutput=1
\documentclass[useAMS,usenatbib,usegraphics]{mn2e}

\usepackage[utf8]{inputenc}
\usepackage{graphicx,epsfig}  
\usepackage{amsmath,amssymb}
\usepackage{subfigure}
\usepackage{longtable}
\usepackage{lscape}
\usepackage{times}
\usepackage{textcomp}

\usepackage{natbib}
\usepackage{txfonts}
\usepackage{multirow}
\usepackage{soul}

\def\ut#1{\mathop{\vtop{\ialign{##\crcr
     $\hfil\displaystyle{#1}\hfil$\crcr\noalign
     {\kern1pt\nointerlineskip}\hbox{$\hfil\sim\hfil$}\crcr
     \noalign{\kern1pt}}}}}

\def\undersymbol#1#2{\mathop{\vtop{\ialign{##\crcr
     $\hfil\displaystyle{#2}\hfil$\crcr\noalign
     {\kern1pt\nointerlineskip}\hbox{$\hfil#1\hfil$}\crcr
     \noalign{\kern1pt}}}}}

\def\second{^{\rm s}}
\def\sat{{\it XMM}-Newton}

\title[A \sat\ observation of a sample of four close dSph galaxies.]{A \sat\ observation of a sample of four close dSph galaxies.}
\author[Manni L., Nucita A.A., De Paolis F.,Testa V, Ingrosso G.]{Manni L.$^{1,2}$, Nucita A.A.$^{1,2}$, De Paolis F.$^{1,2}$,Testa 
V.$^{3}$, Ingrosso G.$^{1,2}$\\
$^{1}$Department of Mathematics and Physics {\it ``E. De Giorgi''}, University of Salento, Via per Arnesano, CP 193, I-73100, 
Lecce, Italy \\ 
$^{2}$INFN, Sez. di Lecce, Via per Arnesano, CP 193, I-73100, Lecce, Italy\\
$^{3}$ INAF, Osservatorio Astronomico di Roma, via di Frascati 33, 00040 Monteporzio, Italy \\}  

\begin{document}

\date{Accepted xxx; Received xxx; in original form xxx}
\pagerange{\pageref{firstpage}--\pageref{lastpage}} \pubyear{2011}

\maketitle
\label{firstpage}

\begin{abstract}
We present the results of the analysis of deep archival \sat\ observations towards 
the dwarf spheroidal galaxies Draco, Leo I, Ursa Major II and Ursa
Minor in the Milky Way neighbourhood.
The X-ray source population is characterized and cross-correlated with 
available databases with the aim to infer their nature. We also investigate if intermediate-mass black holes
are hosted in the center of these galaxies.\\
In the case of Draco, we detect 96 high-energy sources, two of them being possibly local stars, 
while no evidence for any X-ray emitting central compact object is found.\\ 
Towards the  Leo I and UMa II field of view we reveal 116 and 49 X-ray sources, respectively. 
None of them correlates with the putative central black holes and only one is likely associated with a UMa II local source.\\
The study of the UMi dwarf galaxy shows 54 high-energy sources and 
a possible association {with a source at the dSph center}. We put an upper limit to the central compact 
object luminosity of 4.02$\times$10$^{33}$ erg/s. Furthermore, via the correlation with a radio source near the 
galactic center, we get that the putative black hole should have a mass of $\left(2.76^{+32.00}_{-2.54}\right)\times10^6 M_{\odot}$ 
and be radiatively inefficient. This confirms a previous result obtained by using Chandra data alone.
\end{abstract}


\section{Introduction}
Dwarf spheroidal galaxies (dSphs) are very peculiar star systems with mass in the range $10^3$ - $10^{7}$ M$_{\odot}$ 
(\citealt{martin2008}) and relatively poor in stars. Consequently they result very faint and difficult to be detected 
and studied in details. In the past decades, the interest in dSphs has rapidly grown up. 
Indeed, until 2001, only 9 Milky Way (MW) dwarf satellites (at distance between 16 and 250 kpc) were known 
(\citealt{odenkirchen2001}) while, recently, \citet{McConnachie2012} listed over 100 nearby galaxies in and around 
the Local Group, along with their relevant physical properties.

If dSphs orbit close enough to the Galactic center, they may lose mass and be disrupted 
by the tidal forces due to the Galactic potential.
Tidal effects can be easily revealed in the Sagittarius dSph which has a peculiar shape characterized 
by stellar debris along its orbit (see \citealt{mateo1998}, 
\citealt{ibata2001}). Tidal extension signatures for Carina and Ursa Minor dSphs were also detected by \citet{majewski2000} and \citet{martinez2001}.
On the contrary, \citet{odenkirchen2001} showed that 
there is no evidence for the existence of a tail-like extension of the Draco dSph star population beyond its 
tidal radius.

DSph  are characterized by large mass-to-light ratios. This leads scientists to infer that they 
are dark matter (DM) dominated objects (see e.g. \citealt{mateo1997}). \citet{breddels2013} performed a comparison 
between different dSph formation scenarios using a sample of DM profiles 
and argued that no particular model is significantly preferred among others. 
Moreover, some authors suggest different dSph growing processes, without contemplating DM content. In this respect,
\citet{yang2014}, by using numerical simulations, claim that a merger in M31, occurred 8.5 Gyr ago, 
could have ejected a tidal-tail toward our direction. So that, dSph could be generated 
from the interaction between the low-mass tidal dwarf galaxies and our MW.

As far as the origin and evolution of dSphs, 
they can be either remnants of bigger systems, disrupted by tidal forces 
or affected by supernova winds that 
take out the overwhelming majority of gas (\citealt{silk1987}), 
or small mass systems since their origin. In the latter case, they could be 
the building blocks of large galaxies (with mass in the range $10^9$ - $10^{11}$ M$_{\odot}$) while, in the absence of 
interaction (merging or disruption), they can survive until now.
To address these issues, dSphs are nowadays intensively studied by numerical 
simulations to infer their formation and evolution (\citealt{assmann2013a}, 
\citealt{assmann2013b}, \citealt{casas2012}).

{ 
Additionally, as globular clusters, dSphs host old stellar populations and, consequently, X-ray sources are expected 
to be most likely low mass X-ray binaries (LMXBs) or cataclysmic variables (CVs). However, one fundamental difference between 
globular clusters and dSphs is their central stellar density that can be at least two order of magnitude smaller in dSphs (\cite{harris1996,McConnachie2012}). 
Then, at variance with globular clusters
where  it is thought that the high stellar density forms a nursery for LMXB through capture, in the case of dSphs any 
X-ray binary should be primordial. However, due to the old stellar population, LMXBs would turn off in a few hundred million years, making unlikely
finding these systems in dSphs. 

The lower stellar density of dSph offers a contrasting environment with respect to that in globular clusters and a comparison 
of the X-ray source populations in the two cases may help testing the LMXB formation scenario. This is exactly the case 
of the Sculptor dSph which was studied by 
\citet{maccarone2005b} who found five X-ray sources (likely X-ray binaries with a newutron star or a black hole primary) 
with $L_X\geq 6\times 10^{33}$ erg s$^{-1}$ associated to the galaxy. This discovery, from one side proves that LMXBs can exist 
in a old stellar environment with low stellar encounters. From the other side, it is clearly challenging for the LMXB formation theory.

Hence, one of the goals of this paper is to attempt a classification of the X-ray sources identified towards our dSph sample  
and pinpoint local (or candidate local) sources. Detecting X-ray sources in dSph can, moreover, help in the investigation of the dark matter component
in these objects. Indeed, when neutron stars form in SN II explosions they get large kick velocities ($\sim 100$ km s$^{-1}$, \citealt{pods2005}) exceeding 
the escape local velocity (a few km s$^{-1}$) due to the stellar component of the dSph. However, if dSphs have a dark matter halo, LMXBs (and even 
isolated millisecond pulsars which are direct descendant of LMXBs) can be retained within the galaxy \citep{dk2006}. 

DSphs are also the best candidates to host intermediate mass black holes (hereinafter IMBHs) in their gravitational 
cores. This clue derives by extrapolating fundamental $M_{BH} - M_{Bulge}$ relation 
(see e.g. \citealt{magorrian1998}, for the super massive BH case) down to the typical dSph masses. In particular, 
IMBH masses can be evaluated either via dynamical considerations or using the fundamental plane relation at radio and X-ray wavelengths 
(see e.g., \citealt{reines2013}, \citealt{nucita2013a}, \citealt{nucita2013b}). We also mention that, within the galaxy hierarchy scenario, 
IMBHs may be the ground seeds for the formation of super massive BHs hosted in the center of galaxies.

Searches for IMBHs in dSphs have been attempted recently. In a very interesting paper, \citet{lemons2015} have anlayzed a 
sample of $\simeq 44000$ dSph galaxies detected by the Sloan Digital Sky Survey (SDSS) with redshift $z \leq 0.055$ and stellar 
mass content $M_{*}\leq 3\times 10^{9}$ M$_{\odot}$. By a cross-correlation with the Chandra Source Catalogue, it was found that 19 galaxies have at least a 
detectable hard X-ray source within three half light radii. Moreover, for about half of this sample, there is the evidence 
that the X-ray source (possibly a massive black hole candidate) is associated to the optical nucleus of the dSph. Of course, as pointed out by the authors, 
follow-up observations are necessary to disentangle between stellar-mass X-ray binaries and the existence of active galactic nuclei with 
an accreting BH as central engine.

On the basis of these results, we also searched in the selected dSphs for X-ray sources 
with the typical signatures of accreting IMBHs. 
Although the statistics at hand is low, it is interesting  
that we find fraction of galaxies in our sample that appear hosting a central X-ray source is similar to that 
observed by \citet{lemons2015}. Also in our case, only dedicated follow-up observations may allow to unveil the source nature.}

The paper is structured as follows: in Section 2 we briefly describe the sample of objects used in our study.  
Section 3 presents the procedure used to analyze the \sat\ data. 
The results for the high-energy study of dSphs are reported in Section 4 and, finally, 
in Section 5 we address our conclusions.

\section{DSph sample}
Our study is concerned with the high-energy characterization of dSphs.
We select the dSph sample for which \sat\ archival data are available. 
{We did not consider the Fornax dSph galaxy since it was already studied 
in details by \citet{orioproc} and \citet{nucita2013a}. Here, we remind that these authors
found an X-ray source possibly associated to a variable star belonging to the galaxy and 
two more sources at the boundaries of the Fornax globular clusters GC 3 and GC 4. Furthermore, 
following \citet{jardel} who predicted the existence of a central IMBH with mass $\simeq 3\times 10^{4}$ M$_{\odot}$, \citet{nucita2013a} 
searched also for X-ray targets at the galaxy center. In the particular case 
of one of the possible gravitational centers reported in \citet{stetson}, the authors 
found a close X-ray source. The source unabsorbed $0.2-12$ keV flux is $3\times 10^{-15}$ erg s$^{-1}$ cm$^{-2}$ 
corresponding to an intrinsic luminosity of $L_X\simeq 7\times 10^{33}$ erg s$^{-1}$ (assuming a galaxy 
distance of 138 kpc): in the IMBH hypothesis (and assuming a Bondi spherycal accretion or 
in the context of a Keplerian thin disk model), the compact object seems to accrete very inefficiently.

In this paper, we focus our attention on four galaxies\footnote{ 
We mention that the \sat\ observations analysed in this work (see text for details) have been 
already used for different purposes. As an example, \citet{malyshev} searched for a 3.55 keV line as the 
signatures of dark matter (in the form of sterile neutrinos) decays. Their analysis showed
no evidence for the presence of such line in the stacked spectra of the investigated dSphs. 
}: Draco, Leo I, Ursa Major II and Ursa Minor 
(hereafter UMa II and UMi, respectively). 
}

\subsection{Draco}
The Draco dSph (at J2000 coordinates RA = 17$^{\rm h}$ 20$^{\rm m}$ 12.4$\second$ and Dec = 57$^\circ$ 54$'$ 55$''$) 
is a MW companion galaxy. Since its discovery (\citealt{wilson1955}) this galaxy became target of
many observational campaigns. \citet{baade1961} first derived 
the distance to the galaxy to be about {99 kpc} through the identification of variable stars. In addition, others studies on Draco variables were made by \citet{zinn1976}, 
\citet{nemec1985}, \citet{goranskij1982}, \citet{kinemuchi2002} and \citet{bonanos2004}. Recently, \citet{kinemuchi2008}
presented a survey with the photometry of different kind of sources (270 RR Lyrae stars,
9 anomalous Cepheids, 2 eclipsing binaries, 12 slow irregular red variables, 30 background QSOs and 26 
probable double-mode RR Lyrae stars).
Other photometric studies were performed by \citet{bellazzini2002} and \citet{rave2003}.

Draco seems to host a single stellar population older than $\simeq$ 10 Gyr (\citealt{grillmair1998}, \citealt{ora2003}).
According to \citet{bonanos2004}, the Draco luminosity ($\simeq 2 \times 10^5$ L$_{\odot}$) is comparable 
to that of the faintest luminous systems.

By NED\footnote{NASA/IPAC Extragalactic Database is available at http://ned.ipac.caltech.edu.} 
we get not only the galaxy coordinates but also other interesting quantities such as the mean distance of about
{82 kpc} and the {physical} major and minor axes of {1.19 kpc} and {0.72 kpc}, respectively. {\citet{McConnachie2012} reports a position angle of $89^\circ$. 
For our study\footnote{For completeness, we note that Draco dSph was observed also by the {\it Chandra} satellite (IDs 9568 and 9776) with the ACIS-S camera and exposure times of $\simeq 24$ ks and $\simeq 12$ ks, respectively. 
Only four detectors were on during the observations thus limiting the coverage of the Draco galaxy.  The event files were reprocessed using the most updated calibration files and the CIAO tool suite (version 4.6). 
Then, we searched for X-ray sources in the $0.5-7$ keV by 
using the {\it celldetect} code choosing a signal-to-noise ratio of $2.5$ for the source detection. After eliminating a few sources not recognized as such by eye, we were left with $9$ X-ray targets indicated by 
yellow diamonds in Figure \ref{images} We remind that the background corrected rates and fluxes of the identified sources are consistent within the errors with those estimated by using \sat\ .} 
we used five \sat\ observations (IDs $0603190101$, $0603190201$, $0603190301$, $0603190401$, and $0603190501$) 
made in the August 2009 for a total exposure time of about 90 ks. 
}

\subsection{Leo I}
As far as Leo I (at J2000 coordinates RA = 10$^{\rm h}$ 08$^{\rm m}$ 28.1$\second$ and Dec = 12$^\circ$ 18$'$ 23$''$) is 
concerned, NED database gives a mean distance of $\simeq$ {246 kpc} (thus it is one of the furthest MW companions), 
positional angle of 80$^\circ$, physical major and minor axis of {0.86 kpc} and {0.64 kpc}, respectively.
\citet{hw1950} analized this gas poor galaxy whose type II star population and elliptical shape suggested a similarity with the Sculptor 
system. The large distance from the MW and the high 
radial velocity, confirmed by \citet{koch2007}, suggest that Leo I is an isolated dSph, not currently 
affected by Galactic tidal forces. However, the presence of stars beyond the object tidal radius confirms
the hypotesis that some dSphs may be perturbed at least in their outermost regions by the existence of a huge amount
of dark matter at large distances.

Recently \citet{smecker2009} re-analyzed the star population content of this dwarf galaxy and determined the galaxy 
star formation rate as a function of time. In particular, they 
found that Leo I has a stellar population older than previously believed and older than that of the 
irregular Leo A galaxy, although both systems have continuously formed stars. The evidence of a radial age gradient in the red giant branch (RGB) stellar population was shown by 
\citet{gullieuszik2009} who estimated the stellar ages, while {\citet{menzies2010} analyzed the existence of asymptotic giant branch (AGB) variables.\\
In our work we use the \sat\ observation (ID $0555870201$) of about 90 ks made on 2008-11-24,
the same being previously analyzed by \citet{orioproc}. These authors cross-correlated their sample of X-ray detected sources with the 
catalogue of carbon rich AGB stars in Leo I \citealt{held2010} finding no association. As claimed by \citet{orioproc}, a few X-ray sources correlate in position with 
RGB stars (most symbiotic binary systems have red giant companions). This is consistent with our results (see Section 3).

\subsection{Ursa Major II}
UMa II dSph (with J2000 coordinates RA = 08$^{\rm h}$ 51$^{\rm m}$ 30$\second$ and 
Dec = 63$^\circ$ 07$'$ 48$''$) was discovered by \citet{zucker2006} analyzing Sloan Digital Sky Survey (SDSS). They 
noted its irregular and distorted shape, probably due to a tidal disruption process. Evidences supporting this scenario 
are found in morphological studies made by \citet{fellhauer2007} and \citet{munoz2010} to which we refer for more details. In spite of the dominant dark matter contribution scenario, \citet{smith2013} demonstrated, 
via N-Body simulations, that the observed properties can be well reproduced, in absence of DM, by tidal mass loss processes.
So the presence/absence of dark matter in UMa II is a not yet settled issue.

Another still open point is related to the star population of this dwarf galaxy. \citet{ora2012} revealed hints, although not statistically 
significant, of two distinct stellar populations, with different age and metallicity.  {They also detected a RR Lyrae
star in UMa II and evaluated a distance of about 34.7 kpc. We use a distance of 34 kpc, as reported by NED, 
in the analysis of \sat\ data (ID $0650180201$) obtained on 2011-04-21 with an exposure time of about 34 ks. As reported in 
\citet{McConnachie2012}, the galaxy half light radius is $r_h\simeq 0.15$ kpc and a position angle of $98^\circ$.}

\subsection{Ursa Minor}
The fourth target of our sample is UMi dSph at J2000 coordinates 
RA = 15$^{\rm h}$ 09$^{\rm m}$ 10.2$\second$ and Dec = 67$^\circ$ 12$'$ 52$''$ (\citealt{falco1999}).

This dSph is one of the first MW companions revealed in the past century (\citealt{wilson1955}).
\citet{walker2007} analized the stellar velocity dispersion profiles of some dSphs, among which UMi, observing
that it remains approximately constant with the galactocenter distance. The authors also noted that the observed
profiles were well fitted by models of luminous stars systems immersed into a DM halo.

Regarding the star formation history, UMi shows a predominantly old (with age of about 10 Gyr) stellar population (\citealt{carrera2002}). 
In that work, the authors studied also the stellar metallicity claiming the absence of a metallicity radial gradient
throughout the galaxy.
According to NED, UMi has a mean distance of {73 kpc}, positional angle of 69$^\circ$, 
{physical} major and minor axis of {0.85 kpc} and {0.53 kpc}, respectively.\\
We analyzed  two \sat\ observations (ID $0301690301$, and $0301690401$) made between the end of August and the beginning of 
September 2005 for a total exposure time of $\simeq$27 ks.

\citet{nucita2013b} analyzed the X-ray data acquired by the Chandra satellite towards this galaxy and
found an X-ray source spatially coincident (within a few arcseconds) with a radio one. Assuming that the target is an accreting IMBH, the authors 
evaluated the BH mass, which resulted to be $\simeq 2.9\times 10^6$ M$_{\odot}$. However, the detection algorithm used did not allow to 
exclude that one false detection per CCD occurred. Here, to confirm the previous result, here we used \sat\ observations (see the next Section).

\section{{\it XMM}-Newton data reduction and source detection}
We used the {\it XMM}-Science Analysis System (SAS version {see: {\tt http://xmm.esa.int/sas/}}), with the 
{most recent} calibration files, to process the observation data files (ODFs).

We obtained the event lists by {processing}  the raw data via the standard {\it emchain} and {\it epchain} tools, 
and following the standard procedures in screening the data provided by the three European Photon Imaging Cameras 
(MOS 1, MOS 2 and pn) on-board \sat\ satellite. 

We {applied} the analysis procedure described in \citet{nucita2013a}. 
Here we only remind that, in accordance {with} the 2XMM catalogue of serendipitous X-ray sources (\citealt{watson2009}), 
we divided each EPIC camera event list into 5 energy bands, i.e. {:} $B_1: 0.2-0.5$ keV, $B_2: 0.5-1.0$ keV, $B_3: 1.0-2.0$ keV, $B_4: 2.0-4.5$ keV, 
and $B_5: 4.5-12.0$ keV, {producing} one image for each energy band, {plus} a mosaic image in the 0.2-12.0 keV 
energy band (for inspection purposes only), and carried out the source detection using the SAS task {\it edetect$\_$chain}.\\
{The exposure map has been evaluated with the {\it eexpmap} task,} 
for each camera and input image, taking into account the calibration information on spatial quantum efficiency, filter transmission and vignetting. 
Then we {produced appropriate image masks to delimit} the regions in which to perform source detection with the task {\it emldetect}.
This task applies a point spread function (PSF) fitting algorithm simultaneously in the five energy bands giving as output source coordinates, 
energy fluxes and hardness ratios (for details see: {\tt http://xmm.esac.esa.int/sas/current/documentation})}.

The X-ray fluxes are given in units of erg s$^{-1}$ cm$^{-2}$ {through} the formula
\begin{equation}
F_i = \frac{B_{i}}{ECF_i}~,
\end{equation}
where the count rate ($B_i$) and {the} energy conversion factor 
($ECF_i$) {are} in units of $10^{11}$ counts cm$^{2}$ erg$^{-1}$  in the $i$-th band for each EPIC camera.
To obtain the latter factors \footnote{For ECFs and associated correction factors, see the User's 
Guide of the 2XMM catalogue of serendipitous sources available at\\
{\tt http://xmmssc-www.star.le.ac.uk/Catalogue/2XMMi-DR3}. {We note that our results are consistent within a few percent when relaxing the adopted 
assumptions.}} we assumed a power-law model with photon index $\Gamma=1.7$ and
mean Galactic foreground absorption $N_H\simeq 3.0\times 10^{20}$ cm$^{-2}$ (\citealt{watson2009}).

Finally, we purged the candidate source lists by requiring a maximum likelihood threshold equal to 10, 
corresponding to 4$\sigma$, and removing a few spurious sources {identified as false detections} 
or positioned at the borders of the cameras. {Our results are consistent 
with the 3XMMi-DR4 with minor differences due to the fact that we have excluded few spurious sources.} 
 
{At the end of the procedure}, we obtained 89, 116, 49 and 54 sources {for}  Draco, Leo I, UMa II and UMi, respectively
{Previously, \citet{orioproc} analyzed the same Leo I data set detecting 105 X-ray sources. 
We impute the discrepancy to slightly different choices in the data screening procedure and the detection threshold used.}

\begin{figure*}
     \begin{center}
        \subfigure[]{%
            \includegraphics[width=0.6\textwidth]{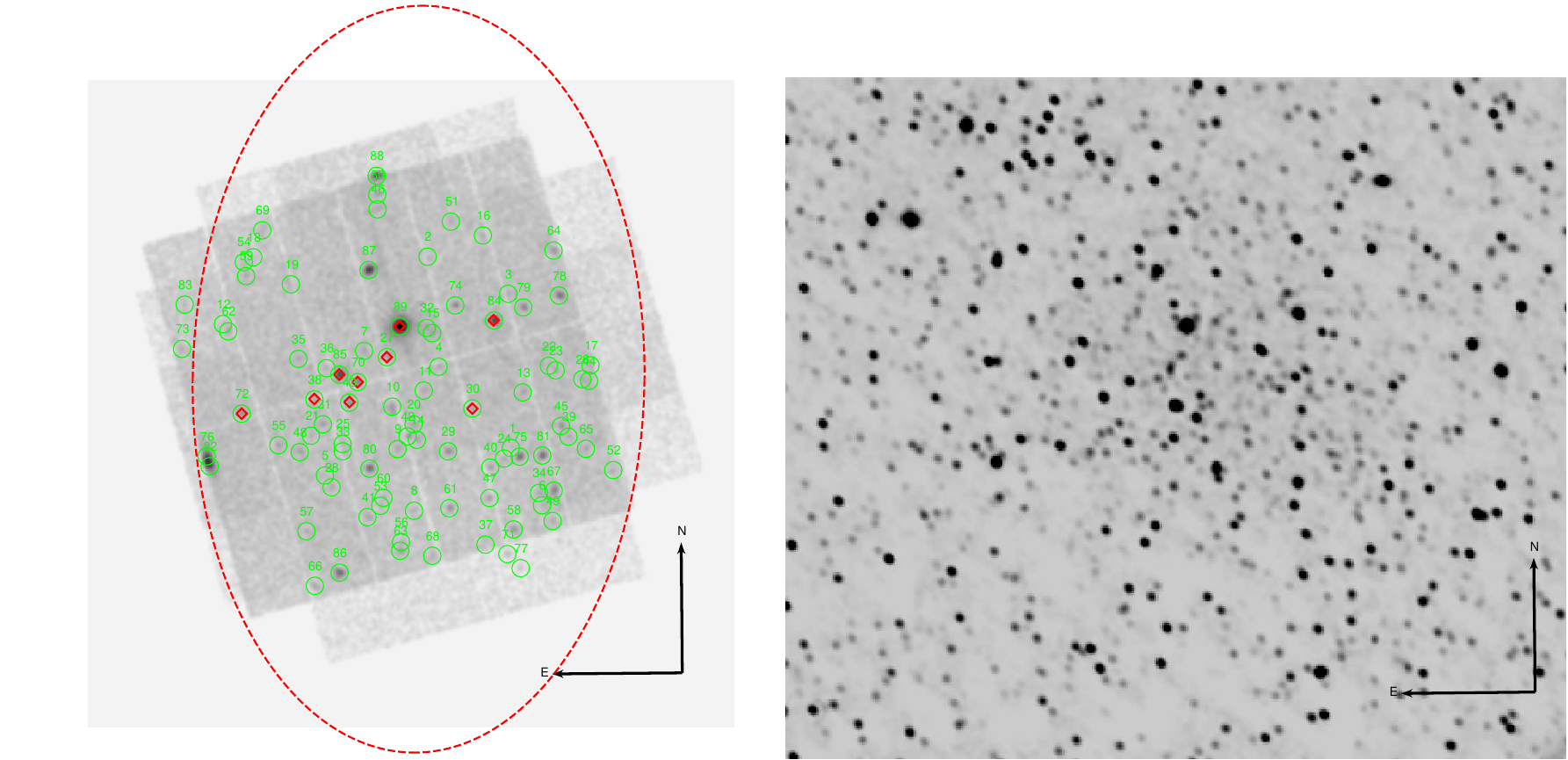}
        }\\%
        \subfigure[]{%
           \includegraphics[width=0.6\textwidth]{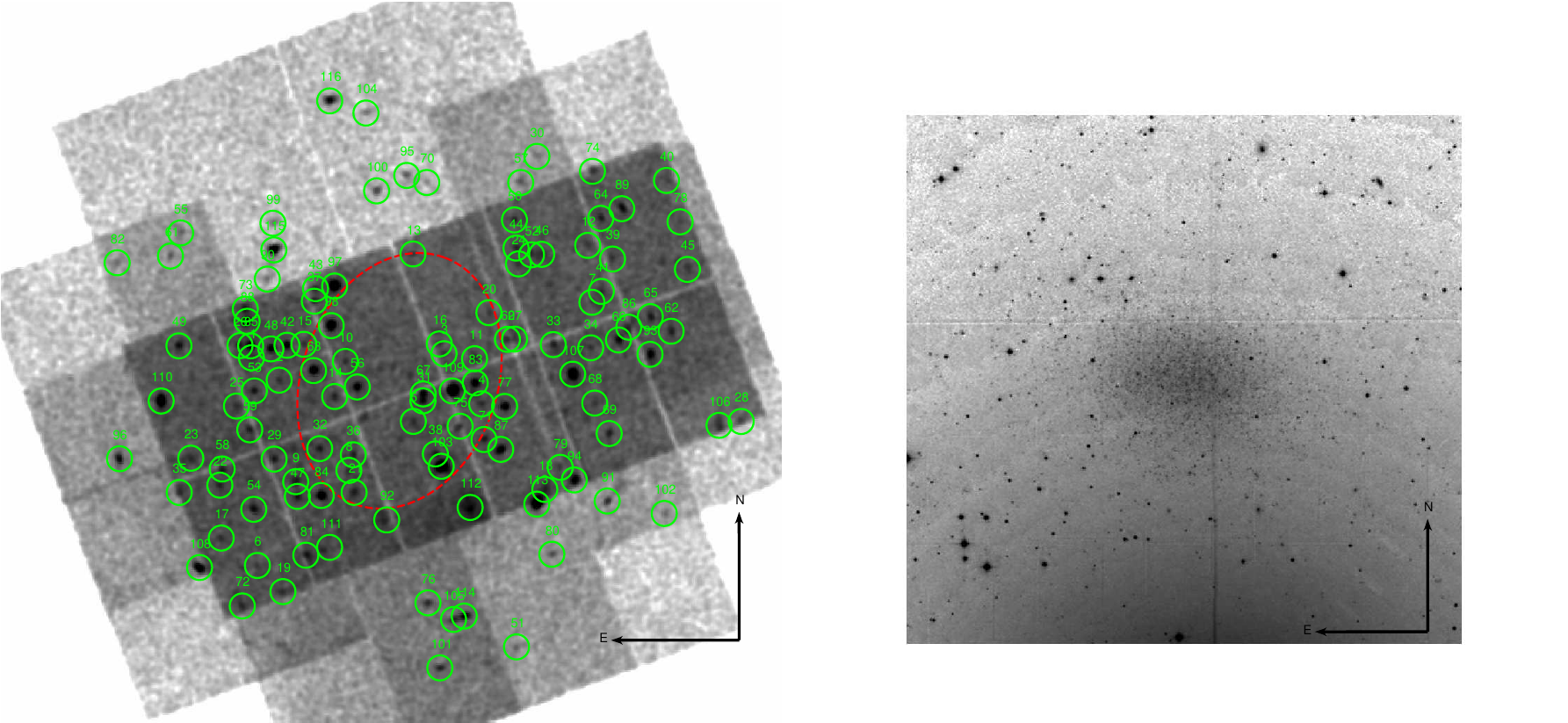}
        }\\
        \subfigure[]{%
            \includegraphics[width=0.6\textwidth]{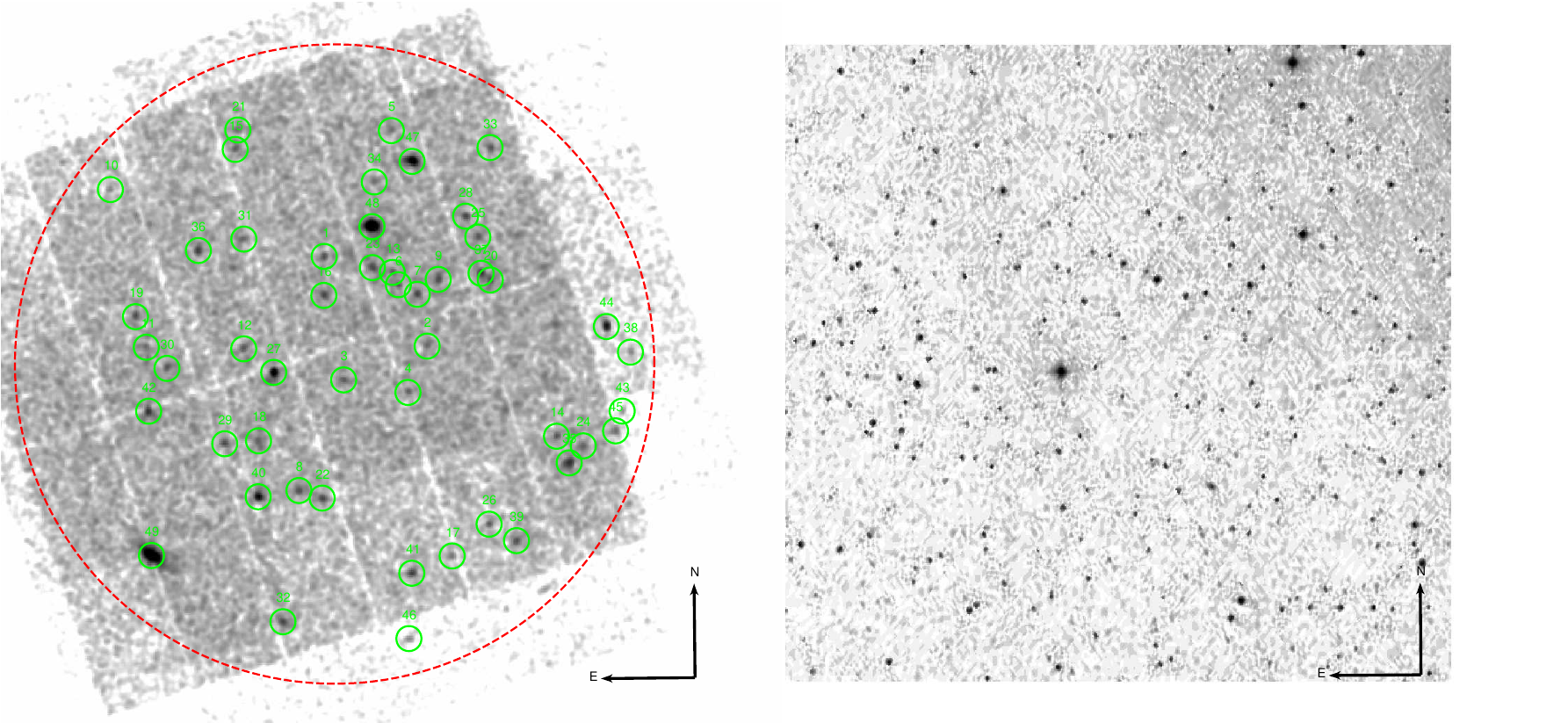}
        }\\%
        \subfigure[]{%
            \includegraphics[width=0.6\textwidth]{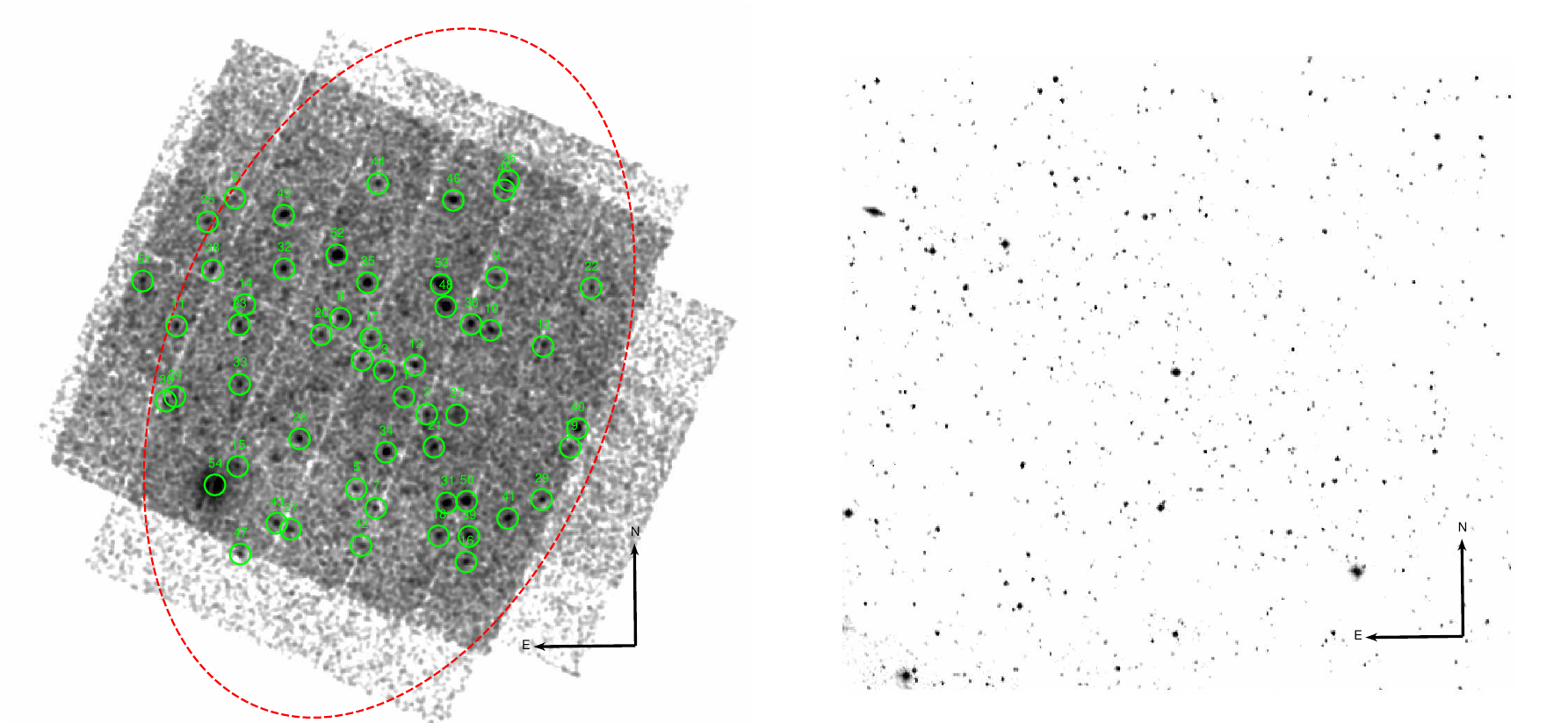}
        }\\%
    \end{center}
\caption{Mosaic images of the \sat\, MOS 1, MOS 2, and pn exposures in the 0.2-12.0 keV energy band of the sample galaxies.}
\label{images}
\end{figure*}

The logarithmically scaled 0.2-12 keV images (smoothed with a 3 pixel gaussian kernel) are shown in Fig. \ref{images}. 
{For Draco, Leo I and UMi the dashed circles represent the extension of the galaxies as suggested by the NASA/IPAC extragalactic database, while in the case of UMa II 
we used the half light radius taken from \citet{McConnachie2012}. In each panel, the  $35''$ radius circles containing $\simeq90\%$ of energy 
at 1.5 keV in the pn camera (see e.g., \citealt{xrps}) indicate the detected sources each of which is labeled with a sequential 
number following the 0.2-12.0 keV increasing flux order.} 

In Tables \ref{DracoSources} - \ref{UMiSources} we report the main results of our analysis, showing the detected sources 
in increasing flux order. For each source, the first two columns {show, respectively} an 
identification number and a label ($\#$) indicating how many times it was revealed in the \sat\ observations, 
i.e.: A - once, B - twice, C - three times, and so on. In Leo I and UMa II cases {there are} only A sources because we used a single observation.
Columns 3-5 {report} the J2000 coordinates with the associated errors. Column 6 reports 
the 0.2-12.0 keV absorbed  flux. 
To get the parameter $log(f_X / f_{opt})$ (column 7), the SIMBAD\footnote{http://simbad.u-strasbg.fr/simbad/}
archive was searched for correlations (within a radius of 3$''$) from which we extracted the relevant magnitude in the V band 
or, if not available, the average between R and B ones \citep{Bartlett2012}.
The high-energy hardness-ratios $HR^*_1$, $HR^*_2$ and $HR_i$ with $i=1,...,4$ (see next section)
are shown in the remaining columns.
In Tables \ref{Dracoclass} - \ref{UMiclass}  we present the correlation with catalogued counterparts, if any. 
For this purpose, we used several databases to correlate our X-ray source catalogue with optical counterparts,
{namely:}  the Two Micron All-Sky Survey (2MASS), 
the Two Micron All-Sky Survey Extended objects (2MASX, \citealt{2mass}),
the United States Naval Observatory all-sky survey (USNO-B1, \citealt{usnob1}),
the position and proper motions extended catalogues (PPMX, \citealt{roeser2008} and PPMXL, \citealt{roeser2010}),
the QSO candidates in the SDSS stripes examined (J/MNRAS/396/223/qsos, \citealt{abrusco2009}),
the candidate AGN objects catalogue (J/MNRAS/437/968, \citealt{cavuoti2014}).
{Regarding} Draco dSph, we also searched for  
long-period, semi-regular red variable stars, carbon stars and eclipsing binaries (J/AJ/136/1921/table9, \citealt{kinemuchi2008}),
QSOs found in the same survey (J/AJ/136/1921/table10, \citealt{kinemuchi2008}),
variable stars (J/AJ/127/861, \citealt{bonanos2004}), 
and late-type stars (J/A+A/442/165, \citealt{cioni2005}).
The quasar-galaxy associations catalogue (J/AZh/78/675, \citealt{bukhmastova2001}) was used for the analysis
of both Draco and UMi. 

In the Leo I case, we used the NIR catalogue 
obtained with the WFCAM wide-field array at the United
Kingdom Infrared Telescope
(J/MNRAS/404/1475/table2, \citealt{held2010}), while 
the Palomar Transient Factory (PTF) photometric catalogue 1.0 (II/313, \citealt{ofek2012}) was {compared}  
with UMa II source list.
Moreover, for UMi dSph we examined the catalogue of quasar candidates from non-parametric Bayes classifier kernel density estimate 
(J/ApJS/180/67, \citealt{richards2010}) and
the atlas of Radio/X-ray associations (V/134/arxa, \citealt{flesch2010}). 

{For details about the limiting magnitude of the used catalogues we remind to the relevant 
publications. Here, for example, we remind that 2MASS has limiting sensitivity of 17.1, 16.4 and 15.3 mag in J, H, and K bands, {respectively},
while the USNO-B1 catalogue is complete {down} to a visual magnitude of $\simeq 21$ and has postional accuracy of $\simeq 0.2''$. 
Note that some catalogues (as PPMX and PPMXL, or 223/qsos and 437/968) refer to other well known 
catalogues (as 2MASS and USNO-B1, or SDSS) and inherit the corresponding limiting magnitudes.}

Therefore, we associated to the coordinates of each of the identified X-ray sources an error resulting from the quadrature sum of the
{\it XMM}-Newton positional accuracy ($\simeq 2''$ at $2\sigma$ confidence level, see \citealt{kirsch2004} and \citealt{guainazzi2010})
and the statistical error as determined by the {\it edetect$\_$chain} tool. 
Since the resulting positional uncertainty is of {the order of} a few arcseconds, we do not over-plot the source error circles in 
any figure of this paper.
In the same way, the error associated with the optical counterpart was derived from the relevant catalogues.\\
Only if an X-ray source is found to be within $1\sigma$ (and $3$ arcsec) from a catalogue counterpart, 
we report the corresponding distance in arcseconds in Tables \ref{Dracoclass} - \ref{UMiclass}. 
{In {the} case of multiple sources satisfying \textst {the} previous condition, {the source having} 
the minimum distance {from} the X-ray target was used .}
We also show the object type (OType) category of the counterpart, as reported in the SIMBAD archive\footnote{The SIMBAD
 object classification (class standard designation, condensed one and extended explanation as well) is available at 
http://simbad.u-strasbg.fr/simbad/sim-display?data=otypes.}, as well as our own source classification (HR Class).

\onecolumn
\renewcommand{\thefootnote}{\fnsymbol{footnote}}
\renewcommand{\arraystretch}{0.81}
\scriptsize
\begin{longtable}{|c|c|c|c|c|c|c|c|c|c|c|c|c|}
\caption{The X-ray sources towards the Draco dSph as detectd by the $XMM$-Newton satellite. The full list is available on-line.}
\label{DracoSources}\\
\hline\hline 
   \multicolumn{1}{c}{\textbf{Src}} &
   \multicolumn{1}{c}{\textbf{\#}} &
   \multicolumn{1}{c}{\textbf{RA}} &
   \multicolumn{1}{c}{\textbf{DEC}} &
   \multicolumn{1}{c}{\textbf{ERR}} &
   \multicolumn{1}{c}{\textbf{F$_{0.2-12 keV}^{Abs}$}} &
   \multicolumn{1}{c}{\textbf{$log\left(\frac{f_X}{f_{opt}}\right)$}} &
   \multicolumn{1}{c}{\textbf{$HR^*_1$}} &
   \multicolumn{1}{c}{\textbf{$HR^*_2$}} &
   \multicolumn{1}{c}{\textbf{$HR_1$}} &
   \multicolumn{1}{c}{\textbf{$HR_2$}} &
   \multicolumn{1}{c}{\textbf{$HR_3$}} &
   \multicolumn{1}{c}{\textbf{$HR_4$}} \\
   \multicolumn{1}{c}{\textbf{}} &
   \multicolumn{1}{c}{\textbf{}} &
   \multicolumn{1}{c}{\textbf{(J2000)}} &
   \multicolumn{1}{c}{\textbf{(J2000)}} &
   \multicolumn{1}{c}{\textbf{arcsec}} &
   \multicolumn{1}{c}{\textbf{$\times 10^{-14}$ erg s$^{-1}$ cm$^{-2}$}} &
   \multicolumn{1}{c}{\textbf{}} &
   \multicolumn{1}{c}{\textbf{}} &
   \multicolumn{1}{c}{\textbf{}} &
   \multicolumn{1}{c}{\textbf{}} &
   \multicolumn{1}{c}{\textbf{}} &
   \multicolumn{1}{c}{\textbf{}} &
   \multicolumn{1}{c}{\textbf{}} \\
\hline\hline
\endfirsthead
\multicolumn{13}{c}{{\tablename} \thetable{} -- Continued}\\
\hline\hline
   \multicolumn{1}{c}{\textbf{Src}} &
   \multicolumn{1}{c}{\textbf{\#}} &
   \multicolumn{1}{c}{\textbf{RA}} &
   \multicolumn{1}{c}{\textbf{DEC}} &
   \multicolumn{1}{c}{\textbf{ERR}} &
   \multicolumn{1}{c}{\textbf{F$_{0.2-12 keV}^{Abs}$}} &
   \multicolumn{1}{c}{\textbf{$log\left(\frac{f_X}{f_{opt}}\right)$}} &
   \multicolumn{1}{c}{\textbf{$HR^*_1$}} &
   \multicolumn{1}{c}{\textbf{$HR^*_2$}} &
   \multicolumn{1}{c}{\textbf{$HR_1$}} &
   \multicolumn{1}{c}{\textbf{$HR_2$}} &
   \multicolumn{1}{c}{\textbf{$HR_3$}} &
   \multicolumn{1}{c}{\textbf{$HR_4$}} \\
   \multicolumn{1}{c}{\textbf{}} &
   \multicolumn{1}{c}{\textbf{}} &
   \multicolumn{1}{c}{\textbf{(J2000)}} &
   \multicolumn{1}{c}{\textbf{(J2000)}} &
   \multicolumn{1}{c}{\textbf{arcsec}} &
   \multicolumn{1}{c}{\textbf{$\times 10^{-14}$ erg s$^{-1}$ cm$^{-2}$}} &
   \multicolumn{1}{c}{\textbf{}} &
   \multicolumn{1}{c}{\textbf{}} &
   \multicolumn{1}{c}{\textbf{}} &
   \multicolumn{1}{c}{\textbf{}} &
   \multicolumn{1}{c}{\textbf{}} &
   \multicolumn{1}{c}{\textbf{}} &
   \multicolumn{1}{c}{\textbf{}} \\
\hline\hline 
\endhead
\multicolumn{13}{c}{{Continued on Next Page\ldots}} \\
\endfoot
\hline\hline
\endlastfoot
  1 & A & 17 19 26.1  &  57  50 18.7  & 2.546   &    $\le$ 1.21       &       &  0.08  $\pm$ 0.58  & -0.66 $\pm$ 0.67  & 0.71  $\pm$ 0.22  & -0.83 $\pm$ 0.20  & -1.00 $\pm$ 1.01  & 1.00  $\pm$ 0.64  \\
  2 & A & 17 20 07.8  &  58  03 5.8   & 2.558   &    $\le$ 1.09       & 0.131 &  -0.07 $\pm$ 0.64  & -0.26 $\pm$ 0.68  & 0.01  $\pm$ 0.30  & 0.29  $\pm$ 0.30  & -1.00 $\pm$ 0.22  & 1.00  $\pm$ 0.23  \\
  3 & A & 17 19 27.2  &  58  00 36.3  & 2.418   & 0.54   $\pm$ 0.45   &-1.314 &  0.28  $\pm$ 0.55  & -0.44 $\pm$ 0.56  & -0.11 $\pm$ 0.29  & -0.35 $\pm$ 0.29  & -0.76 $\pm$ 0.47  & 1.00  $\pm$ 0.14  \\
  4 & C & 17 20 02.3  &  57  55 44.5  & 2.510   & 0.55   $\pm$ 0.27   &       &  -0.08 $\pm$ 0.63  & -0.13 $\pm$ 0.69  & 0.18  $\pm$ 0.25  & 0.01  $\pm$ 0.25  & -0.24 $\pm$ 0.31  & -0.64 $\pm$ 0.37  \\
  5 & A & 17 20 59.4  &  57  48 28.7  & 2.567   &   $\le$ 1.29        &       &  -0.22 $\pm$ 0.73  & -0.09 $\pm$ 0.64  & -0.04 $\pm$ 0.20  & 0.27  $\pm$ 0.24  & -0.46 $\pm$ 0.27  & -0.94 $\pm$ 0.94  \\
  6 & A & 17 19 10.6  &  57  46 28.3  & 2.591   &   $\le$ 1.31        &       &  -0.13 $\pm$ 0.79  & -0.02 $\pm$ 0.59  & 0.37  $\pm$ 0.31  & 0.28  $\pm$ 0.25  & -0.42 $\pm$ 0.26  & -1.00 $\pm$ 0.93  \\
  7 & A & 17 20 39.8  &  57  56 48.2  & 2.427   & 0.66   $\pm$ 0.37   &       &  0.09  $\pm$ 0.65  & 0.23  $\pm$ 0.56  & -0.10 $\pm$ 0.31  & 0.68  $\pm$ 0.21  & -0.13 $\pm$ 0.25  & -0.23 $\pm$ 0.31  \\
  8 & A & 17 20 14.7  &  57  46 07.5  & 2.578   & 0.70   $\pm$ 0.62   &       &  -0.14 $\pm$ 0.68  & 0.14  $\pm$ 0.59  & 0.35  $\pm$ 0.31  & 0.28  $\pm$ 0.22  & -0.20 $\pm$ 0.23  & -1.00 $\pm$ 0.48  \\
  9 & B & 17 20 22.9  &  57  50 14.8  & 3.287   & 0.71   $\pm$ 0.33   &       &   0.00 $\pm$ 0.63  & 0.20  $\pm$ 0.57  & 0.67  $\pm$ 0.44  & 0.53  $\pm$ 0.23  & 0.01  $\pm$ 0.21  & -0.75 $\pm$ 0.21  \\
\end{longtable}
\normalsize
\renewcommand{\thefootnote}{\arabic{footnote}}
\renewcommand{\arraystretch}{1.0}
%
%
\renewcommand{\thefootnote}{\fnsymbol{footnote}}
\renewcommand{\arraystretch}{0.81}
\scriptsize
\begin{longtable}{|c|c|c|c|c|c|c|c|c|c|c|c|c|}
\caption{The X-ray sources towards the LeoI dSph as detectd by the $XMM$-Newton satellite. The full list is available on-line. }
\label{LeoISources}\\
\hline\hline 
   \multicolumn{1}{c}{\textbf{Src}} &
   \multicolumn{1}{c}{\textbf{\#}} &
   \multicolumn{1}{c}{\textbf{RA}} &
   \multicolumn{1}{c}{\textbf{DEC}} &
   \multicolumn{1}{c}{\textbf{ERR}} &
   \multicolumn{1}{c}{\textbf{F$_{0.2-12 keV}^{Abs}$}} &
   \multicolumn{1}{c}{\textbf{$log\left(\frac{f_X}{f_{opt}}\right)$}} &
   \multicolumn{1}{c}{\textbf{$HR^*_1$}} &
   \multicolumn{1}{c}{\textbf{$HR^*_2$}} &
   \multicolumn{1}{c}{\textbf{$HR_1$}} &
   \multicolumn{1}{c}{\textbf{$HR_2$}} &
   \multicolumn{1}{c}{\textbf{$HR_3$}} &
   \multicolumn{1}{c}{\textbf{$HR_4$}} \\
   \multicolumn{1}{c}{\textbf{}} &
   \multicolumn{1}{c}{\textbf{}} &
   \multicolumn{1}{c}{\textbf{(J2000)}} &
   \multicolumn{1}{c}{\textbf{(J2000)}} &
   \multicolumn{1}{c}{\textbf{arcsec}} &
   \multicolumn{1}{c}{\textbf{$\times 10^{-14}$ erg s$^{-1}$ cm$^{-2}$}} &
   \multicolumn{1}{c}{\textbf{}} &
   \multicolumn{1}{c}{\textbf{}} &
   \multicolumn{1}{c}{\textbf{}} &
   \multicolumn{1}{c}{\textbf{}} &
   \multicolumn{1}{c}{\textbf{}} &
   \multicolumn{1}{c}{\textbf{}} &
   \multicolumn{1}{c}{\textbf{}} \\
\hline\hline
\endfirsthead
\multicolumn{13}{c}{{\tablename} \thetable{} -- Continued}\\
\hline\hline
   \multicolumn{1}{c}{\textbf{Src}} &
   \multicolumn{1}{c}{\textbf{\#}} &
   \multicolumn{1}{c}{\textbf{RA}} &
   \multicolumn{1}{c}{\textbf{DEC}} &
   \multicolumn{1}{c}{\textbf{ERR}} &
   \multicolumn{1}{c}{\textbf{F$_{0.2-12 keV}^{Abs}$}} &
   \multicolumn{1}{c}{\textbf{$log\left(\frac{f_X}{f_{opt}}\right)$}} &
   \multicolumn{1}{c}{\textbf{$HR^*_1$}} &
   \multicolumn{1}{c}{\textbf{$HR^*_2$}} &
   \multicolumn{1}{c}{\textbf{$HR_1$}} &
   \multicolumn{1}{c}{\textbf{$HR_2$}} &
   \multicolumn{1}{c}{\textbf{$HR_3$}} &
   \multicolumn{1}{c}{\textbf{$HR_4$}} \\
   \multicolumn{1}{c}{\textbf{}} &
   \multicolumn{1}{c}{\textbf{}} &
   \multicolumn{1}{c}{\textbf{(J2000)}} &
   \multicolumn{1}{c}{\textbf{(J2000)}} &
   \multicolumn{1}{c}{\textbf{arcsec}} &
   \multicolumn{1}{c}{\textbf{$\times 10^{-14}$ erg s$^{-1}$ cm$^{-2}$}} &
   \multicolumn{1}{c}{\textbf{}} &
   \multicolumn{1}{c}{\textbf{}} &
   \multicolumn{1}{c}{\textbf{}} &
   \multicolumn{1}{c}{\textbf{}} &
   \multicolumn{1}{c}{\textbf{}} &
   \multicolumn{1}{c}{\textbf{}} &
   \multicolumn{1}{c}{\textbf{}} \\
\hline\hline 
\endhead
\multicolumn{13}{c}{{Continued on Next Page\ldots}} \\
\endfoot
\hline\hline
\endlastfoot
  1 & A &  10 08 56.2  &  12 19 27.2  & 2.480 & 0.15  $\pm$ 0.14 &        & -0.16 $\pm$ 0.60 & -0.01 $\pm$ 0.49 & 0.27  $\pm$ 0.26  & 0.09  $\pm$ 0.22  & -0.76 $\pm$ 0.24  & 0.42  $\pm$ 0.38  \\
  2 & A &  10 08 50.9  &  12 18 24.8  & 2.403 & 0.20  $\pm$ 0.15 &        & -0.39 $\pm$ 0.54 & -0.03 $\pm$ 0.49 & 0.20  $\pm$ 0.22  & 0.22  $\pm$ 0.17  & -0.95 $\pm$ 0.14  & 0.76  $\pm$ 1.14  \\
  3 & A &  10 08 19.5  &  12 19 38.6  & 2.344 & 0.21  $\pm$ 0.13 & -0.870 & -0.12 $\pm$ 0.59 & -0.17 $\pm$ 0.57 & -0.03 $\pm$ 0.22  & -0.03 $\pm$ 0.24  & -0.40 $\pm$ 0.33  & 0.06  $\pm$ 0.51  \\
  4 & A &  10 08 12.4  &  12 17 18.7  & 2.358 & 0.21  $\pm$ 0.14 &        & -0.32 $\pm$ 0.53 &  0.04 $\pm$ 0.46 & 0.63  $\pm$ 0.17  & -0.08 $\pm$ 0.16  & -0.35 $\pm$ 0.21  & -0.99 $\pm$ 0.71  \\
  5 & A &  10 08 25.4  &  12 16 29.9  & 2.358 & 0.22  $\pm$ 0.11 &        & -0.15 $\pm$ 0.37 & -0.53 $\pm$ 0.46 & 0.27  $\pm$ 0.16  & -0.44 $\pm$ 0.18  & -0.86 $\pm$ 0.35  & 0.34  $\pm$ 0.85  \\
  6 & A &  10 08 55.0  &  12 09 47.9  & 2.475 &       $\le$ 0.50 &        & -0.06 $\pm$ 0.77 & -0.52 $\pm$ 0.77 & -0.22 $\pm$ 0.20  & -0.19 $\pm$ 0.26  & -0.22 $\pm$ 0.40  & -1.00 $\pm$ 1.39  \\
  7 & A &  10 07 51.4  &  12 22 01.9  & 2.408 &       $\le$ 0.58 &        &  0.08 $\pm$ 0.77 &  0.02 $\pm$ 0.54 & -0.98 $\pm$ 0.12  & 0.98  $\pm$ 0.16  & -0.20 $\pm$ 0.30  & 0.09  $\pm$ 0.34  \\
  8 & A &  10 08 37.6  &  12 14 13.2  & 2.350 & 0.26  $\pm$ 0.17 &        & -0.18 $\pm$ 0.52 & -0.05 $\pm$ 0.45 & 0.26  $\pm$ 0.22  & 0.11  $\pm$ 0.20  & -0.47 $\pm$ 0.24  & 0.33  $\pm$ 0.51  \\
  9 & A &  10 08 47.8  &  12 13 41.6  & 2.651 & 0.30  $\pm$ 0.12 &        & -0.14 $\pm$ 0.38 & -0.08 $\pm$ 0.36 & 0.49  $\pm$ 0.18  & 0.06  $\pm$ 0.15  & -0.28 $\pm$ 0.18  & -0.99 $\pm$ 0.40  \\
 10 & A &  10 08 38.3  &  12 19 17.1  & 2.321 & 0.33  $\pm$ 0.15 &        & -0.21 $\pm$ 0.52 & -0.12 $\pm$ 0.46 & -0.09 $\pm$ 0.17  & 0.32  $\pm$ 0.19  & -0.94 $\pm$ 0.16  & 0.81  $\pm$ 0.50  \\
\end{longtable}
\normalsize
\renewcommand{\thefootnote}{\arabic{footnote}}
\renewcommand{\arraystretch}{1.0}
%
\renewcommand{\thefootnote}{\fnsymbol{footnote}}
\renewcommand{\arraystretch}{0.81}
\scriptsize
\begin{longtable}{|c|c|c|c|c|c|c|c|c|c|c|c|c|}
\caption{The X-ray sources towards the UMaII dSph as detected by the $XMM$-Newton satellite. The full list is available on-line.}
 \label{UMaIISources} \\
\hline\hline 
   \multicolumn{1}{c}{\textbf{Src}} &
   \multicolumn{1}{c}{\textbf{\#}} &
   \multicolumn{1}{c}{\textbf{RA}} &
   \multicolumn{1}{c}{\textbf{DEC}} &
   \multicolumn{1}{c}{\textbf{ERR}} &
   \multicolumn{1}{c}{\textbf{F$_{0.2-12 keV}^{Abs}$}} &
   \multicolumn{1}{c}{\textbf{$log\left(\frac{f_X}{f_{opt}}\right)$}} &
   \multicolumn{1}{c}{\textbf{$HR^*_1$}} &
   \multicolumn{1}{c}{\textbf{$HR^*_2$}} &
   \multicolumn{1}{c}{\textbf{$HR_1$}} &
   \multicolumn{1}{c}{\textbf{$HR_2$}} &
   \multicolumn{1}{c}{\textbf{$HR_3$}} &
   \multicolumn{1}{c}{\textbf{$HR_4$}} \\
   \multicolumn{1}{c}{\textbf{}} &
   \multicolumn{1}{c}{\textbf{}} &
   \multicolumn{1}{c}{\textbf{(J2000)}} &
   \multicolumn{1}{c}{\textbf{(J2000)}} &
   \multicolumn{1}{c}{\textbf{arcsec}} &
   \multicolumn{1}{c}{\textbf{$\times 10^{-14}$ erg s$^{-1}$ cm$^{-2}$}} &
   \multicolumn{1}{c}{\textbf{}} &
   \multicolumn{1}{c}{\textbf{}} &
   \multicolumn{1}{c}{\textbf{}} &
   \multicolumn{1}{c}{\textbf{}} &
   \multicolumn{1}{c}{\textbf{}} &
   \multicolumn{1}{c}{\textbf{}} &
   \multicolumn{1}{c}{\textbf{}} \\
\hline\hline
\endfirsthead
\multicolumn{13}{c}{{\tablename} \thetable{} -- Continued}\\
\hline\hline
   \multicolumn{1}{c}{\textbf{Src}} &
   \multicolumn{1}{c}{\textbf{\#}} &
   \multicolumn{1}{c}{\textbf{RA}} &
   \multicolumn{1}{c}{\textbf{DEC}} &
   \multicolumn{1}{c}{\textbf{ERR}} &
   \multicolumn{1}{c}{\textbf{F$_{0.2-12 keV}^{Abs}$}} &
   \multicolumn{1}{c}{\textbf{$log\left(\frac{f_X}{f_{opt}}\right)$}} &
   \multicolumn{1}{c}{\textbf{$HR^*_1$}} &
   \multicolumn{1}{c}{\textbf{$HR^*_2$}} &
   \multicolumn{1}{c}{\textbf{$HR_1$}} &
   \multicolumn{1}{c}{\textbf{$HR_2$}} &
   \multicolumn{1}{c}{\textbf{$HR_3$}} &
   \multicolumn{1}{c}{\textbf{$HR_4$}} \\
   \multicolumn{1}{c}{\textbf{}} &
   \multicolumn{1}{c}{\textbf{}} &
   \multicolumn{1}{c}{\textbf{(J2000)}} &
   \multicolumn{1}{c}{\textbf{(J2000)}} &
   \multicolumn{1}{c}{\textbf{arcsec}} &
   \multicolumn{1}{c}{\textbf{$\times 10^{-14}$ erg s$^{-1}$ cm$^{-2}$}} &
   \multicolumn{1}{c}{\textbf{}} &
   \multicolumn{1}{c}{\textbf{}} &
   \multicolumn{1}{c}{\textbf{}} &
   \multicolumn{1}{c}{\textbf{}} &
   \multicolumn{1}{c}{\textbf{}} &
   \multicolumn{1}{c}{\textbf{}} &
   \multicolumn{1}{c}{\textbf{}} \\
\hline\hline 
\endhead
\multicolumn{13}{c}{{Continued on Next Page\ldots}} \\
\endfoot
\hline\hline
\endlastfoot
  1 & A &  08 51 34.3  &  63 12 48.2  & 2.414 & 0.67   $\pm$ 0.48  &       & -0.24 $\pm$  0.41 & -0.45 $\pm$  0.50 & 0.58  $\pm$ 0.17  & -0.38 $\pm$ 0.17  & -0.90 $\pm$ 0.23  & -1.00 $\pm$ 2.82  \\
  2 & A &  08 50 51.9  &  63 08 37.6  & 2.578 & 0.68   $\pm$ 0.25  &       & -0.08 $\pm$  0.65 &  0.30 $\pm$  0.65 & 0.15  $\pm$ 0.42  & 0.64  $\pm$ 0.23  & -0.16 $\pm$ 0.23  & -1.00 $\pm$ 0.19  \\
  3 & A &  08 51 26.3  &  63 07 03.7  & 2.431 & 0.69   $\pm$ 0.32  &       & -0.08 $\pm$  0.52 & -0.08 $\pm$  0.55 & 0.56  $\pm$ 0.27  & 0.28  $\pm$ 0.22  & -0.32 $\pm$ 0.22  & -0.38 $\pm$ 0.35  \\
  4 & A &  08 50 59.9  &  63 06 28.6  & 2.496 & 0.80   $\pm$ 0.48  &       & -0.23 $\pm$  0.71 &  0.22 $\pm$  0.63 & 0.25  $\pm$ 0.39  & 0.54  $\pm$ 0.24  & -0.47 $\pm$ 0.22  & -0.93 $\pm$ 0.49  \\
  5 & A &  08 51 06.5  &  63 18 38.9  & 2.708 & 0.83   $\pm$ 0.56  &       & -0.31 $\pm$  0.51 & -0.24 $\pm$  0.59 & 0.67  $\pm$ 0.19  & -0.17 $\pm$ 0.22  & -0.90 $\pm$ 0.15  & -1.00 $\pm$ 0.73  \\
  6 & A &  08 51 03.7  &  63 11 29.1  & 2.449 & 0.87   $\pm$ 0.32  &       & -0.30 $\pm$  0.57 &  0.06 $\pm$  0.56 & 0.53  $\pm$ 0.28  & 0.15  $\pm$ 0.20  & -0.41 $\pm$ 0.24  & -0.89 $\pm$ 0.30  \\
  7 & A &  08 50 55.9  &  63 11 02.2  & 2.246 & 1.01   $\pm$ 0.37  &       & -0.21 $\pm$  0.32 & -0.44 $\pm$  0.41 & 0.48  $\pm$ 0.14  & -0.44 $\pm$ 0.15  & -0.85 $\pm$ 0.25  & -0.29 $\pm$ 0.82  \\
  8 & A &  08 51 44.6  &  63 01 55.2  & 2.303 & 1.18   $\pm$ 0.51  &       & -0.29 $\pm$  0.39 & -0.19 $\pm$  0.45 & 0.43  $\pm$ 0.17  & -0.04 $\pm$ 0.17  & -0.81 $\pm$ 0.17  & 0.62  $\pm$ 0.45  \\
  9 & A &  08 50 47.3  &  63 11 43.6  & 2.323 & 1.20   $\pm$ 0.50  &       & -0.21 $\pm$  0.41 & -0.08 $\pm$  0.43 & 0.14  $\pm$ 0.23  & 0.34  $\pm$ 0.16  & -0.66 $\pm$ 0.16  & -0.49 $\pm$ 0.46  \\
 10 & A &  08 53 02.7  &  63 15 53.0  & 2.505 &        $\le$ 3.14  &       & -0.10 $\pm$  0.50 & -0.46 $\pm$  0.59 & 0.51  $\pm$ 0.27  & -0.21 $\pm$ 0.27  & -0.70 $\pm$ 0.27  & -1.00 $\pm$ 0.92  \\
\end{longtable}
\normalsize
\renewcommand{\thefootnote}{\arabic{footnote}}
\renewcommand{\arraystretch}{1.0}
%
%
\renewcommand{\thefootnote}{\fnsymbol{footnote}}
\renewcommand{\arraystretch}{0.81}
\scriptsize
\begin{longtable}{|c|c|c|c|c|c|c|c|c|c|c|c|c|}
\caption{The X-ray sources towards the UMi dSph as detectd by the $XMM$-Newton satellite. The full list is available on-line.}
\label{UMiSources}\\
\hline\hline 
   \multicolumn{1}{c}{\textbf{Src}} &
   \multicolumn{1}{c}{\textbf{\#}} &
   \multicolumn{1}{c}{\textbf{RA}} &
   \multicolumn{1}{c}{\textbf{DEC}} &
   \multicolumn{1}{c}{\textbf{ERR}} &
   \multicolumn{1}{c}{\textbf{F$_{0.2-12 keV}^{Abs}$}} &
   \multicolumn{1}{c}{\textbf{$log\left(\frac{f_X}{f_{opt}}\right)$}} &
   \multicolumn{1}{c}{\textbf{$HR^*_1$}} &
   \multicolumn{1}{c}{\textbf{$HR^*_2$}} &
   \multicolumn{1}{c}{\textbf{$HR_1$}} &
   \multicolumn{1}{c}{\textbf{$HR_2$}} &
   \multicolumn{1}{c}{\textbf{$HR_3$}} &
   \multicolumn{1}{c}{\textbf{$HR_4$}} \\
   \multicolumn{1}{c}{\textbf{}} &
   \multicolumn{1}{c}{\textbf{}} &
   \multicolumn{1}{c}{\textbf{(J2000)}} &
   \multicolumn{1}{c}{\textbf{(J2000)}} &
   \multicolumn{1}{c}{\textbf{arcsec}} &
   \multicolumn{1}{c}{\textbf{$\times 10^{-14}$ erg s$^{-1}$ cm$^{-2}$}} &
   \multicolumn{1}{c}{\textbf{}} &
   \multicolumn{1}{c}{\textbf{}} &
   \multicolumn{1}{c}{\textbf{}} &
   \multicolumn{1}{c}{\textbf{}} &
   \multicolumn{1}{c}{\textbf{}} &
   \multicolumn{1}{c}{\textbf{}} &
   \multicolumn{1}{c}{\textbf{}} \\
\hline\hline
\endfirsthead
\multicolumn{13}{c}{{\tablename} \thetable{} -- Continued}\\
\hline\hline
   \multicolumn{1}{c}{\textbf{Src}} &
   \multicolumn{1}{c}{\textbf{\#}} &
   \multicolumn{1}{c}{\textbf{RA}} &
   \multicolumn{1}{c}{\textbf{DEC}} &
   \multicolumn{1}{c}{\textbf{ERR}} &
   \multicolumn{1}{c}{\textbf{F$_{0.2-12 keV}^{Abs}$}} &
   \multicolumn{1}{c}{\textbf{$log\left(\frac{f_X}{f_{opt}}\right)$}} &
   \multicolumn{1}{c}{\textbf{$HR^*_1$}} &
   \multicolumn{1}{c}{\textbf{$HR^*_2$}} &
   \multicolumn{1}{c}{\textbf{$HR_1$}} &
   \multicolumn{1}{c}{\textbf{$HR_2$}} &
   \multicolumn{1}{c}{\textbf{$HR_3$}} &
   \multicolumn{1}{c}{\textbf{$HR_4$}} \\
   \multicolumn{1}{c}{\textbf{}} &
   \multicolumn{1}{c}{\textbf{}} &
   \multicolumn{1}{c}{\textbf{(J2000)}} &
   \multicolumn{1}{c}{\textbf{(J2000)}} &
   \multicolumn{1}{c}{\textbf{arcsec}} &
   \multicolumn{1}{c}{\textbf{$\times 10^{-14}$ erg s$^{-1}$ cm$^{-2}$}} &
   \multicolumn{1}{c}{\textbf{}} &
   \multicolumn{1}{c}{\textbf{}} &
   \multicolumn{1}{c}{\textbf{}} &
   \multicolumn{1}{c}{\textbf{}} &
   \multicolumn{1}{c}{\textbf{}} &
   \multicolumn{1}{c}{\textbf{}} &
   \multicolumn{1}{c}{\textbf{}} \\
\hline\hline 
\endhead
\multicolumn{13}{c}{{Continued on Next Page\ldots}} \\
\endfoot
\hline\hline
\endlastfoot
  1 & A &  15 09 01.8  &  67 11 31.9  & 2.582   & 0.59   $\pm$ 0.34 & -0.923 & -0.32 $\pm$ 0.66 &  0.16 $\pm$ 0.70 &  -0.20 $\pm$ 0.35  & 0.38  $\pm$ 0.28  & -0.38 $\pm$ 0.24  & -1.00 $\pm$ 0.31  \\
  2 & A &  15 08 48.6  &  67 10 34.6  & 2.708   & 0.72   $\pm$ 0.59 &        & -0.34 $\pm$ 0.79 &  0.20 $\pm$ 0.83 &  0.17  $\pm$ 0.44  & 0.37  $\pm$ 0.28  & -0.28 $\pm$ 0.23  & -0.66 $\pm$ 0.40  \\
  3 & A &  15 09 13.1  &  67 12 59.4  & 3.052   & 0.73   $\pm$ 0.29 &        & -0.18 $\pm$ 0.52 & -0.33 $\pm$ 0.67 &  0.42  $\pm$ 0.24  & -0.33 $\pm$ 0.24  & -0.66 $\pm$ 0.19  & -1.00 $\pm$ 0.29  \\
  4 & A &  15 09 25.9  &  67 13 35.3  & 2.477   & 0.85   $\pm$ 0.45 &        & -0.25 $\pm$ 0.73 &  0.27 $\pm$ 0.71 &  0.71  $\pm$ 0.23  & 0.37  $\pm$ 0.26  & -0.51 $\pm$ 0.24  & -0.17 $\pm$ 0.38  \\
  5 & A &  15 09 29.2  &  67 06 24.6  & 2.779   & 0.88   $\pm$ 0.64 & -0.402 & -0.19 $\pm$ 0.45 & -0.23 $\pm$ 0.54 &  0.24  $\pm$ 0.31  & -0.01 $\pm$ 0.28  & -0.39 $\pm$ 0.32  & -1.00 $\pm$ 1.03  \\
  6 & A &  15 10 39.6  &  67 22 36.6  & 2.912   &        $\le$ 3.59 &  0.717 & -0.01 $\pm$ 0.98 & -0.49 $\pm$ 0.96 &  -0.01 $\pm$ 0.29  & -0.32 $\pm$ 0.39  & -0.09 $\pm$ 0.90  & -1.00 $\pm$ 2.72  \\
  7 & A &  15 09 18.0  &  67 05 20.3  & 2.607   & 0.98   $\pm$ 0.67 &        & -0.06 $\pm$ 0.33 & -0.67 $\pm$ 0.45 &  0.51  $\pm$ 0.22  & -0.74 $\pm$ 0.17  & -0.29 $\pm$ 0.58  & -1.00 $\pm$ 1.63  \\
  8 & A &  15 09 38.4  &  67 15 56.0  & 2.414   & 1.12   $\pm$ 0.51 &  0.507 & -0.42 $\pm$ 0.58 &  0.28 $\pm$ 0.56 &  -0.01 $\pm$ 0.24  & 0.61  $\pm$ 0.15  & -0.76 $\pm$ 0.18  & -0.49 $\pm$ 0.47  \\
  9 & A &  15 08 08.1  &  67 18 10.8  & 2.573   & 1.17   $\pm$ 0.89 &  0.478 & -0.14 $\pm$ 0.55 & -0.39 $\pm$ 0.63 &  0.11  $\pm$ 0.24  & -0.01 $\pm$ 0.25  & -0.78 $\pm$ 0.26  & 0.62  $\pm$ 0.53  \\
 10 & A &  15 08 11.7  &  67 15 14.4  & 2.339   & 1.24   $\pm$ 0.82 &  0.407 & -0.19 $\pm$ 0.47 & -0.26 $\pm$ 0.49 &  0.13  $\pm$ 0.20  & -0.04 $\pm$ 0.20  & -0.36 $\pm$ 0.23  & -1.00 $\pm$ 0.82  \\
\end{longtable}
\normalsize
\renewcommand{\thefootnote}{\arabic{footnote}}
\renewcommand{\arraystretch}{1.0}

\nopagebreak

\begin{landscape}
\renewcommand{\thefootnote}{\fnsymbol{footnote}}
\renewcommand{\arraystretch}{0.81}
\scriptsize
\begin{longtable}{|c|c|c|c|c|c|c|c|c|c|c|c|c|c|c|c|c|}
\caption{The correlation of the X-ray sources towards the Draco dSph. The full list is available on-line.}
\label{Dracoclass}\\
\hline\hline 
   \multicolumn{1}{c}{\textbf{Src}} &
   \multicolumn{1}{c}{\textbf{RA}} &
   \multicolumn{1}{c}{\textbf{DEC}} &
   \multicolumn{1}{c}{\textbf{223 qsos}} &
   \multicolumn{1}{c}{\textbf{2MASS}} &
   \multicolumn{1}{c}{\textbf{2MASX}} &
   \multicolumn{1}{c}{\textbf{PPMX}} &
   \multicolumn{1}{c}{\textbf{PPMXL}} &
   \multicolumn{1}{c}{\textbf{USNO-B1}} &
   \multicolumn{1}{c}{\textbf{table9}} &
   \multicolumn{1}{c}{\textbf{table10}} &
   \multicolumn{1}{c}{\textbf{437 968}} &
   \multicolumn{1}{c}{\textbf{127 861}} &
   \multicolumn{1}{c}{\textbf{78 675}} &
   \multicolumn{1}{c}{\textbf{442 165}} &
   \multicolumn{1}{c}{\textbf{SIMBAD}} &
   \multicolumn{1}{c}{\textbf{HR}} \\
   \multicolumn{1}{c}{\textbf{}} &
   \multicolumn{1}{c}{\textbf{(J2000)}} &
   \multicolumn{1}{c}{\textbf{(J2000)}} &
   \multicolumn{1}{c}{\textbf{(arcsec)}} &
   \multicolumn{1}{c}{\textbf{(arcsec)}} &
   \multicolumn{1}{c}{\textbf{(arcsec)}} &
   \multicolumn{1}{c}{\textbf{(arcsec)}} &
   \multicolumn{1}{c}{\textbf{(arcsec)}} &
   \multicolumn{1}{c}{\textbf{(arcsec)}} &
   \multicolumn{1}{c}{\textbf{(arcsec)}} &
   \multicolumn{1}{c}{\textbf{(arcsec)}} &
   \multicolumn{1}{c}{\textbf{(arcsec)}} &
   \multicolumn{1}{c}{\textbf{(arcsec)}} &
   \multicolumn{1}{c}{\textbf{(arcsec)}} &
   \multicolumn{1}{c}{\textbf{(arcsec)}} &
    \multicolumn{1}{c}{\textbf{OType}} &
    \multicolumn{1}{c}{\textbf{Class}} \\
\hline\hline
\endfirsthead
\multicolumn{17}{c}{{\tablename} \thetable{} -- Continued}\\
\hline\hline
   \multicolumn{1}{c}{\textbf{Src}} &
   \multicolumn{1}{c}{\textbf{RA}} &
   \multicolumn{1}{c}{\textbf{DEC}} &
   \multicolumn{1}{c}{\textbf{223 qsos}} &
   \multicolumn{1}{c}{\textbf{2MASS}} &
   \multicolumn{1}{c}{\textbf{2MASX}} &
   \multicolumn{1}{c}{\textbf{PPMX}} &
   \multicolumn{1}{c}{\textbf{PPMXL}} &
   \multicolumn{1}{c}{\textbf{USNO-B1}} &
   \multicolumn{1}{c}{\textbf{table9}} &
   \multicolumn{1}{c}{\textbf{table10}} &
   \multicolumn{1}{c}{\textbf{437 968}} &
   \multicolumn{1}{c}{\textbf{127 861}} &
   \multicolumn{1}{c}{\textbf{78 675}} &
   \multicolumn{1}{c}{\textbf{442 165}} &
   \multicolumn{1}{c}{\textbf{SIMBAD}} &
   \multicolumn{1}{c}{\textbf{HR}} \\
   \multicolumn{1}{c}{\textbf{}} &
   \multicolumn{1}{c}{\textbf{(J2000)}} &
   \multicolumn{1}{c}{\textbf{(J2000)}} &
   \multicolumn{1}{c}{\textbf{(arcsec)}} &
   \multicolumn{1}{c}{\textbf{(arcsec)}} &
   \multicolumn{1}{c}{\textbf{(arcsec)}} &
   \multicolumn{1}{c}{\textbf{(arcsec)}} &
   \multicolumn{1}{c}{\textbf{(arcsec)}} &
   \multicolumn{1}{c}{\textbf{(arcsec)}} &
   \multicolumn{1}{c}{\textbf{(arcsec)}} &
   \multicolumn{1}{c}{\textbf{(arcsec)}} &
   \multicolumn{1}{c}{\textbf{(arcsec)}} &
   \multicolumn{1}{c}{\textbf{(arcsec)}} &
   \multicolumn{1}{c}{\textbf{(arcsec)}} &
   \multicolumn{1}{c}{\textbf{(arcsec)}} &
    \multicolumn{1}{c}{\textbf{OType}} &
    \multicolumn{1}{c}{\textbf{Class}} \\
\hline\hline 
\endhead
\multicolumn{17}{c}{{Continued on Next Page\ldots}} \\
\endfoot
\hline\hline
\endlastfoot
  1 &  17 19 26.1  &  57  50 18.7&       & 1.46 &      & 1.46 & 1.46 &      &      &      &      &      &      &       &    & f  \\
  2 &  17 20 07.8  &  58  03 5.8 &       &      &      &      & 0.13 & 0.58 &      &      &      &      &      &       & *  & fg \\
  3 &  17 19 27.2  &  58  00 36.3&       & 1.86 &      &      & 1.82 & 1.65 &      &      &      &      &      &       & *  & FG  \\
  4 &  17 20 02.3  &  57  55 44.5&       &      &      &      &      &      &      &      &      &      &      &       &    &   \\
  5 &  17 20 59.4  &  57  48 28.7&       &      &      &      &      &      &      &      &      &      &      &       &    & f  \\
  6 &  17 19 10.6  &  57  46 28.3&       &      &      &      &      &      &      &      &      &      &      &       &    & f  \\
  7 &  17 20 39.8  &  57  56 48.2&       &      &      &      &      &      &      &      &      &      &      &       &    & h  \\
  8 &  17 20 14.7  &  57  46 07.5&       &      &      &      &      &      &      &      &      &      &      &       &    & h  \\
  9 &  17 20 22.9  &  57  50 14.8&       &      &      &      &      &      &      &      &      &      &      &       &    & h  \\
 10 &  17 20 25.6  &  57  53 04.7&       & 0.32 &      & 0.37 & 0.37 & 0.38 &      &      & 0.40 &      &      &       & *  & FG  \\
\end{longtable}
\normalsize
\renewcommand{\thefootnote}{\arabic{footnote}}
\renewcommand{\arraystretch}{1.0}


\nopagebreak

\renewcommand{\thefootnote}{\fnsymbol{footnote}}
\renewcommand{\arraystretch}{0.81}
\scriptsize
\begin{longtable}{|c|c|c|c|c|c|c|c|c|c|c|c|}
\caption{The correlation of the X-ray sources towards the LeoI dSph. The full list is available on-line.}
\label{LeoIclass}\\
\hline\hline 
   \multicolumn{1}{c}{\textbf{Src}} &
   \multicolumn{1}{c}{\textbf{RA}} &
   \multicolumn{1}{c}{\textbf{DEC}} &
   \multicolumn{1}{c}{\textbf{223 qsos}} &
   \multicolumn{1}{c}{\textbf{2MASS}} &
   \multicolumn{1}{c}{\textbf{PPMX}} &
   \multicolumn{1}{c}{\textbf{PPMXL}} &
   \multicolumn{1}{c}{\textbf{USNO-B1}} &
   \multicolumn{1}{c}{\textbf{437 968}} &
   \multicolumn{1}{c}{\textbf{1475/table2}} &
   \multicolumn{1}{c}{\textbf{SIMBAD}} &
   \multicolumn{1}{c}{\textbf{HR}} \\
   \multicolumn{1}{c}{\textbf{ }} &
   \multicolumn{1}{c}{\textbf{(J2000)}} &
   \multicolumn{1}{c}{\textbf{(J2000)}} &
   \multicolumn{1}{c}{\textbf{(arcsec)}} &
   \multicolumn{1}{c}{\textbf{(arcsec)}} &
   \multicolumn{1}{c}{\textbf{(arcsec)}} &
   \multicolumn{1}{c}{\textbf{(arcsec)}} &
   \multicolumn{1}{c}{\textbf{(arcsec)}} &
   \multicolumn{1}{c}{\textbf{(arcsec)}} &
   \multicolumn{1}{c}{\textbf{(arcsec)}} &
    \multicolumn{1}{c}{\textbf{OType}} &
    \multicolumn{1}{c}{\textbf{Class}} \\
\hline\hline
\endfirsthead
\multicolumn{12}{c}{{\tablename} \thetable{} -- Continued}\\
\hline\hline
   \multicolumn{1}{c}{\textbf{Src}} &
   \multicolumn{1}{c}{\textbf{RA}} &
   \multicolumn{1}{c}{\textbf{DEC}} &
   \multicolumn{1}{c}{\textbf{223 qsos}} &
   \multicolumn{1}{c}{\textbf{2MASS}} &
   \multicolumn{1}{c}{\textbf{PPMX}} &
   \multicolumn{1}{c}{\textbf{PPMXL}} &
   \multicolumn{1}{c}{\textbf{USNO-B1}} &
   \multicolumn{1}{c}{\textbf{437 968}} &
   \multicolumn{1}{c}{\textbf{1475/table2}} &
   \multicolumn{1}{c}{\textbf{SIMBAD}} &
   \multicolumn{1}{c}{\textbf{HR}} \\
   \multicolumn{1}{c}{\textbf{}} &
   \multicolumn{1}{c}{\textbf{(J2000)}} &
   \multicolumn{1}{c}{\textbf{(J2000)}} &
   \multicolumn{1}{c}{\textbf{(arcsec)}} &
   \multicolumn{1}{c}{\textbf{(arcsec)}} &
   \multicolumn{1}{c}{\textbf{(arcsec)}} &
   \multicolumn{1}{c}{\textbf{(arcsec)}} &
   \multicolumn{1}{c}{\textbf{(arcsec)}} &
   \multicolumn{1}{c}{\textbf{(arcsec)}} &
   \multicolumn{1}{c}{\textbf{(arcsec)}} &
    \multicolumn{1}{c}{\textbf{OType}} &
    \multicolumn{1}{c}{\textbf{Class}} \\
\hline\hline 
\endhead
\multicolumn{12}{c}{{Continued on Next Page\ldots}} \\
\endfoot
\hline\hline
\endlastfoot
  1 & 10 08 56.2 & 12 19 27.2 & 	&      &      & 	&           &      &       &    & f  \\
  2 & 10 08 50.9 & 12 18 24.8 & 	&      &      & 	&           &      &       &    & f  \\
  3 & 10 08 19.5 & 12 19 38.6 & 	&      &      & 	&           &      & 1.43  & RG*& l  \\
  4 & 10 08 12.4 & 12 17 18.7 & 	&      &      & 	&           &      & 2.23  &    &   \\
  5 & 10 08 25.4 & 12 16 29.9 & 	&      &      & 	&           &      &       &    & f   \\
  6 & 10 08 55.0 & 12 09 47.9 & 	& 2.99 &      & 	&           &      &       &    &    \\
  7 & 10 07 51.4 & 12 22 01.9 & 	&      &      & 	&           &      &       &    & h  \\
  8 & 10 08 37.6 & 12 14 13.2 & 	&      &      & 	&           &      &       &    & f  \\
  9 & 10 08 47.8 & 12 13 41.6 & 	&      &      & 	&           &      &       &    & h  \\
 10 & 10 08 38.3 & 12 19 17.1 & 	&      &      & 	&           &      & 2.39  &    & h  \\
\end{longtable}
\normalsize
\renewcommand{\thefootnote}{\arabic{footnote}}
\renewcommand{\arraystretch}{1.0}

\renewcommand{\thefootnote}{\fnsymbol{footnote}}
\renewcommand{\arraystretch}{0.81}
\scriptsize
\begin{longtable}{|c|c|c|c|c|c|c|c|c|c|c|c|c|}
\caption{The correlation of the X-ray sources towards the UMaII dSph. The full list is available on-line.}
\label{UMaIIclass}\\
\hline\hline 
   \multicolumn{1}{c}{\textbf{Src}} &
   \multicolumn{1}{c}{\textbf{RA}} &
   \multicolumn{1}{c}{\textbf{DEC}} &
   \multicolumn{1}{c}{\textbf{223 qsos}} &
   \multicolumn{1}{c}{\textbf{2MASS}} &
   \multicolumn{1}{c}{\textbf{2MASX}} &
   \multicolumn{1}{c}{\textbf{PPMX}} &
   \multicolumn{1}{c}{\textbf{PPMXL}} &
   \multicolumn{1}{c}{\textbf{USNO-B1}} &
   \multicolumn{1}{c}{\textbf{I1 313}} &
   \multicolumn{1}{c}{\textbf{437 968}} &
   \multicolumn{1}{c}{\textbf{SIMBAD}} &
   \multicolumn{1}{c}{\textbf{HR}} \\
   \multicolumn{1}{c}{\textbf{}} &
   \multicolumn{1}{c}{\textbf{(J2000)}} &
   \multicolumn{1}{c}{\textbf{(J2000)}} &
   \multicolumn{1}{c}{\textbf{(arcsec)}} &
   \multicolumn{1}{c}{\textbf{(arcsec)}} &
   \multicolumn{1}{c}{\textbf{(arcsec)}} &
   \multicolumn{1}{c}{\textbf{(arcsec)}} &
   \multicolumn{1}{c}{\textbf{(arcsec)}} &
   \multicolumn{1}{c}{\textbf{(arcsec)}} &
   \multicolumn{1}{c}{\textbf{(arcsec)}} &
   \multicolumn{1}{c}{\textbf{(arcsec)}} &
    \multicolumn{1}{c}{\textbf{OType}} &
    \multicolumn{1}{c}{\textbf{Class}} \\
\hline\hline
\endfirsthead
\multicolumn{13}{c}{{\tablename} \thetable{} -- Continued}\\
\hline\hline
   \multicolumn{1}{c}{\textbf{Src}} &
   \multicolumn{1}{c}{\textbf{RA}} &
   \multicolumn{1}{c}{\textbf{DEC}} &
   \multicolumn{1}{c}{\textbf{223 qsos}} &
   \multicolumn{1}{c}{\textbf{2MASS}} &
   \multicolumn{1}{c}{\textbf{2MASX}} &
   \multicolumn{1}{c}{\textbf{PPMX}} &
   \multicolumn{1}{c}{\textbf{PPMXL}} &
   \multicolumn{1}{c}{\textbf{USNO-B1}} &
   \multicolumn{1}{c}{\textbf{I1 313}} &
   \multicolumn{1}{c}{\textbf{437 968}} &
   \multicolumn{1}{c}{\textbf{SIMBAD}} &
   \multicolumn{1}{c}{\textbf{HR}} \\
   \multicolumn{1}{c}{\textbf{}} &
   \multicolumn{1}{c}{\textbf{(J2000)}} &
   \multicolumn{1}{c}{\textbf{(J2000)}} &
   \multicolumn{1}{c}{\textbf{(arcsec)}} &
   \multicolumn{1}{c}{\textbf{(arcsec)}} &
   \multicolumn{1}{c}{\textbf{(arcsec)}} &
   \multicolumn{1}{c}{\textbf{(arcsec)}} &
   \multicolumn{1}{c}{\textbf{(arcsec)}} &
   \multicolumn{1}{c}{\textbf{(arcsec)}} &
   \multicolumn{1}{c}{\textbf{(arcsec)}} &
   \multicolumn{1}{c}{\textbf{(arcsec)}} &
    \multicolumn{1}{c}{\textbf{OType}} &
    \multicolumn{1}{c}{\textbf{Class}} \\
\hline\hline 
\endhead
\multicolumn{13}{c}{{Continued on Next Page\ldots}} \\
\endfoot
\hline\hline
\endlastfoot
  1 & 08 51 34.3 & 63 12 48.2 &      & 1.86 &      &      & 2.63 & 0.77 &      &      &    & f  \\
  2 & 08 50 51.9 & 63 08 37.6 &      &      &      &      &      &      &      &      &    & h  \\
  3 & 08 51 26.3 & 63 07 03.7 &      &      &      &      &      &      &      &      &    & h  \\
  4 & 08 50 59.9 & 63 06 28.6 &      &      &      &      &      &      &      &      &    & h  \\
  5 & 08 51 06.5 & 63 18 38.9 &      &      &      &      &      &      &      &      &    & f  \\
  6 & 08 51 03.7 & 63 11 29.1 &      &      &      &      &      &      &      &      &    & f  \\
  7 & 08 50 55.9 & 63 11 02.2 &      & 0.95 &      &      & 0.96 & 0.74 & 0.87 &      &    & f  \\
  8 & 08 51 44.6 & 63 01 55.2 &      & 1.27 &      &      & 1.33 & 1.77 & 1.31 &      &    & f  \\
  9 & 08 50 47.3 & 63 11 43.6 &      &      &      &      &      &      &      &      &    & h  \\
 10 & 08 53 02.7 & 63 15 53.0 &      &      &      &      &      &      &      &      &    & f  \\
\end{longtable}
\normalsize
\renewcommand{\thefootnote}{\arabic{footnote}}
\renewcommand{\arraystretch}{1.0}

\nopagebreak

%
\renewcommand{\thefootnote}{\fnsymbol{footnote}}
\renewcommand{\arraystretch}{0.81}
\scriptsize
\begin{longtable}{|c|c|c|c|c|c|c|c|c|c|c|c|}
\caption{The correlation of the X-ray sources towards the UMi dSph. The full list is available on-line.}
\label{UMiclass}\\
\hline\hline 
   \multicolumn{1}{c}{\textbf{Src}} &
   \multicolumn{1}{c}{\textbf{RA}} &
   \multicolumn{1}{c}{\textbf{DEC}} &
   \multicolumn{1}{c}{\textbf{223 qsos}} &
   \multicolumn{1}{c}{\textbf{2MASS}} &
   \multicolumn{1}{c}{\textbf{PPMX}} &
   \multicolumn{1}{c}{\textbf{PPMXL}} &
   \multicolumn{1}{c}{\textbf{USNO-B1}} &
   \multicolumn{1}{c}{\textbf{180 67}} &
   \multicolumn{1}{c}{\textbf{V 134}} &
   \multicolumn{1}{c}{\textbf{SIMBAD}} &
   \multicolumn{1}{c}{\textbf{HR}} \\
   \multicolumn{1}{c}{\textbf{}} &
   \multicolumn{1}{c}{\textbf{(J2000)}} &
   \multicolumn{1}{c}{\textbf{(J2000)}} &
   \multicolumn{1}{c}{\textbf{(arcsec)}} &
   \multicolumn{1}{c}{\textbf{(arcsec)}} &
   \multicolumn{1}{c}{\textbf{(arcsec)}} &
   \multicolumn{1}{c}{\textbf{(arcsec)}} &
   \multicolumn{1}{c}{\textbf{(arcsec)}} &
   \multicolumn{1}{c}{\textbf{(arcsec)}} &
   \multicolumn{1}{c}{\textbf{(arcsec)}} &
    \multicolumn{1}{c}{\textbf{OType}} &
    \multicolumn{1}{c}{\textbf{Class}} \\
\hline\hline
\endfirsthead
\multicolumn{12}{c}{{\tablename} \thetable{} -- Continued}\\
\hline\hline
   \multicolumn{1}{c}{\textbf{Src}} &
   \multicolumn{1}{c}{\textbf{RA}} &
   \multicolumn{1}{c}{\textbf{DEC}} &
   \multicolumn{1}{c}{\textbf{223 qsos}} &
   \multicolumn{1}{c}{\textbf{2MASS}} &
   \multicolumn{1}{c}{\textbf{PPMX}} &
   \multicolumn{1}{c}{\textbf{PPMXL}} &
   \multicolumn{1}{c}{\textbf{USNO-B1}} &
   \multicolumn{1}{c}{\textbf{180 67}} &
   \multicolumn{1}{c}{\textbf{V 134}} &
   \multicolumn{1}{c}{\textbf{SIMBAD}} &
   \multicolumn{1}{c}{\textbf{HR}} \\
   \multicolumn{1}{c}{\textbf{}} &
   \multicolumn{1}{c}{\textbf{(J2000)}} &
   \multicolumn{1}{c}{\textbf{(J2000)}} &
   \multicolumn{1}{c}{\textbf{(arcsec)}} &
   \multicolumn{1}{c}{\textbf{(arcsec)}} &
   \multicolumn{1}{c}{\textbf{(arcsec)}} &
   \multicolumn{1}{c}{\textbf{(arcsec)}} &
   \multicolumn{1}{c}{\textbf{(arcsec)}} &
   \multicolumn{1}{c}{\textbf{(arcsec)}} &
   \multicolumn{1}{c}{\textbf{(arcsec)}} &
    \multicolumn{1}{c}{\textbf{OType}} &
    \multicolumn{1}{c}{\textbf{Class}} \\
\hline\hline 
\endhead
\multicolumn{12}{c}{{Continued on Next Page\ldots}} \\
\endfoot
\hline\hline
\endlastfoot
  1 & 15 09 01.8 & 67 11 31.9 &      & 1.10 &      & 0.96 & 0.54 &      &       & *  & l   \\                                                                                      
  2 & 15 08 48.6 & 67 10 34.6 &      &      &      &      &      &      &       &    & f   \\                                                                                      
  3 & 15 09 13.1 & 67 12 59.4 &      &      &      &      &      &      &       &    &    \\                                                                                      
  4 & 15 09 25.9 & 67 13 35.3 &      &      &      &      &      &      &       &    & f   \\                                                                                      
  5 & 15 09 29.2 & 67 06 24.6 & 2.28 &      &      & 2.22 & 2.41 & 2.29 &       & *  & l   \\                                                                                      
  6 & 15 10 39.6 & 67 22 36.6 &      &      &      &      &      &      &       & *  & l   \\                                                                                      
  7 & 15 09 18.0 & 67 05 20.3 &      &      &      &      &      &      &       &    &    \\                                                                                      
  8 & 15 09 38.4 & 67 15 56.0 &      &      &      &      &      &      &       & *  & snr   \\                                                                                      
  9 & 15 08 08.1 & 67 18 10.8 &      &      &      &      &      &      &       & *  & l   \\                                                                                      
 10 & 15 08 11.7 & 67 15 14.4 &      &      &      &      &      &      &       & *  & fg   \\                                                                                      
\end{longtable}
\normalsize
\renewcommand{\thefootnote}{\arabic{footnote}}
\renewcommand{\arraystretch}{1.0}
\end{landscape}

\twocolumn

\section{High-energy view}
In the following section we address the problem of the membership of our X-ray sources to the dSph galaxies, 
mainly through the analysis of high-energy data.

\subsection{Hardness-Ratios and X-ray-to-NIR Flux Diagrams}
\begin{figure*}
\centering
\subfigure[Draco dSph]
   {\includegraphics[width=8cm]{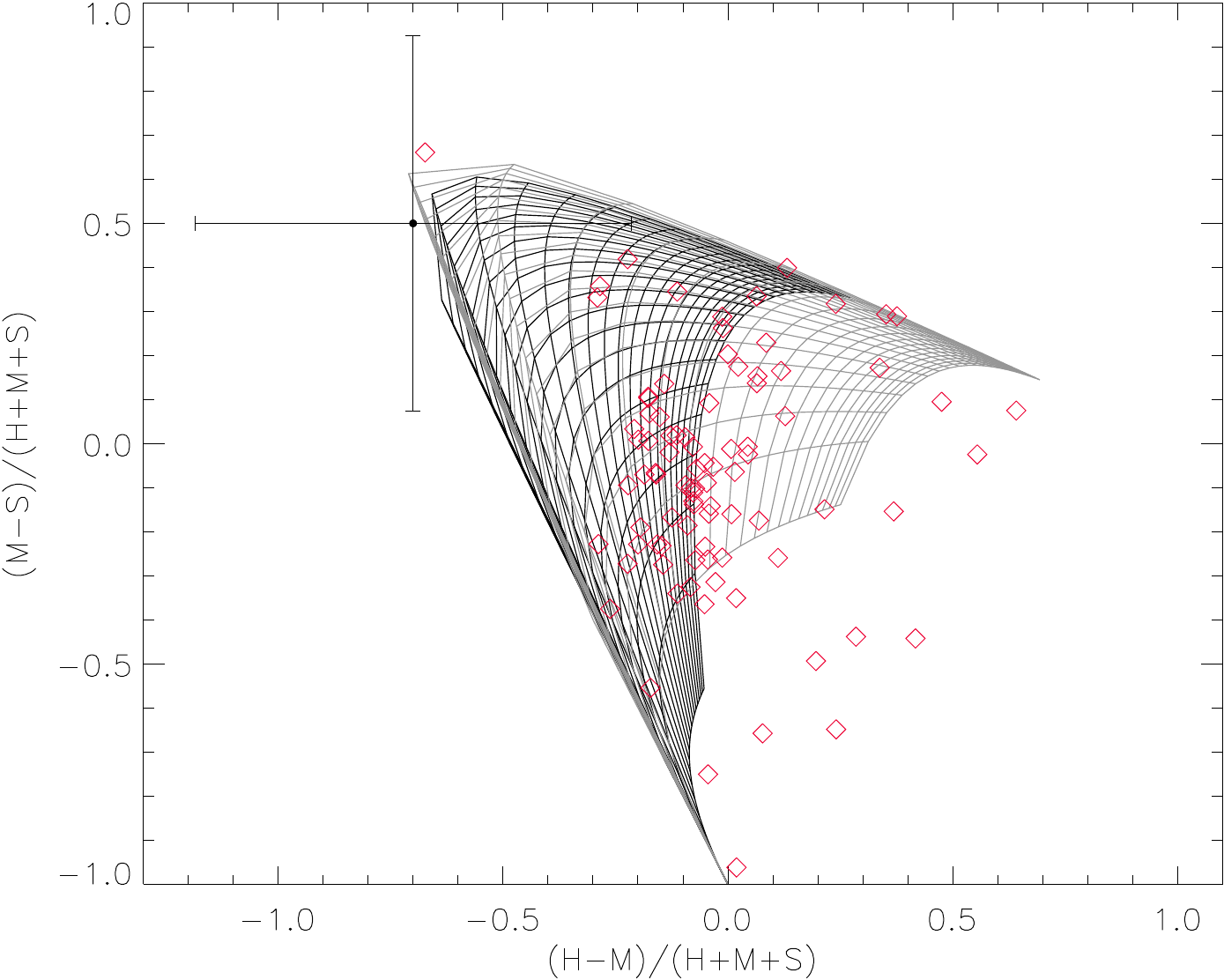}}\qquad\qquad
\subfigure[Leo I dSph]
   {\includegraphics[width=8cm]{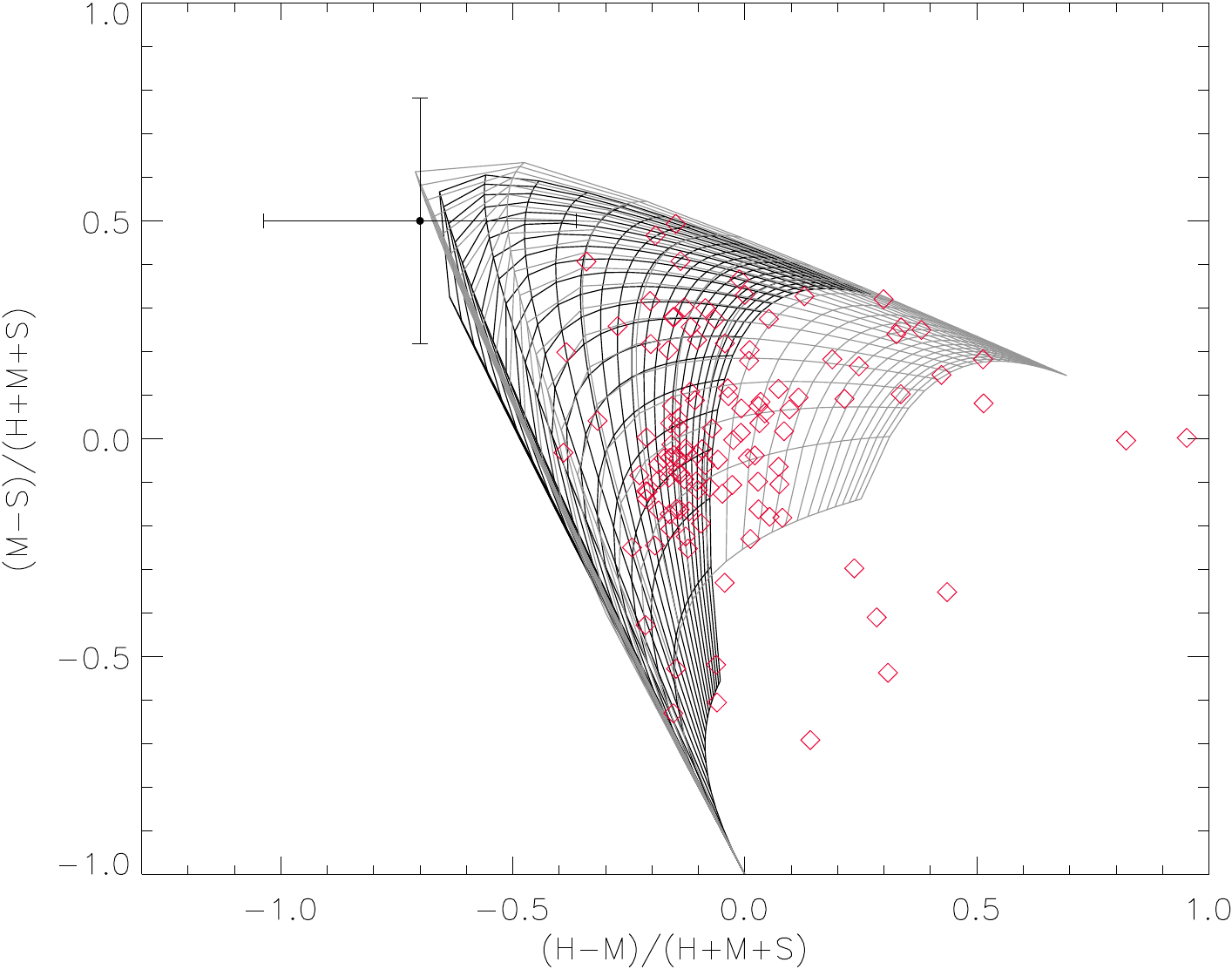}}\qquad\qquad
\subfigure[UMa II dSph]
   {\includegraphics[width=8cm]{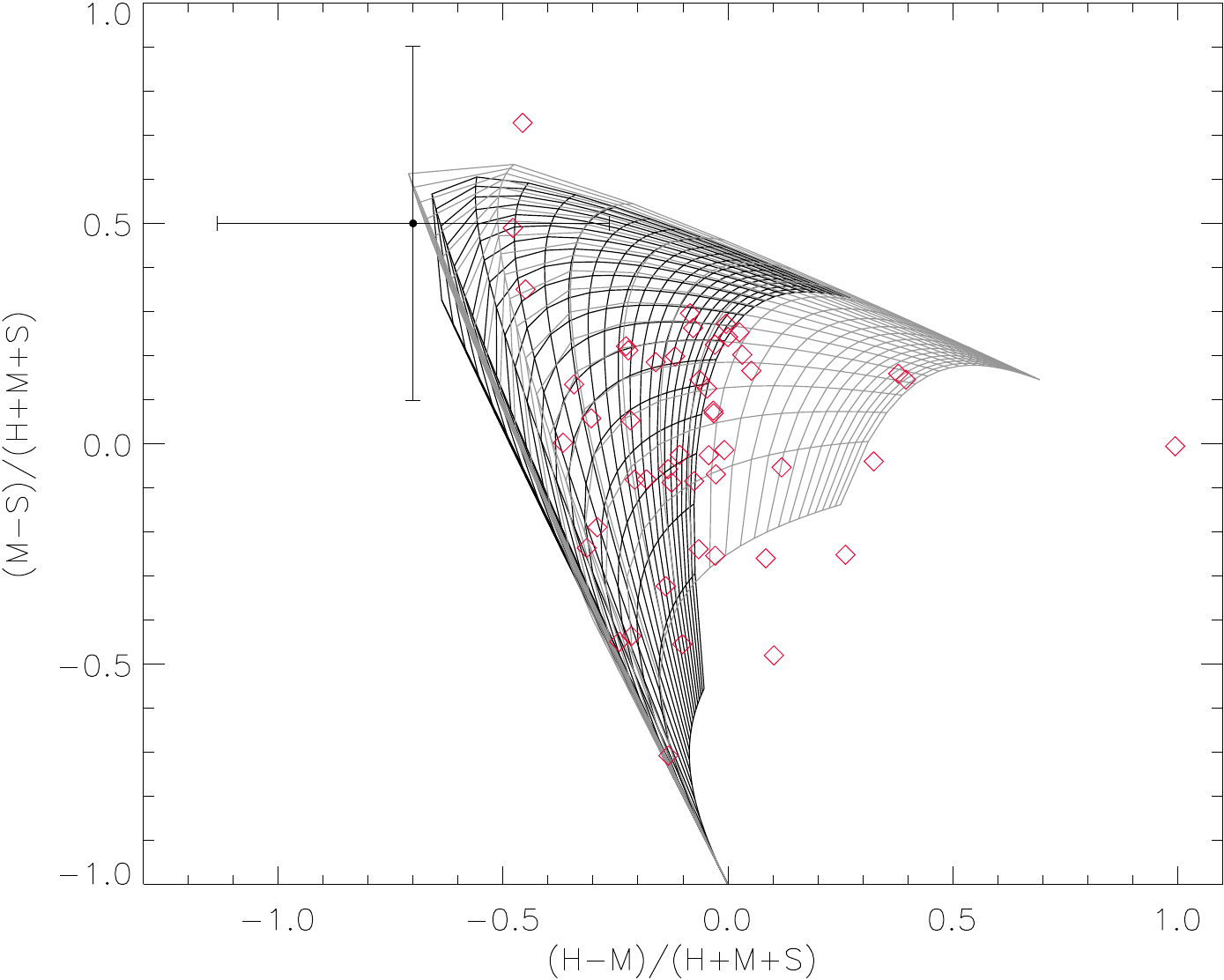}}\qquad\qquad
\subfigure[UMi dSph]
   {\includegraphics[width=8cm]{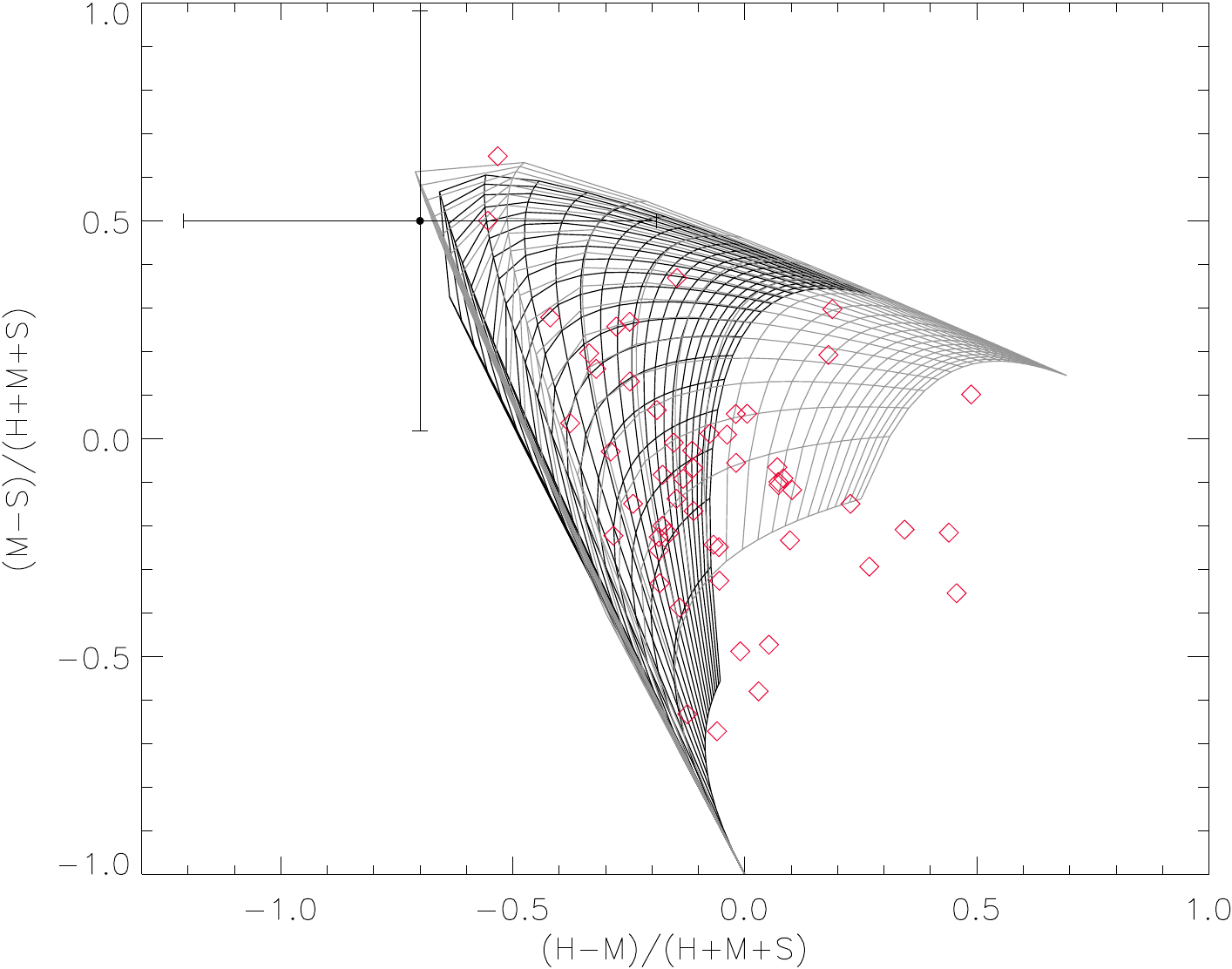}}
\caption{The color-color diagram of the sources detected by the EPIC cameras towards the dSphs. The solid lines represent the theoretical 
tracks expected for different emitting models (see text for details). A representative error bar (obtained averaging all 
data point error bars) is also shown in each panel.}
\label{fig2}
\end{figure*}

With the aim to attempt a classification of the high-energy sources detected towards our dSph sample, we followed 
\citet{ramsay2006} and calculated the hardness-ratios ($HR^*$) as
\begin{equation}
HR^*_1 = \frac{H-M}{S+M+H}~~{\rm and}~~HR^*_2 = \frac{M-S}{S+M+H}~.
\end{equation}
Here, S, M, and H correspond to the count rates in the 0.2-1.0 keV,
1.0-2.0 keV, and 2.0-12.0 keV energy bands. 
From these {$HR^*$} values we constructed the color-color diagrams shown in Fig. \ref{fig2},
where the obtained values are compared with two spectral models. We used bremsstrahlung (grey tracks) 
and power-law (black lines) models in order to simulate the {$HR^*_{1}$} and {$HR^*_{2}$} of Cataclysmic Variables (CVs), 
Active Galactic Nuclei (AGN) or X-ray binaries\footnote{The reader can also see \citet{ramsay2006}.}, respectively.

We used the XSPEC package (\citealt{arnaud}) version 12.0.0 to obtain the color-color contours.
In both cases, we vary the equivalent hydrogen column density $N_H$ from $10^{19}$cm$^{-2}$ to $10^{22}$ cm$^{-2}$ so
 each of the almost horizontal lines corresponds to models with equal $N_H$ which increases from bottom to top. 
The temperature associated to each bremsstrahlung model (kT, taken in the range  0.1 - 3.0 keV) and
the power-law index $\Gamma$ (in the range 0.1 - 3.0) are associated with
primarily vertical lines: the values of kT and $\Gamma$ increase from left to right and from
right to left, respectively.

In the upper left side of each panel of Fig. \ref{fig2} we give a representative error bar, obtained by averaging all the
data point error bars. Some of the detected sources have colors consistent with those of the absorbed power-law or
absorbed bremsstrahlung models, others seem to require combined spectra or fall outside the pattern area.
Even if many sources appear to have spectra consistent with that of a typical AGN (squared dots close black 
tracks) we cannot rule out that some of the sources have a different nature.
{In fact, due to the large error bars affecting {$HR^*$} values, a classification based only on the 
hardness-ratio cannot constrain the nature of the objects in our sample, 
allowing to distinguish among AGNs, X-ray binaries, CVs and X-ray active stars.}

\begin{figure*}
 \centering
\subfigure[Draco dSph]
   {\includegraphics[width=8cm]{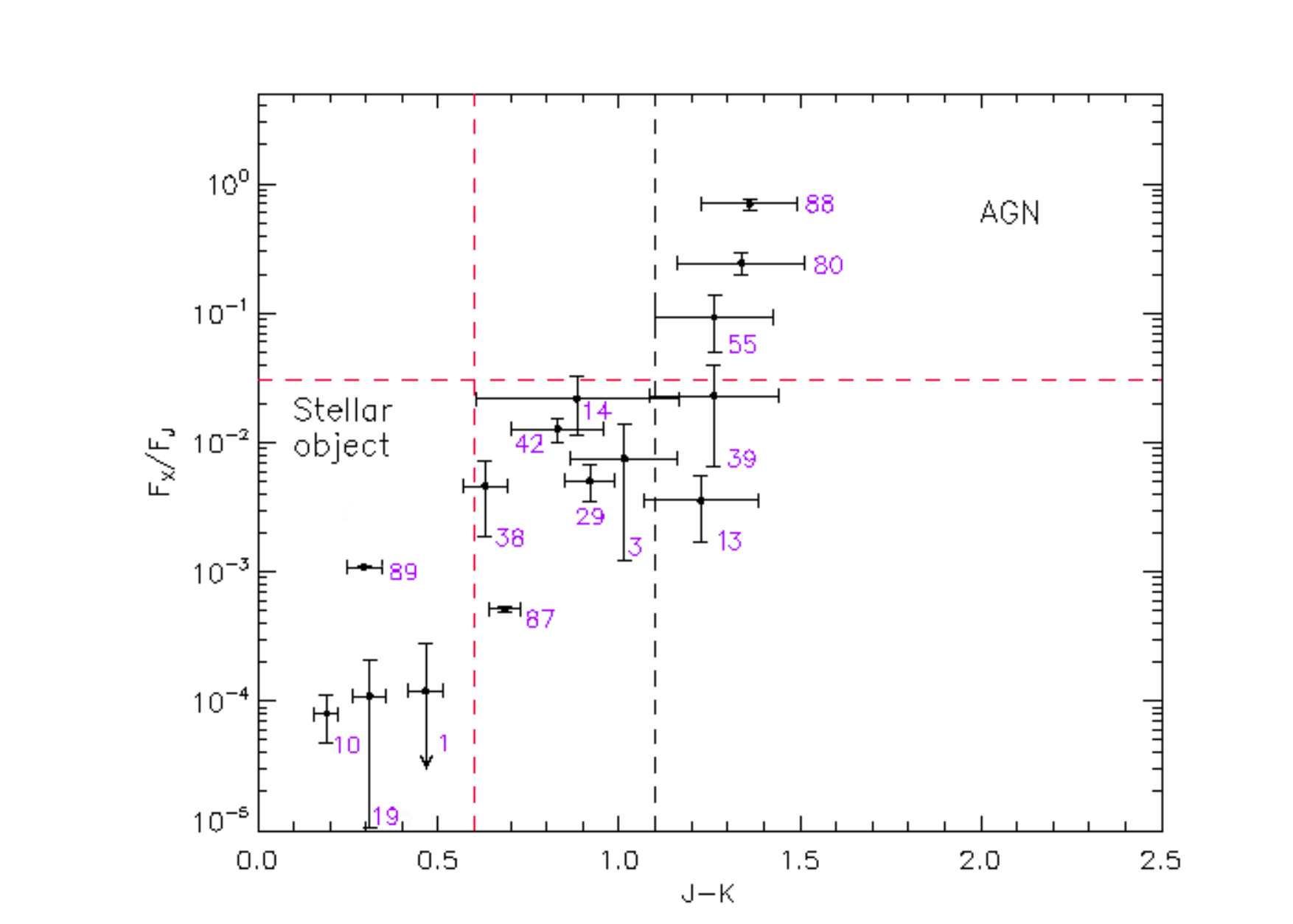}}\qquad\qquad
\subfigure[Leo I dSph]
   {\includegraphics[width=8cm]{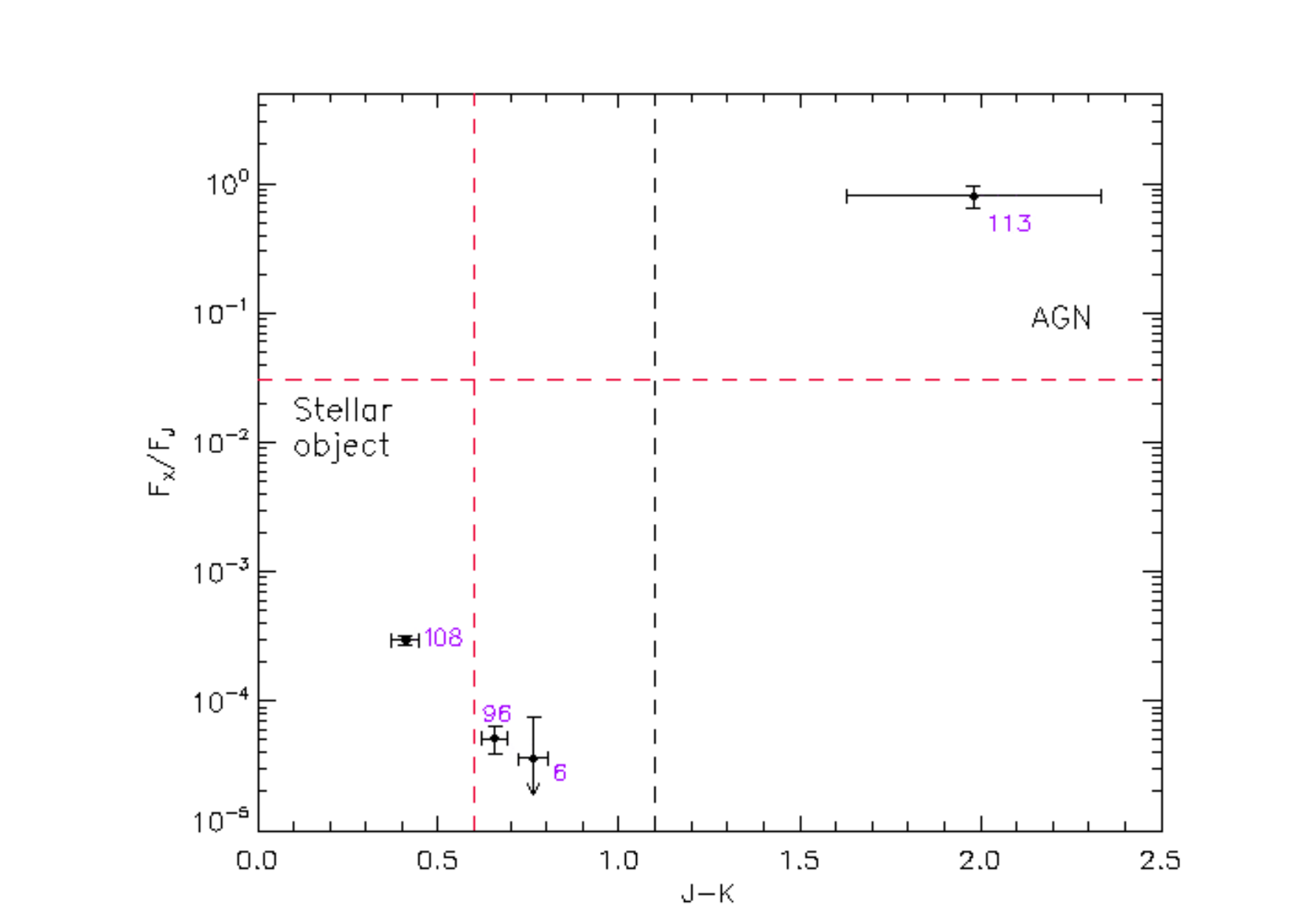}}\qquad\qquad
\subfigure[UMa II dSph]
   {\includegraphics[width=8cm]{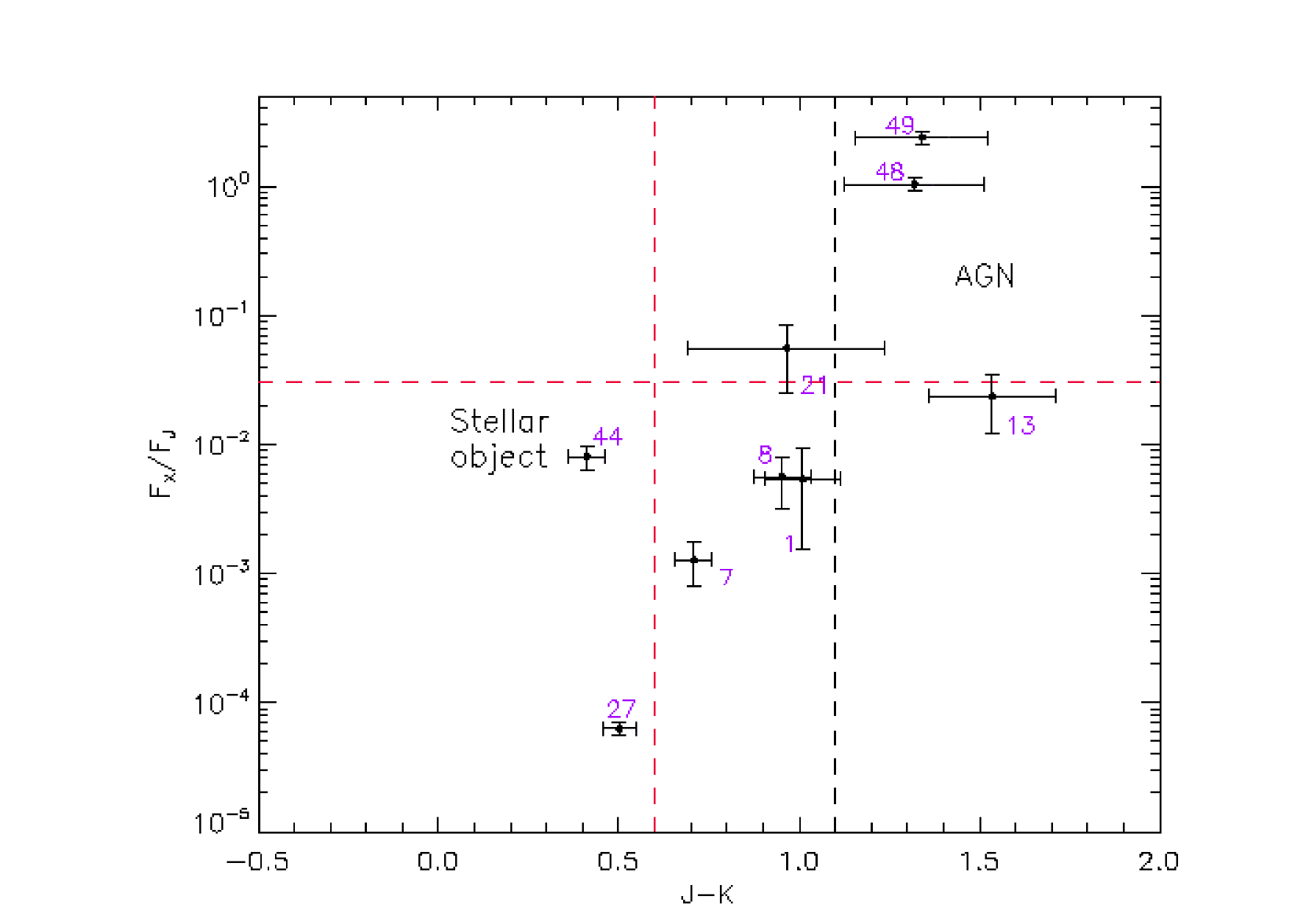}}\qquad\qquad
\subfigure[UMi dSph]
   {\includegraphics[width=8cm]{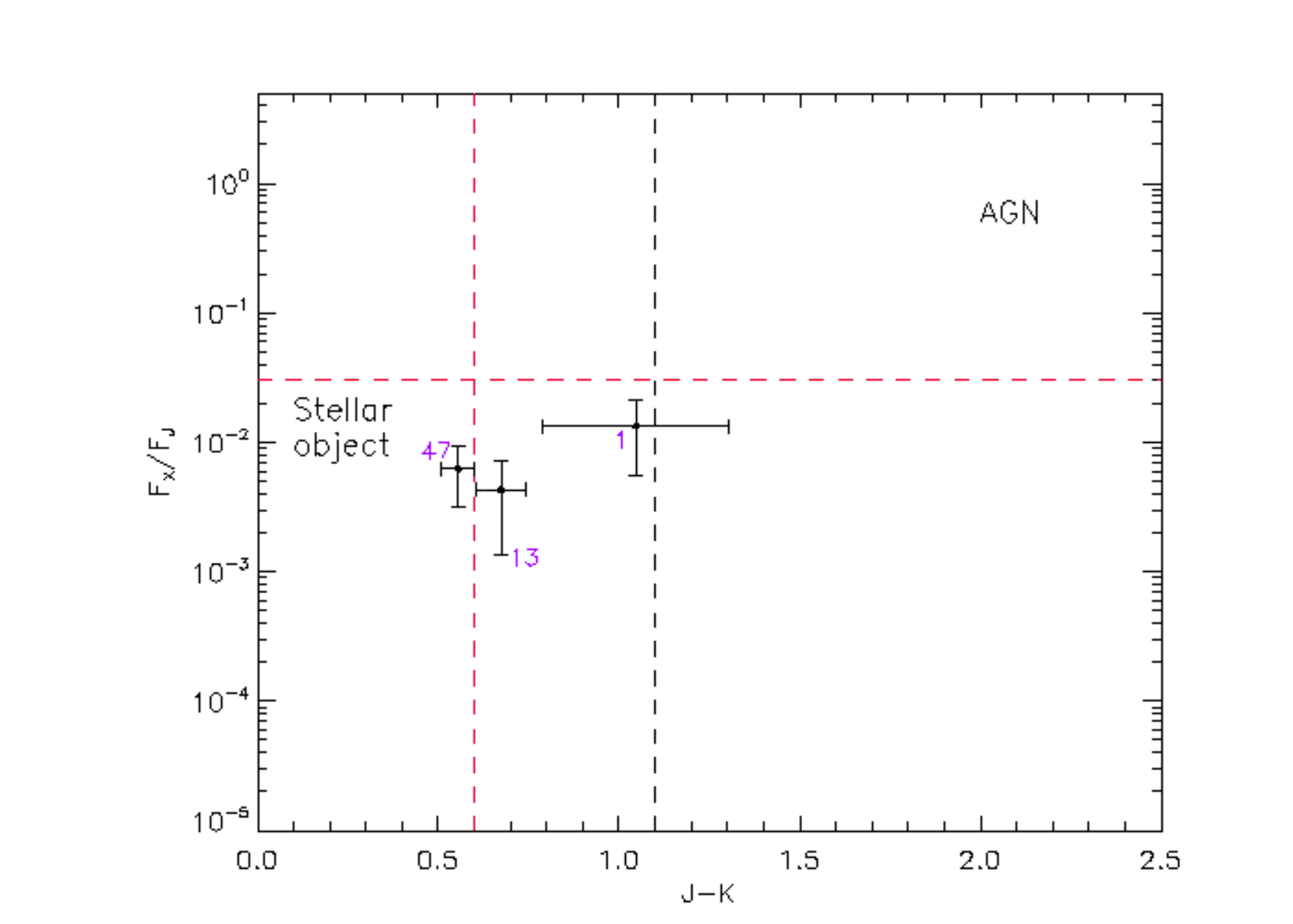}}
\caption{The color-color diagram for our X-ray sources with a counterpart in the 2MASS catalogue (see text for details).}
\label{figclass}
\end{figure*}

In order to improve our classification, we adopted the method of \citet{haakonsen2009} 
that uses a color-color diagram based on the ratio between the
$0.2-2.4$ keV flux ($F_X$) and the NIR flux in the J band ($F_J$) versus the J-K color. 
In particular, QSO and Seyfert 1 objects lie in the upper right part having $(J-K)\ut >0.6$ and $F_X/F_J\ut >3\times 10^{-2}$ 
(dashed red lines in Fig. \ref{figclass}) and the coronally active stars (including pre-main sequence and main-sequence
 stars, high proper-motion objects and binary systems) the lower left side with $(J-K)\ut <1.1$ 
(dotted black line in Fig. \ref{figclass}) and $F_X/F_J\ut <3\times 10^{-2}$.

As one can see in Tables \ref{DracoSources} - \ref{UMiSources} we found many correlations between the detected sources 
and 2MASS catalogue. For these sources, we give in Tables \ref{table_FxFnir} 
the 0.2-2.4 keV and NIR J band fluxes, as well as the J and K magnitudes. 
\begin{table*}
\footnotesize{
\subtable[Draco dSph]{
\begin{tabular}{|c|c|c|c|c|}
\hline
Src & F$_X$                    & F$_{J}$                 & J & K                    \\
    & ($\times 10^{-14}$ erg s$^{-1}$ cm$^{-2}$)  & ($\times 10^{-13}$ erg s$^{-1}$ cm$^{-2}$) &   &                    \\
\hline
 1 &   $\leq$ 0.48                    &   173.6   $\pm$   3.7 &  11.17 $\pm$  0.02 &   10.70   $\pm$ 0.02 \\
 3 &   0.21           $\pm$      0.18 &   2.87    $\pm$ 0.19  &  15.62 $\pm$  0.07 &   14.61   $\pm$ 0.08 \\
10 &   0.28           $\pm$      0.12 &   351.6   $\pm$   5.5 &  10.40 $\pm$  0.02 &   10.21   $\pm$ 0.02 \\
13 &   0.33           $\pm$      0.17 &   9.11    $\pm$ 0.69  &  14.37 $\pm$  0.08 &   13.14   $\pm$ 0.08 \\
14 &   0.33           $\pm$      0.16 &   1.52    $\pm$ 0.15  &  16.31 $\pm$  0.10 &   15.42   $\pm$ 0.18 \\
19 &   0.37           $\pm$      0.34 &   343.3   $\pm$   8.2 &  10.43 $\pm$  0.03 &   10.12   $\pm$ 0.02 \\
29 &   0.45           $\pm$      0.14 &   8.97    $\pm$ 0.26  &  14.38 $\pm$  0.03 &   13.46   $\pm$ 0.04 \\
38 &   0.61           $\pm$      0.36 &   13.29   $\pm$  0.38 &  13.96 $\pm$  0.03 &   13.33   $\pm$ 0.03 \\
39 &   0.61           $\pm$      0.44 &   2.67    $\pm$ 0.20  &  15.70 $\pm$  0.08 &   14.44   $\pm$ 0.10 \\
42 &   0.68           $\pm$      0.14 &   5.32    $\pm$ 0.24  &  14.95 $\pm$  0.05 &   14.12   $\pm$ 0.08 \\
55 &    1.04          $\pm$      0.46 &   1.12    $\pm$ 0.17  &  16.64 $\pm$  0.16 &   15.38   $\pm$ 0.02 \\
80 &    5.19          $\pm$      0.86 &   2.13    $\pm$ 0.18  &  15.94 $\pm$  0.09 &   14.60   $\pm$ 0.09 \\
87 &    10.70         $\pm$      0.56 &   2086    $\pm$  37   &  8.47  $\pm$  0.02 &   7.78    $\pm$ 0.02 \\
88 &    28.9          $\pm$      2.3  &   4.14    $\pm$ 0.24  &  15.22 $\pm$  0.06 &   13.86   $\pm$ 0.07 \\
89 &    44.48         $\pm$      0.55 &   4090    $\pm$  115  &  7.74  $\pm$  0.03 &   7.44    $\pm$ 0.02 \\
\hline                                                                                                 
\end{tabular}

} \qquad

\subtable[Leo I dSph]{
\begin{tabular}{|c|c|c|c|c|}
\hline
Src & F$_X$                    & F$_{J}$                 & J & K                    \\
    & ($\times 10^{-14}$ erg s$^{-1}$ cm$^{-2}$)  & ($\times 10^{-13}$ erg s$^{-1}$ cm$^{-2}$) &   &                    \\
\hline
  6 &  $\leq$0.20           &     262      $\pm$   5   &  10.72  $\pm$   0.02  &   9.96   $\pm$  0.02  \\ 
 96 &  1.90 $\pm$   0.46    &    3716      $\pm$   62  &   7.84  $\pm$   0.02  &   7.18   $\pm$  0.02  \\ 
108 &  3.29 $\pm$   0.27    &    1117      $\pm$   22  &   9.15  $\pm$   0.02  &   8.74   $\pm$  0.02  \\ 
113 &  5.70 $\pm$   0.45    &    0.71      $\pm$  0.15 &  17.14  $\pm$   0.21  &  15.16   $\pm$  0.15  \\                                                     
\hline
\end{tabular}
} \qquad

\subtable[UMa II dSph]{
\begin{tabular}{|c|c|c|c|c|}
\hline
Src & F$_X$                    & F$_{J}$                 & J & K                    \\
    & ($\times 10^{-14}$ erg s$^{-1}$ cm$^{-2}$)  & ($\times 10^{-13}$ erg s$^{-1}$ cm$^{-2}$) &   &                    \\
\hline
 1 &  0.27  $\pm$  0.19    &    4.94    $\pm$   0.23  &  15.03  $\pm$   0.05  &   14.02   $\pm$  0.06  \\
 7 &  0.40  $\pm$  0.15    &    31.54   $\pm$   0.78  &  13.02  $\pm$   0.03  &   12.31   $\pm$  0.03  \\
 8 &  0.47  $\pm$  0.20    &    8.36    $\pm$   0.27  &  14.46  $\pm$   0.04  &   13.51   $\pm$  0.04  \\
13 &  0.64  $\pm$  0.30    &    2.70    $\pm$   0.13  &  15.69  $\pm$   0.09  &   14.15   $\pm$  0.08  \\
21 &  0.92  $\pm$  0.49    &    1.66    $\pm$   0.16  &  16.22  $\pm$   0.10  &   15.25   $\pm$  0.17  \\
26 &   1.23 $\pm$  0.59    &    0.83    $\pm$   0.12  &  16.96  $\pm$   0.15  &   --                  \\
27 &   1.24 $\pm$  0.14    &    1973    $\pm$   42    &   8.53  $\pm$   0.02  &    8.03   $\pm$  0.02  \\
44 &   3.92 $\pm$  0.77    &    49.2    $\pm$   1.2   &  12.54  $\pm$   0.03  &   12.13   $\pm$  0.02  \\
48 &   22.1 $\pm$   1.3    &    2.14    $\pm$   0.20  &  15.94  $\pm$   0.10  &   14.62   $\pm$  0.10  \\
49 &   63.7 $\pm$   4.8    &    2.67    $\pm$   0.23  &  15.70  $\pm$   0.09  &   14.36   $\pm$  0.09  \\
\hline
\end{tabular}
} \qquad

\subtable[UMi dSph]{
\begin{tabular}{|c|c|c|c|c|}
\hline
Src & F$_X$                    & F$_{J}$                 & J & K                    \\
    & ($\times 10^{-14}$ erg s$^{-1}$ cm$^{-2}$)  & ($\times 10^{-13}$ erg s$^{-1}$ cm$^{-2}$) &   &                    \\
\hline
 1 &  0.23 $\pm$   0.14    &    1.75      $\pm$   0.15  &  16.16  $\pm$   0.09  &   15.11   $\pm$  0.17  \\
13 &  0.52 $\pm$   0.36    &    12.26     $\pm$   0.30  &  14.04  $\pm$   0.03  &   13.37   $\pm$  0.04  \\
47 &  2.4  $\pm$   1.2     &    37.47     $\pm$   0.76  &  12.83  $\pm$   0.02  &   12.28   $\pm$  0.02  \\
\hline
\end{tabular}
}
\caption{The detected X-ray sources that correlate with the 2MASS catalog. Following
\citet{haakonsen2009}, we try to constrain their nature by using the X-ray (in the 0.2-2.4 keV band) and NIR (J and K bands) fluxes (see text and Fig. \ref{figclass} for details). 
{We remind that the 2MASS data reach the $3\sigma$ limiting sensitivity of 17.1, and 15.3 mag in the J and K bands, respectively. A long dash means that the corresponding source was not detected 
in the relevant band.}
\label{table_FxFnir}}
}
\end{table*}
To obtain the 0.2-2.4 keV band flux we used the 0.2-12.0 keV one assuming, in webPIMMS\footnote{WebPIMMS is available at \\ \tt http://heasarc.gsfc.nasa.gov/Tools/w3pimms.html}, a power-law model 
with spectral index $\Gamma$ and absorption column density $N_H$ inferred by NASA on-line tool
\footnote{This tool is available at \\ \tt http://heasarc.gsfc.nasa.gov/cgi-bin/Tools/w3nh/w3nh.pl}. Thus we obtained
$\Gamma$=1.7 and $N_H$=2.77$\times 10^{20} cm^{-2}$ for Draco, $\Gamma$=1.7 and $N_H$=3.75$\times 10^{20} cm^{-2}$ for Leo I,
{$\Gamma$=1.7 and $N_H$=4.65$\times 10^{20} cm^{-2}$ for UMa II} and
$\Gamma$=1.7 and $N_H$=2.20$\times 10^{20} cm^{-2}$ for UMi. In Fig. \ref{figclass}, we present the color-color diagrams 
for the sources listed in Table \ref{table_FxFnir}.

In particular, Fig. \ref{figclass} (a) shows the fifteen sources of Draco dSph with the detected counterparts. 
Among these, three (55, 80 and 88) clearly reside in the AGN part of the diagram. The likely galactic nature of source 55 is 
also inferred by the correlation with a QSO (see catalogues 223/qsos and 1921/table10). Instead sources 1, 3, 10, 14, 
19, 29, 38, 42, 87, and  89 can be stellar objects. Many of them (1, 10, 19, 38, 87, 89) correlate with PPMX 
sources so are probably Milky Way sources. Scr 38 is also associated with an object belonging to a late-type stars catalogue (442/165), 
while Src 29 is realistically a ``carbon star (C1)'', as also reported in the Draco stellar catalogue 
1921/table9. Src 10 may be either a star (PPMX catalogue) or a background 
(437/968) object. In this case, the use of the diagram can be useful to settle the querelle in favour of a 
stellar type.
We cannot use the same method for Src 13, correlating with both PPMX (stellar) and 78/675 
(quasar-galaxy association) catalogues, because of its not clear identification in the aforementioned regions 
of the diagram. Finally, Src 39 is not catalogued with our diagram although it seems correlated 
with an AGN candidate (437/968 catalogue).

In the Leo I case we find four sources correlating with the 2MASS catalogue. The position of Scr 113 in
Fig. \ref{figclass} (b) denotes its background nature. The sources 96 and 108, recognized as stellar sources,
are also associated with two objects in the PPMX catalogue, so they can be foreground stars.

Fig. \ref{figclass} (c) shows five (1, 7, 8, 27, 44) stellar and three (21, 48, 49) galactic sources toward UMa II dSph. 
We correctly obtain a correlation of Src 27 with a source in the PPMX catalogue. In the same catalogue 
we find a source correlating with Src 44 that has an association with an AGN candiate (437/968 catalogue) so
the diagram is useful in order to fix the possible stellar nature of the source.

In the last panel of Fig. \ref{figclass}, named (d), we present the analysis for UMi dSph. The 
relevant sources (1, 13 and 47) seem to have characteristics similar to X-ray active stars and binary sources.
In particular, Src 13 correlates with a Radio/X-ray source (V/134 catalogue) with Red [Blue] magnitude
 of 17.0 [20.2] while Scr 47, correlating also with PPMX catalogue, is probably a foreground star. 
 
    We also used the SIMBAD database to search for correlations between our X-ray sources and 
    available catalogues, getting more information (e.g., magnitude, redshift, source interpretation) afterwards 
    used.

\begin{figure*}
 \centering
\subfigure[Draco dSph HRs]
   {\includegraphics[width=5cm]{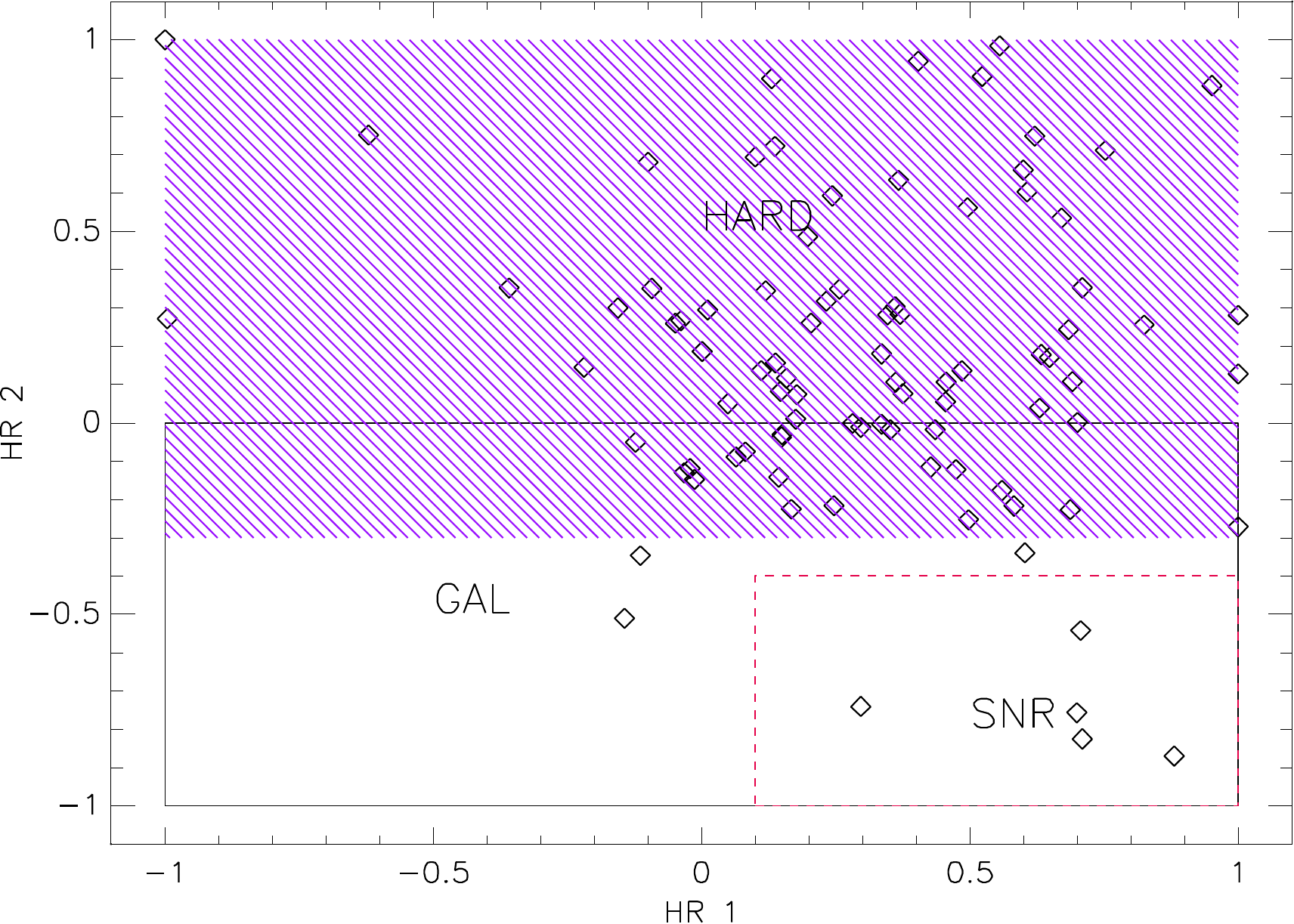}}\qquad\qquad
   {\includegraphics[width=5cm]{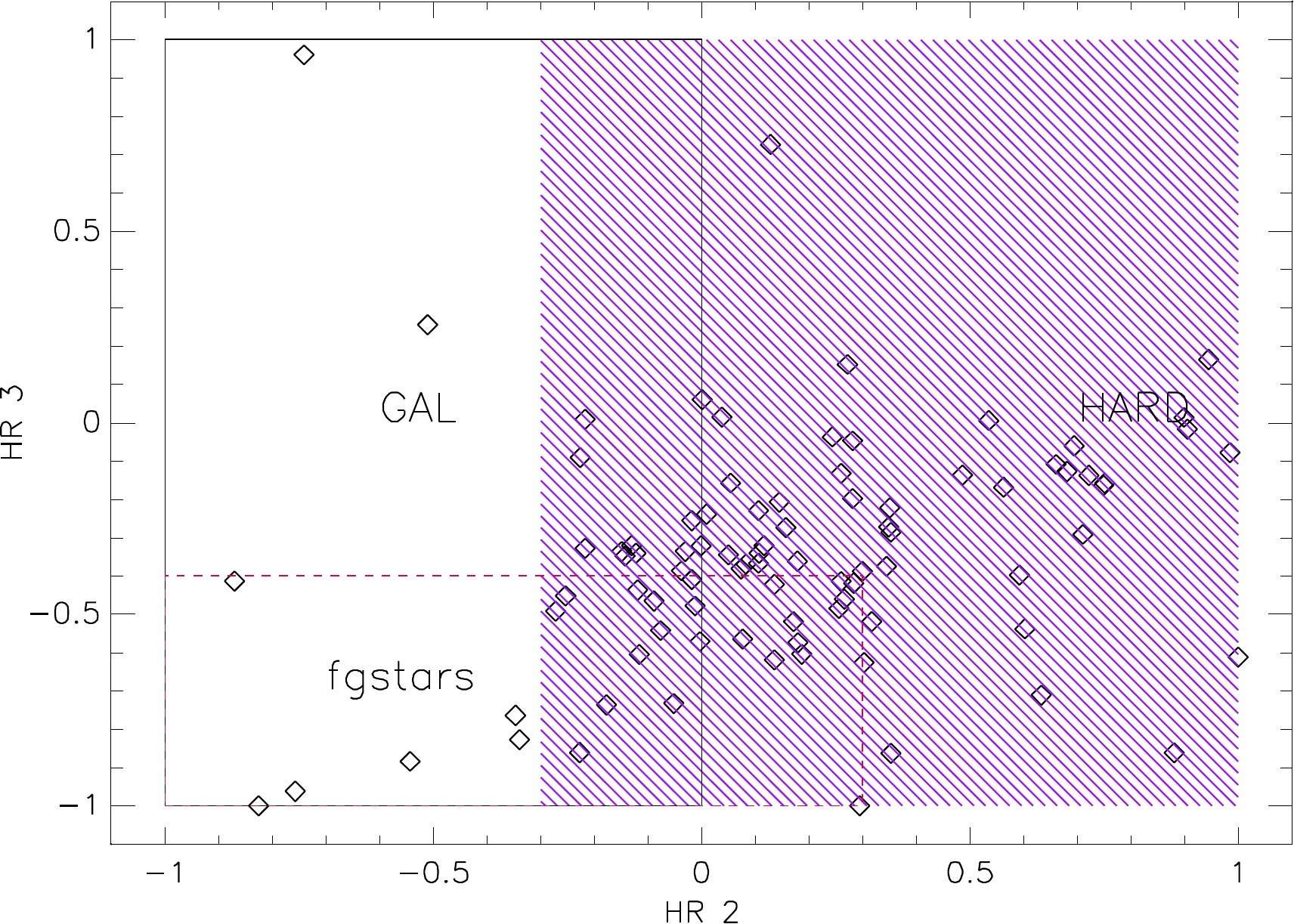}}\\
\subfigure[Leo I dSph HRs]
   {\includegraphics[width=5cm]{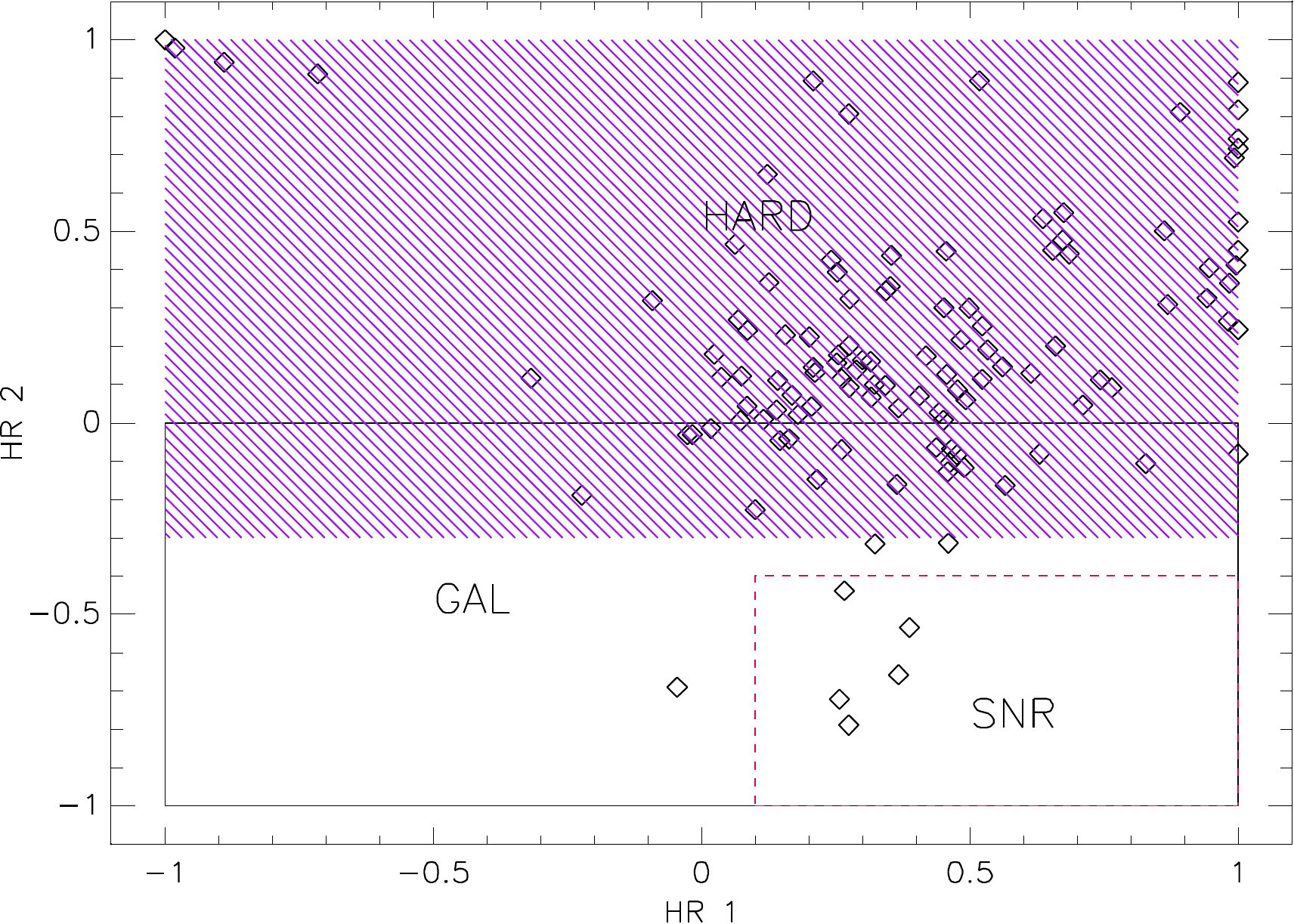}}\qquad\qquad
   {\includegraphics[width=5cm]{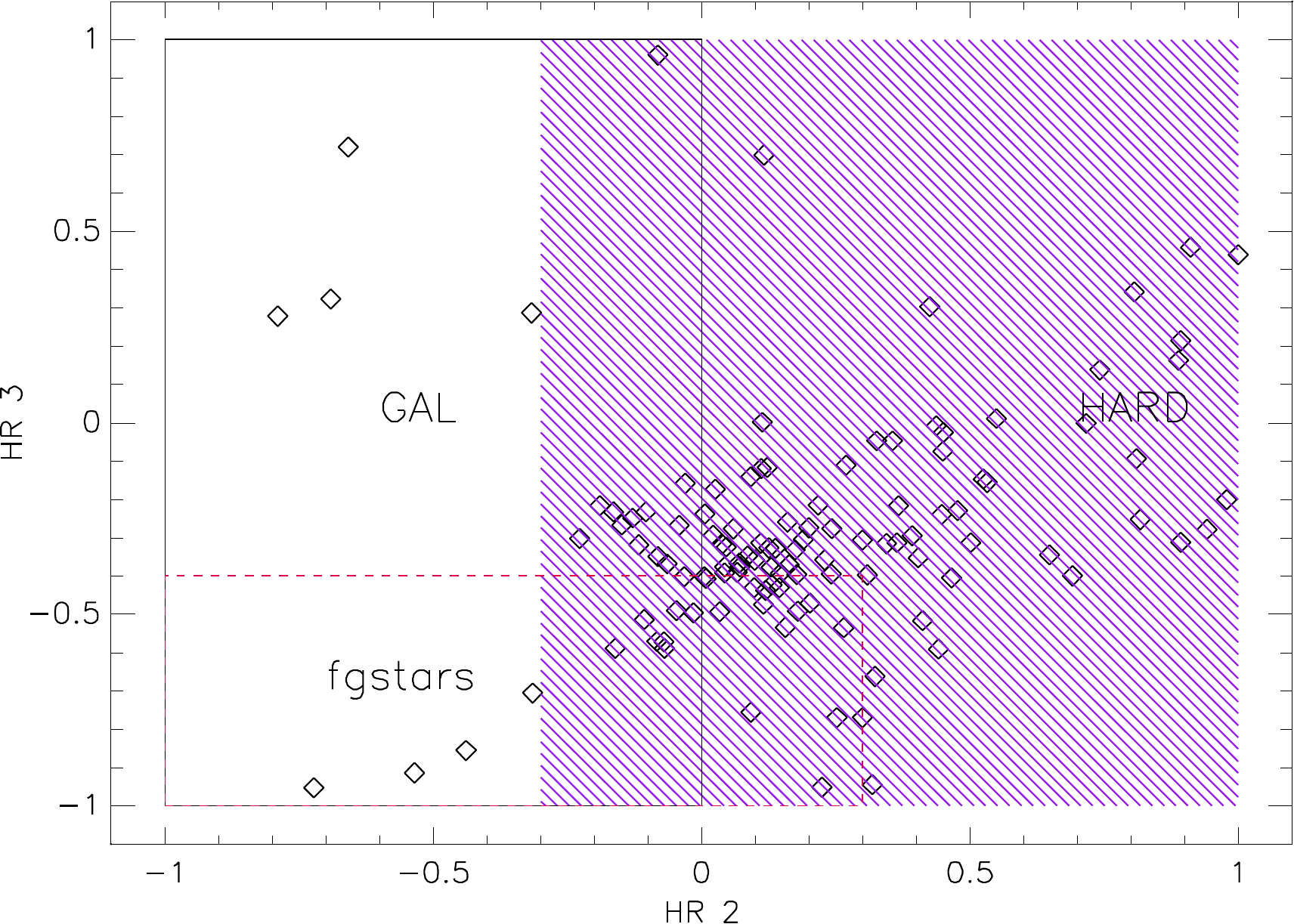}}\\
\subfigure[UMa II dSph HRs]
   {\includegraphics[width=5cm]{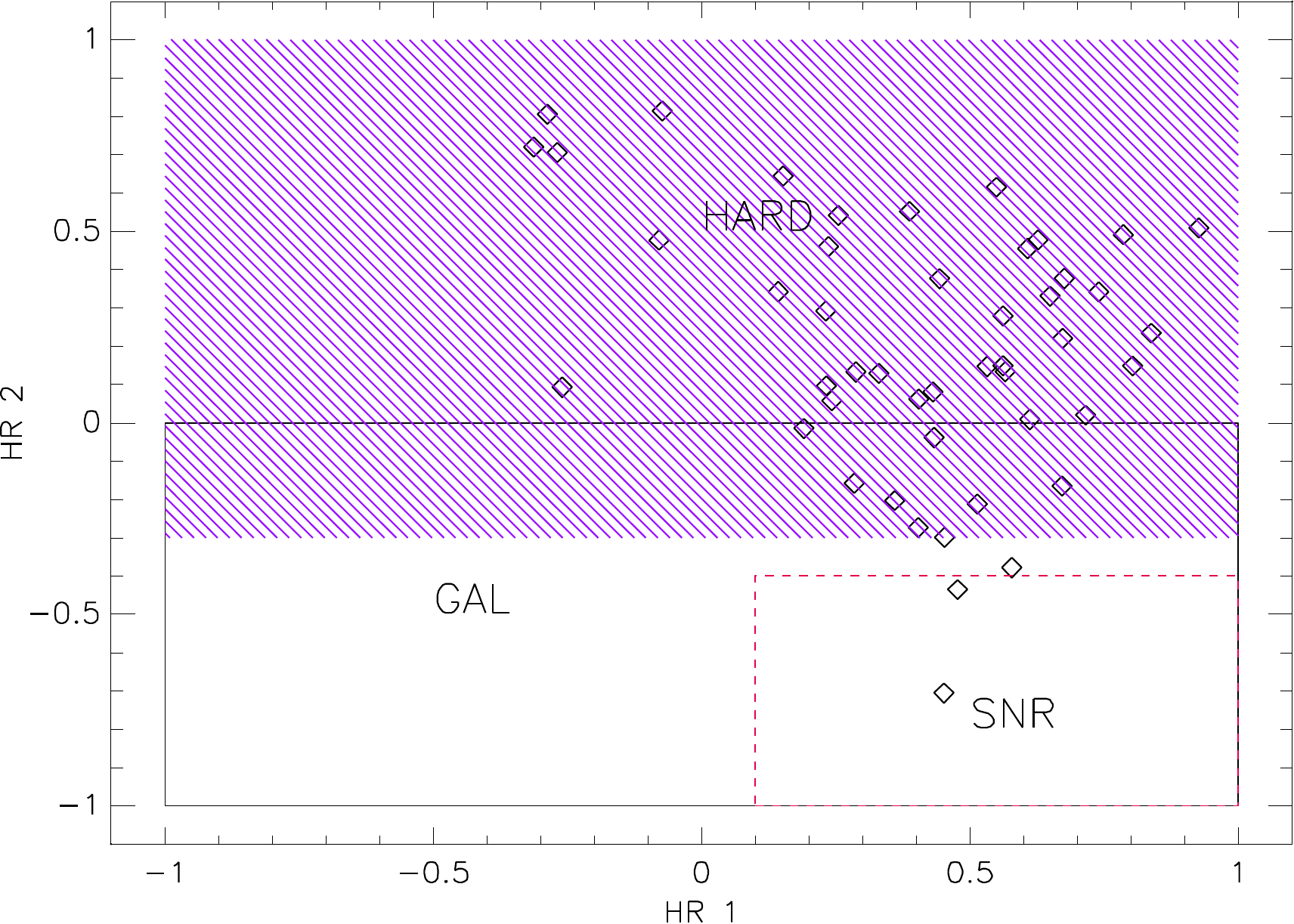}}\qquad\qquad
   {\includegraphics[width=5cm]{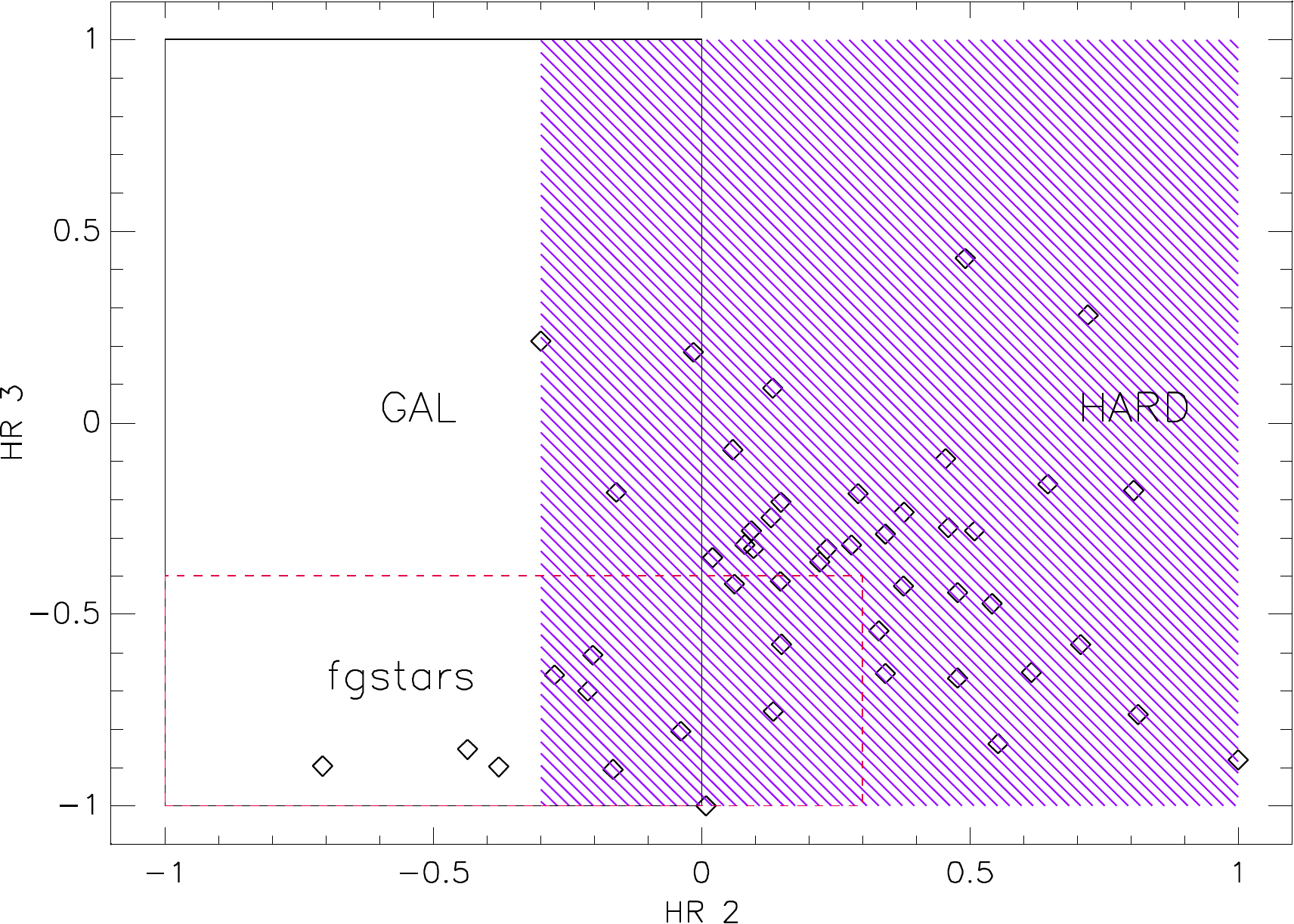}}\\
\subfigure[UMi dSph HRs]
   {\includegraphics[width=5cm]{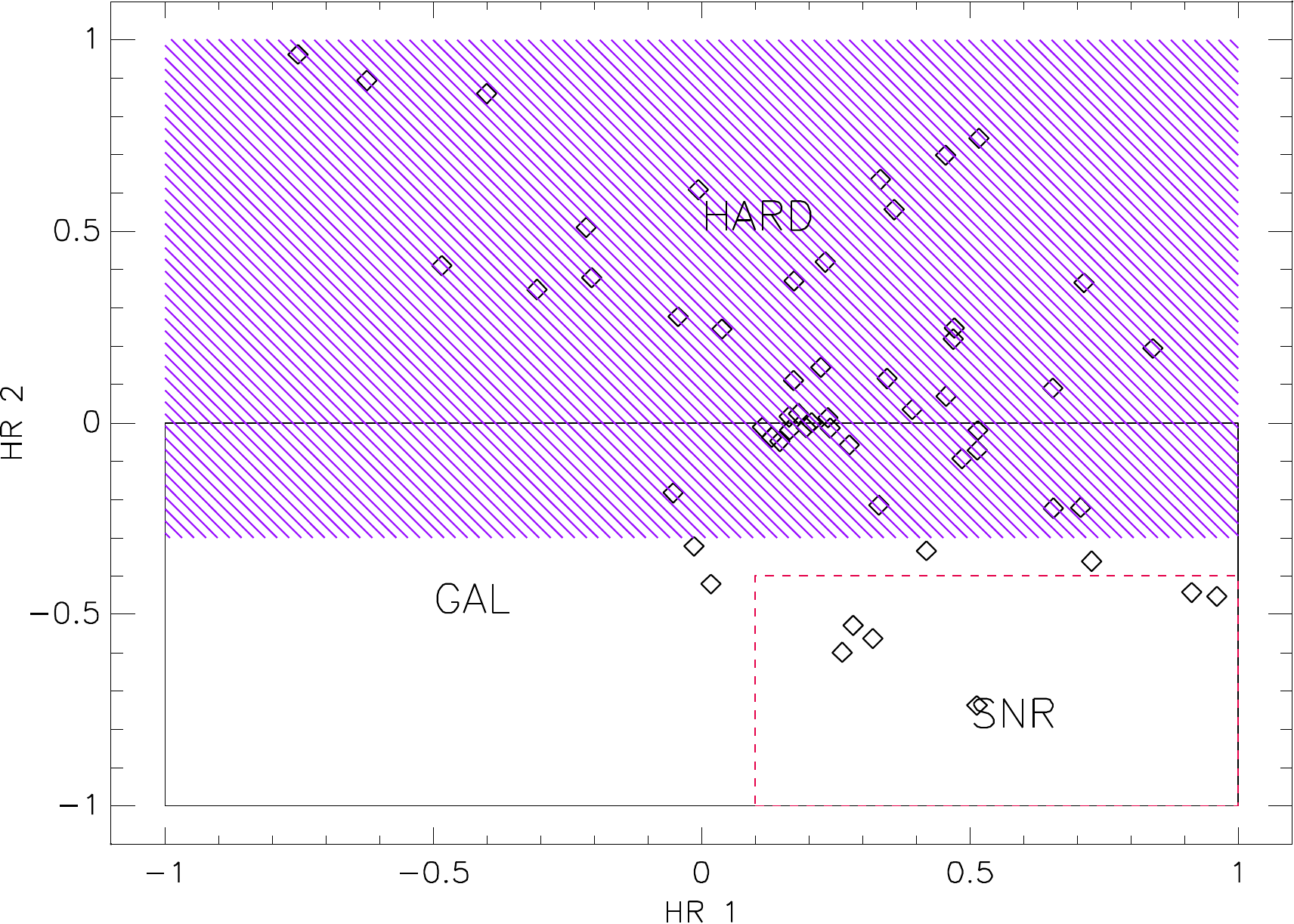}}\qquad\qquad
   {\includegraphics[width=5cm]{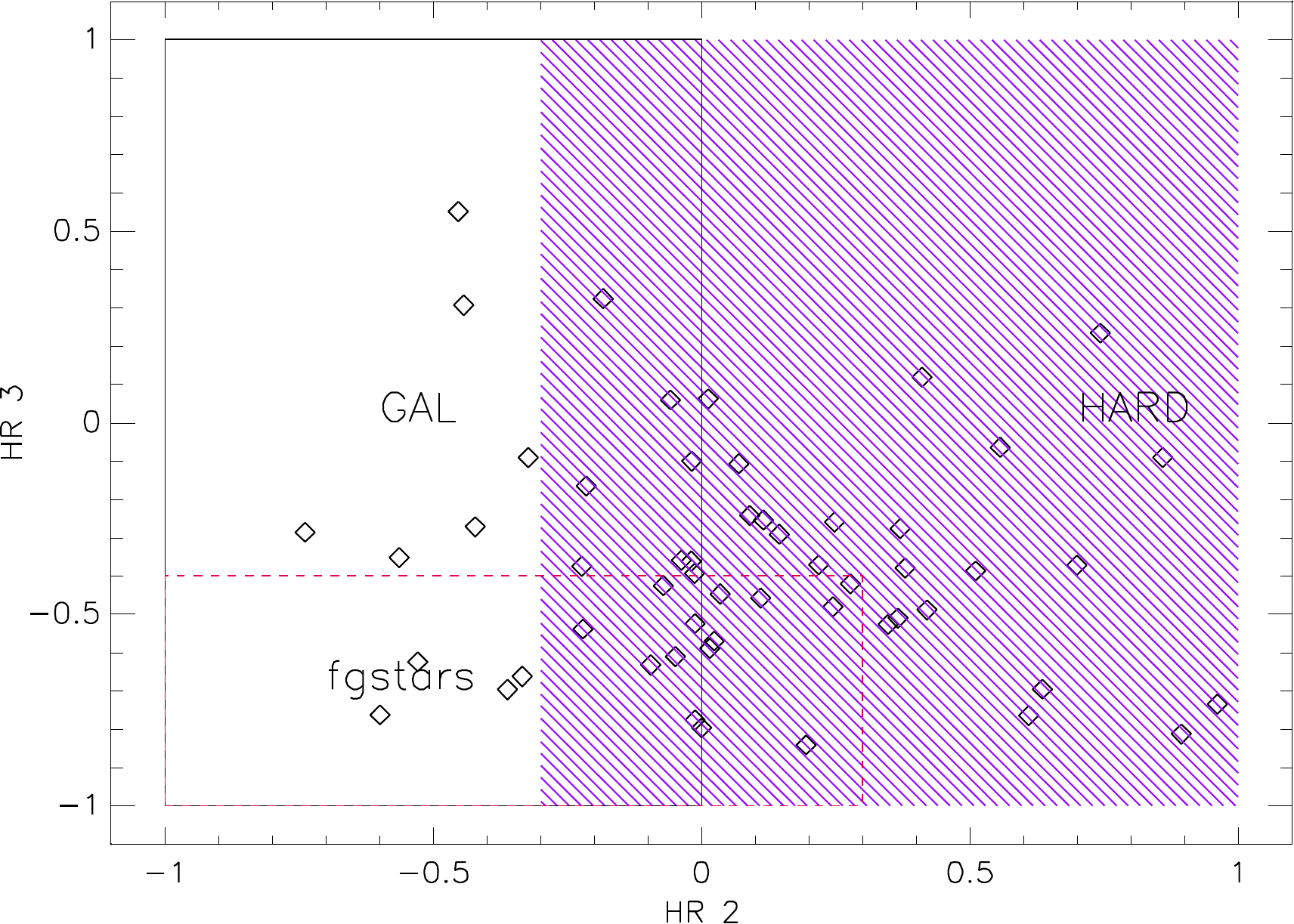}}\\
\caption{The HRs color-color diagram for our X-ray sources (see text for details).}
\label{figHR}
\end{figure*}

    Moreover we follow the method of \cite{Pietsch2004}, already used by \cite{Bartlett2012}
    to classify high-energy sources in Phoenix dSph. 
    These authors statistically evaluated the nature of the sources by considering some criteria based on HR values 
    definied as $HR_i=(B_{i+1}-B_{i})/(B_{i+1}+B_{i})$, with the count rates $B_i$ of the bands already given in S 3. 
    We get our own criteria, summarized in Table \ref{mycriteria}, using the criteria fixed by \cite{Bartlett2012} and 
    requesting the relevant classification in SIMBAD (for results of our classification, see Fig. \ref{figHR} and Table \ref{total}). 
    The sources are labelled identified (i), classified (cl) or candidate (ca) if they fulfill all, the mayority or only few criteria, respectively.
 
    Furthermore, in the last column of Tables \ref{Dracoclass}-\ref{UMiclass} we report the source classification (HR Class.) 
    based on these criteria. 
    Here, FG [fg/f] point out the identified [classified/candidate] foreground stars, snr the classified Super Nova Remnants,
    A the identified Active Galactic Nuclei, G [g] the identified [classified] galaxies, sss the classified 
    super soft sources, h the classified hard sources and Loc [loc/l] the identified [classified/candidate] 
    local X-ray sources.

    As the reader can see, the obtained results allow to get a wider X-ray source classification and show a good agreement 
    with the previous inferences.

\begin{table*}
\begin{center}
\begin{tabular}{cc}
\hline
Src Type & Criteria                    \\
\hline
fg star &  classified as star in SIMBAD, log$\left( \frac{f_X}{f_{opt}}\right)<-1.0$ and $HR_2<0.3$ and $HR_3<-0.4$ or not defined    \\ 
SNR &  classified as SNR in SIMBAD, $HR_1>0.1$ and $HR_2<-0.4$ and not a fg star   \\                            
AGN &  classified as AGN, Sy1 or QSO in SIMBAD, not classification as SNR from $HR_2$          \\ 
GAL &  classified as G in SIMBAD, Optical id with galaxy and $HR_2<0.0$   \\                            
SSS &  $HR_1<-0.2$, $HR_2$ - $EHR_2<-0.99$  or $HR_2$ not defined, $HR_3$ and $HR_4$ not defined       \\ 
HARD & $HR_2 - EHR_2>-0.2$ or only $HR_3$ and $HR_4$ defined and no other classification  \\ 
LX & not a fg star, classified as star in SIMBAD, redshift compatible with dSph's one \\                             
\hline
\end{tabular}
\caption{Our source classification criteria. Here LX stands for "Local X-ray source". \label{mycriteria}}
\end{center}
\end{table*}

\begin{table*}
\begin{tabular}{|c|c|c|c|c|}
\hline
Type      & Draco                     & Leo I             & UMa II      & UMi   \\
\hline
fg star   &  5$^i$+7$^{cl}$+13$^{ca}$ &  2$^i$+24$^{ca}$  &  11$^{ca}$  &  1$^i$+8$^{cl}$+8$^{ca}$  \\ 
SNR       &  1$^{cl}$                 &  2$^{cl}$         &   0         &  4$^{cl}$\\                            
AGN       &  11$^i$                   &  1$^i$            &  1$^i$      &  2$^i$ \\ 
GAL       &  1$^{cl}$                 &  1$^i$+1$^{cl}$   &  0          &  0 \\                            
SSS       &  0                        &  0                &  0          &  0 \\ 
HARD      &  28$^{cl}$                &  72$^{cl}$        &  31$^{cl}$  &  10$^{cl}$ \\
LX        &  1$^i$+1$^{cl}$+13$^{ca}$ &  1$^{ca}$         &  1$^{cl}$   &  16$^{ca}$ \\
not class &  8                          &  12               &  5          &  5 \\                             
\hline
\end{tabular}
\caption{Results of dSphs sources classification made using Table \ref{mycriteria} (see the text for details).\label{total}}
\end{table*}

\subsection{Background sources}
As suggested in the previous discussion, some of the detected sources may be AGNs or background galaxies. We estimate 
the expected number of background sources towards our targets through the logN-logS diagram (\citealt{Hasinger2005}).
Starting from the minimum absorbed fluxes (in the 0.2-12.0 keV energy band) and assuming  
$\Gamma$ and $N_H$ values as in the previous section, we estimated, via 
webPIMMS v3.9, 
the unabsorbed fluxes in the 0.5-2.0 keV energy band to be\\ 
$F^{Unabs}_{0.5-2.0}=1.67\,\times\,10^{-15} {\rm erg\, s^{-1}\, cm^{-2}}$ for Draco,\\
$F^{Unabs}_{0.5-2.0}=4.73\,\times\,10^{-16} {\rm erg\, s^{-1}\, cm^{-2}}$ for Leo I,\\
$F^{Unabs}_{0.5-2.0}=2.15\,\times\,10^{-15} {\rm erg\, s^{-1}\, cm^{-2}}$ for UMa II,\\
$F^{Unabs}_{0.5-2.0}=1.80\,\times\,10^{-15} {\rm erg\, s^{-1}\, cm^{-2}}$ for UMi.\\
Using these values as input parameters in the Hasinger relation, we find the
expected number of background AGNs as a function of the angular distance from the galaxy center. 
In order to compare the theoretically estimated source number with the observed one, 
we divided each dSph field of view (FOV) 
into five rings and performed therein the comparison. 
As it can be seen in Table \ref{hasinger}, the number of X-ray sources in each annulus 
is generally consistent with the expected one, although this is not the case for the external annuli since the logN-logS does 
not account for border effects.

Nevertheless, we cannot rule out that some sources actually belong to MW or to the dSphs. 
This claim is supported by the presence of X-ray sources in the bottom-left region of the panels in Fig. \ref{figclass} and  
the correlations with variable stars catalogues and the analysis summarized in the following section as well.

\begin{table*}
\footnotesize
\centering
\subtable[Draco dSph\label{hasinger_Draco}]{
  \begin{tabular}{|c|c|c|c|c|}
  \hline
  Annulus & R$_{in}$ & R$_{ex}$ & \# Exp & \# Obs       \\
          & (arcmin) & (arcmin) &        &              \\
  \hline
  1 & 0.00 &  0.76  &  0.2 $\pm$  0.1  &  0 \\
  2 & 0.76 &  3.60  &  5.6 $\pm$  1.0  &  7 \\
  3 & 3.60 &  6.50  & 13.3 $\pm$  2.4  & 13 \\
  4 & 6.50 &  9.00  & 17.6 $\pm$  3.1  & 22 \\
  5 & 9.00 & 16.00  & 79.4 $\pm$ 14.1  & 46 \\
  \hline
  \end{tabular}} \qquad \qquad \qquad \qquad
\subtable[Leo I dSph\label{hasinger_LeoI}]{
  \begin{tabular}{|c|c|c|c|c|}
  \hline
  Annulus & R$_{in}$ & R$_{ex}$ & \# Exp & \# Obs       \\
          & (arcmin) & (arcmin) &        &              \\
  \hline
  1 & 0.00 &  0.76  &  0.6 $\pm$  0.1  &  0 \\
  2 & 0.76 &  3.60  & 13.4 $\pm$  2.2  & 11 \\
  3 & 3.60 &  6.50  & 31.4 $\pm$  5.0  & 26 \\
  4 & 6.50 &  9.00  & 41.7 $\pm$  6.8  & 25 \\
  5 & 9.00 & 16.00  & 188.3$\pm$ 30.7  & 53 \\
  \hline
  \end{tabular}}\\
\subtable[UMa II dSph\label{hasinger_UMaII}]{
  \begin{tabular}{|c|c|c|c|c|}
  \hline
  Annulus & R$_{in}$ & R$_{ex}$ & \# Exp & \# Obs       \\
          & (arcmin) & (arcmin) &        &              \\
  \hline
  1 & 0.00 &  0.76  &  0.2 $\pm$ 0.1  &  0 \\
  2 & 0.76 &  3.60  &  4.5 $\pm$  0.9  &  3 \\
  3 & 3.60 &  6.50  & 10.9 $\pm$  2.0  & 13 \\
  4 & 6.50 &  9.00  & 14.4 $\pm$  2.6  & 11 \\
  5 & 9.00 & 16.00  & 65.1 $\pm$ 11.6  & 22 \\                                       
  \hline              
  \end{tabular}} \qquad \qquad \qquad \qquad
\subtable[UMi dSph\label{hasinger_UMi}]{
  \begin{tabular}{|c|c|c|c|c|}
  \hline
  Annulus & R$_{in}$ & R$_{ex}$ & \# Exp & \# Obs       \\
          & (arcmin) & (arcmin) &        &              \\
  \hline
  1 & 0.00 &  0.76  &  0.3 $\pm$  0.1  &   1 \\
  2 & 0.76 &  3.60  &  6.1 $\pm$  1.8  &   5 \\
  3 & 3.60 &  6.50  & 14.6 $\pm$  4.3  &  10 \\
  4 & 6.50 &  9.00  & 19.2 $\pm$  5.6  &  12 \\
  5 & 9.00 & 16.00  & 86.8 $\pm$  25.4 &  26 \\
  \hline
  \end{tabular}}
\caption {List of sources expected through the logN-logS diagram and observed in annuli around the galaxy centers of our sample. 
Here, $R_{in}$ and $R_{ex}$ represent the interior and exterior annulus radii, respectively.}
\label{hasinger}
\end{table*}

\subsection{Possible local stars}
As next we investigate the local nature of some of the detected high-energy sources.

Draco sources 3, 14 and 42, recognized as stellar objects in Fig. \ref{figclass} (a), 
are not associated with any of the sources in the PPMX catalogue and so it cannot be excluded that they are local high-energy stellar systems.
Src 30 has a ``carbon like star counterpart'' at a distance of 0.38$''$, as reported by the 1921/table9 catalogue that collects many kinds 
of stars belonging to Draco. Sources 60, 64 and 77 correlate with late-type stars in Draco but Scr 64 and 77 have closer 
counterparts in two background object catalogues as well. 
Finally, Src 84 is associated with a Draco variable star (within 1$''$) of the 127/861 catalogue but also to a QSO 
at distance of 0.52$''$ [0.55$''$] in the 1921/table10 [1223/qsos] catalogue. Its local nature is
not so clear and we must conclude that Src 84 is likely linked to a background source, owing to the smaller distance from this kind of counterpart.
The background nature of the last three sources is confirmed by the SIMBAD database. 
In the same data collection we find clues for stellar nature of X-ray sources toward/in Draco dSph.

As regards Leo I dSph, we find twenty correlations between our catalogue and 1475/table2 one, a NIR catalogue of Leo I stars.
One of these sources (Scr 103) also correlates (within $\simeq2.27''$) with a QSO candidate, so we infer a possible background nature for it.
Instead Scr 6 seems to have a stellar nature (as already seen in Sect. 4.1) and no association with PPMX sources, therefore we argue that it belongs to Leo I galaxy.\\

In the UMa II case, sources 1, 7 and 8 may belong to this dSph because of their stellar nature (see the aforementioned Sect. 4.1) and no association with any 
PPMX object. Furthermore sources 7 and 8 correlate with objects in II/313 catalogue, a list of not variable sources. According to their
``stellarity index'', a star/galaxy discriminator defined by \citet{ofek2012}, these two sources seem not to have an extended 
profile and could be UMa II stars, lacking foreground hints.

As for UMi, sources 1 and 13 are probably stars that belong to UMi 
because they are recognized as likely stellar objects in Fig. \ref{figclass} (d) 
but don't show any correlations with the PPMX catalogue.

We also perform a more restrictive source analysis, using the method deduced from \cite{Pietsch2004} and 
\cite{Bartlett2012} and considering as local sources only those identified or classified as such. Therefore, 
in the Draco dSph case we have two local sources (the identified 29 and the classified 14), only one (the classified 14) 
for UMa II and none as for Leo I and UMi. A subsequent follow-up (also in other wavelengths) would be 
useful to fully probe the nature of such objects. In addition, it could have some implications in the 
formation and evolution scenarios of these galactic systems.

Apart from the aforementioned sources, about thirty of them remain at the candidate stage. So, a further study could 
settle the issue if they really belong to the dSphs or not.

\subsection{Clues for IMBHs}
\begin{figure}[!t]
\centering
\subfigure[Draco dSph\label{center_Draco}]
{\includegraphics[width=8.5cm]{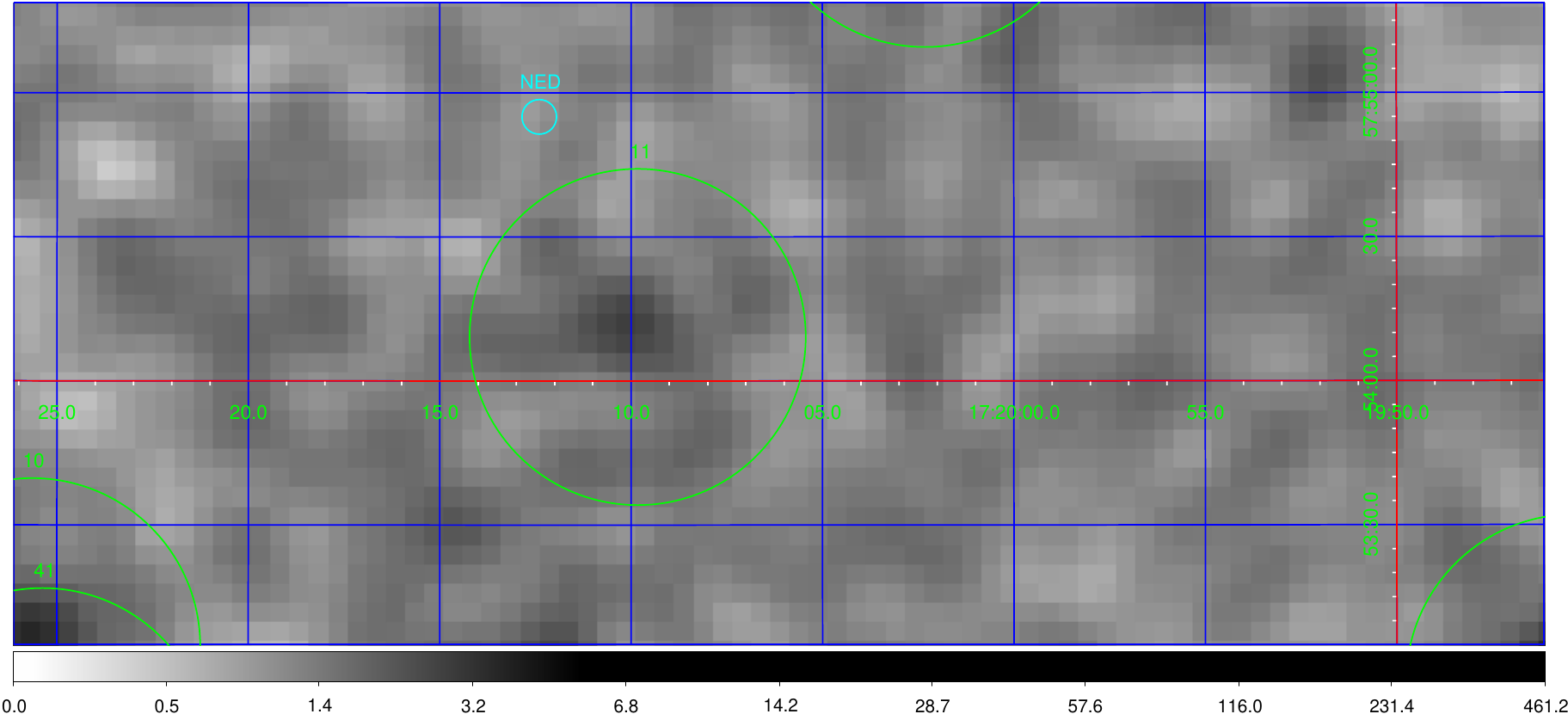}}\qquad\qquad
\subfigure[Leo I dSph\label{center_LeoI}]
   {\includegraphics[width=8.5cm]{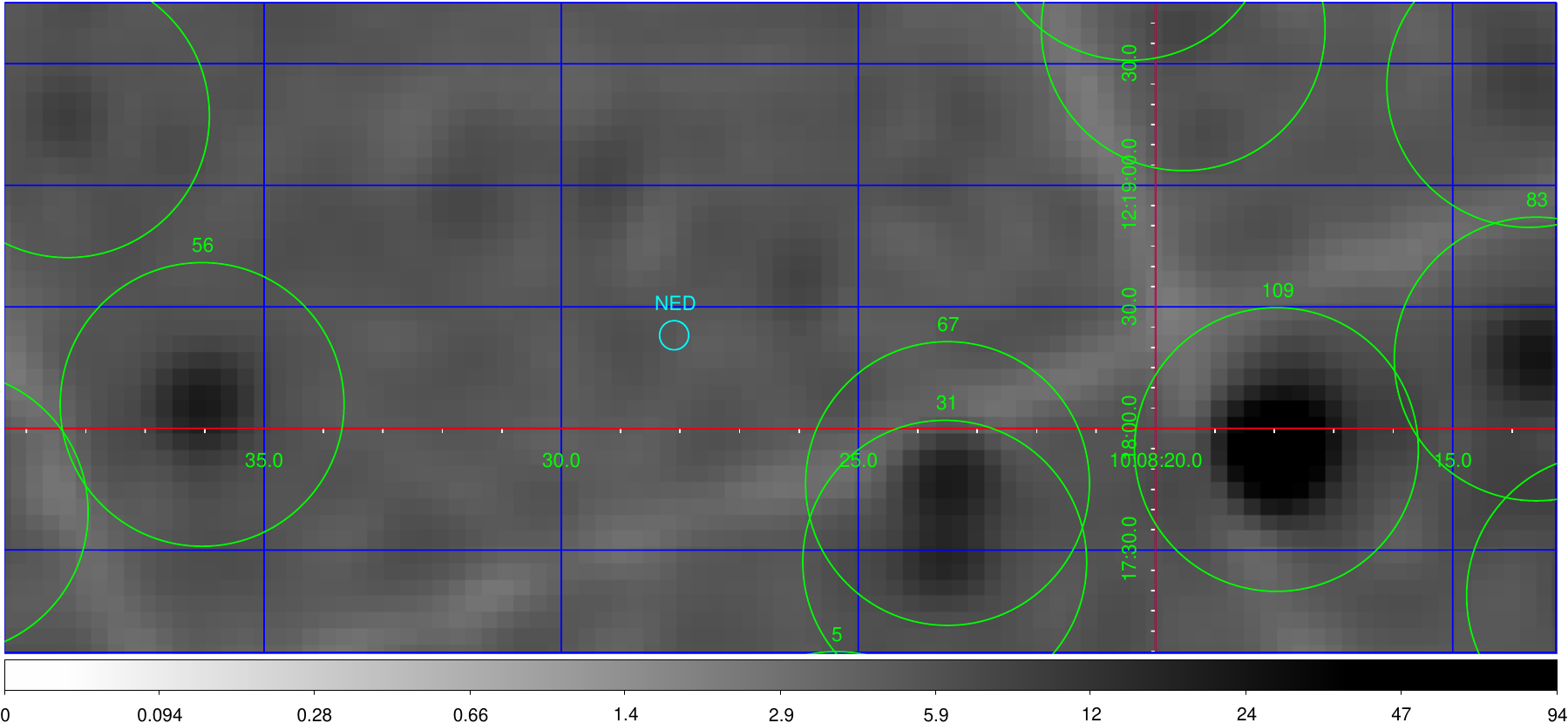}}\qquad\qquad
\subfigure[UMa II dSph\label{center_UMaII}]
{\includegraphics[width=8.5cm]{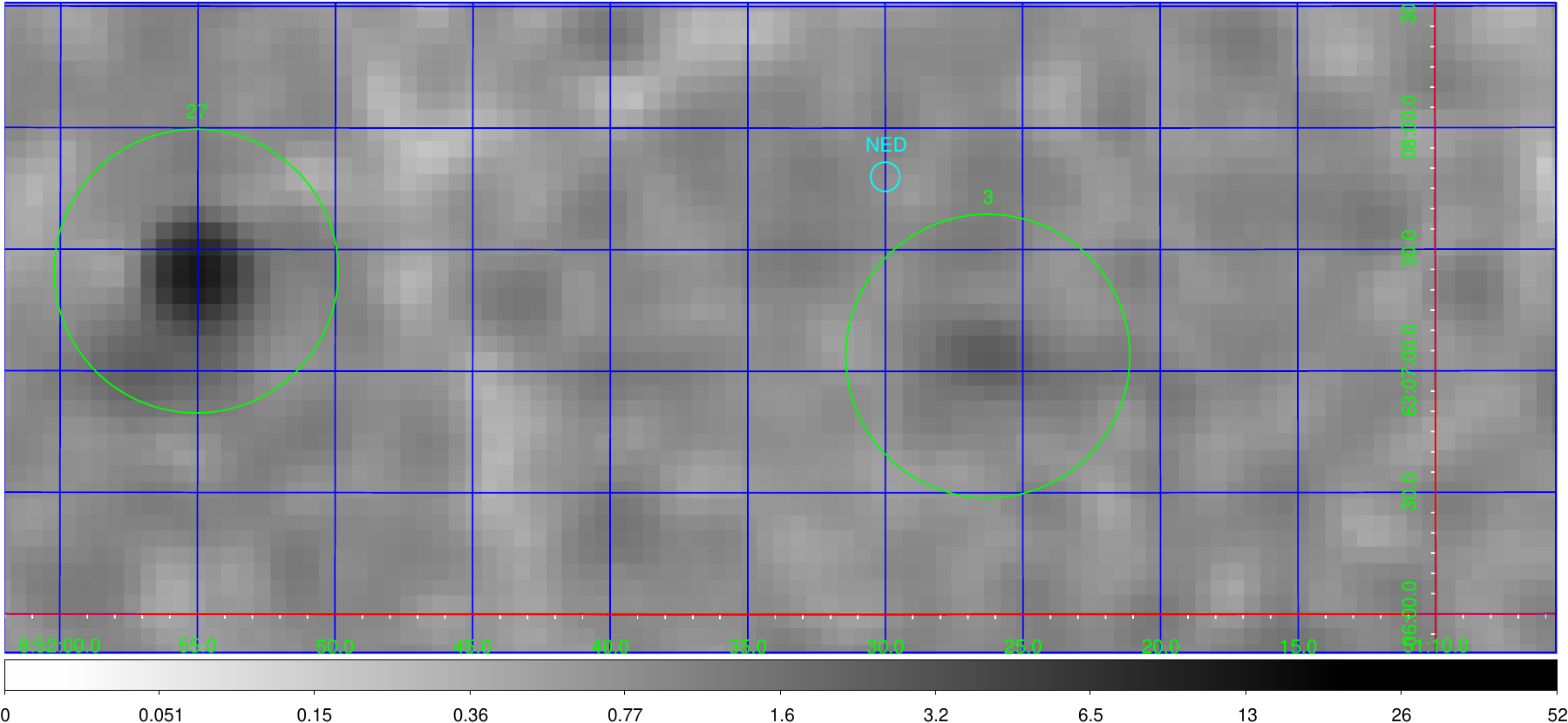}}\qquad\qquad
\subfigure[UMi dSph\label{center_UMi}]
   {\includegraphics[width=8.5cm]{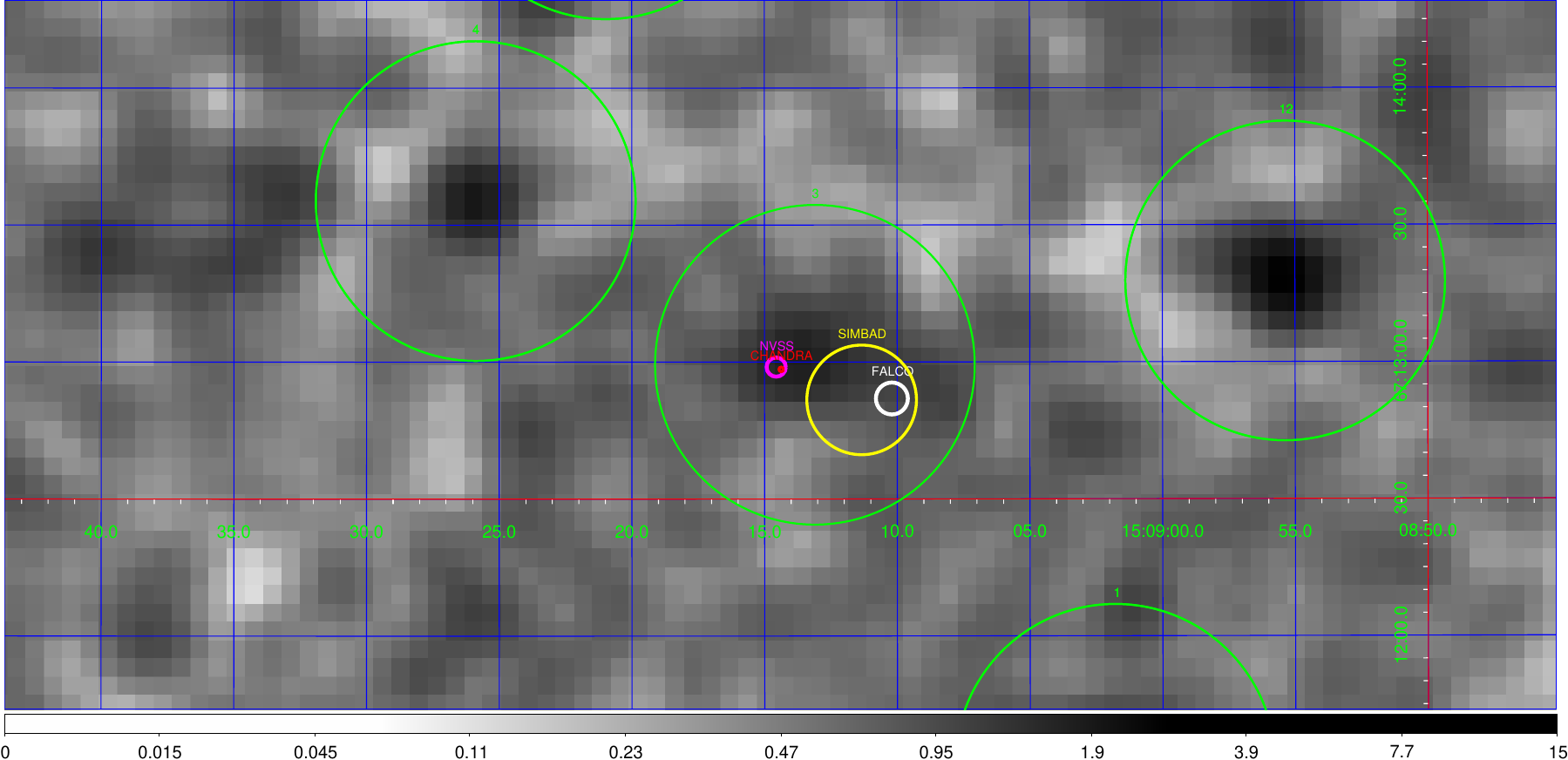}}
\caption{Zoom view of the four dSph centers. In the last panel, the green, red and magenta circles represent the position of the 
X-ray source identified at the center of UMi and its counterparts for Chandra and NVSS, respectively. The yellow and white circles 
show the centers of the galaxy as reported in \citet{falco1999} and SIMBAD database. }
\label{center}
\end{figure}

Both the M$_{BH}$-$\sigma$ and M$_{BH}$-M$_{Bulge}$ relations (e.g., \citealt{gebhardt2000} and
\citealt{ferrarese2000}) suggest to search for compact objects, belonging to IMBH's
range, toward dSphs. For this reason we investigate the presence of BHs in their cores,
where these X-ray sources are likely located.\\
The closest source (Src 11) to the Draco center is at distance of about 51$''$,
Src 67 [Src 3] at 76$''$ [51$''$] from Leo I [UMa II] center (see Fig. \ref{center}).
These distances are well above the source error boxes (1.2$''$,
0.7$''$ and 1.4$''$, respectively) so we infer that none of the high-energy sources is
located at the center of these dSphs.
{Nevertheless, using the minimum unabsorbed fluxes and the distances reported by NED, we estimated an upper limit
to the luminosity (in the 0.2-12.0 keV energy band) of the compact central object (if any),
getting about 4.75$\times$10$^{33}$ erg s$^{-1}$,
1.21 $\times$10$^{34}$ erg s$^{-1}$ and 1.05$\times$10$^{33}$ erg s$^{-1}$ for Draco, Leo I and UMa II, respectively.\\}

Scaling the previous minimum fluxes to the 0.5-2.0 keV energy band, we obtained, via the logN-logS relation, the expected
number of the background sources (within 25$''$): 0.08$^{+0.18}_{-0.03}$, 0.18$^{+0.07}_{-0.06}$ and 0.06$^{+0.10}_{-0.02}$, respectively.
Being well below the unity, we infer that if any source exist in such a region, it should be located within the dwarf galaxies.\\
Despite any clear identification of central object, we also put an upper limit to the putative central BH mass (M$_{BH}$) by
assuming it accretes through the Bondi-Hoyle spherical model. \citet{bondi1944} claimed that when a BH moves with
velocity $v$ through a gaseus medium, characterized by an hydrogen number density $n$, it accretes at a rate given by
\begin{equation}
\dot{M}\simeq \frac{4\pi(GM_{BH})^2\,m_p\,n}{(v^2+c_s^2)^{3/2}}~,
\end{equation}
where $m_p$ is the proton mass and $c_s$ the sound speed in the medium. Then, the subsequent X-ray luminosity
$L_X\simeq \epsilon \eta \dot{M} c^2$ can be rewritten as
\begin{equation}
L_X\simeq \epsilon \eta 8.8\times10^{36}\left(\frac{M_{BH}}{10^3 ~{\rm M}_{\odot}}\right)^2 \left(\frac{V}{15~{\rm km~s^{-1}}}\right)^{-3}
\left(\frac{n}{0.1 ~{\rm cm^{-3}}}\right)~{\rm erg~s^{-1}}~,\nonumber
\label{explum}
\end{equation}
with $V=(v^2+c_s^2)^{1/2}$, while $\epsilon$ is the efficiency in converting mass to radiant energy and $\eta$ 
the fraction of the Bondi-Hoyle accretion rate onto the BH.

Assuming $\nu\simeq c_s\simeq$ 10 km s$^{-1}$ (consequently V $\simeq$ 15 km s$^{-1}$) and $n$ in the range
10$^{-3}$ - 10$^{-1}$ cm$^{-3}$ as typical values for this kind of system, we obtain that the IMBHs (if any)
at the center of these galaxies must have an upper limit mass of $\simeq 100$ M$_{\odot}$. Of course, this estimate scales as
$(\epsilon\eta)^{-0.5}$.
We also searched for radio sources in the NRAO VLA Sky Survey (NVSS) (within 100$''$ from the dSph centers)
with the aim of obtaining a better BH mass estimate. However, no radio source was found.

\subsection{The case of UMi dSph}
{The possibility that UMi dSph hosts a BH in its center was already studied quite intensely. For example, 
\citet{maccarone2005} found a radio source within 3 $\sigma$ error box from the galaxy center and, assuming the existence of an accreting 
IMBH and a gas density as low as  1/30-1/100 of the typical density in globular clusters, these authors realized that a central black hole with mass 
of few 10$^5$ M$_{\odot}$  would be necessary in order to explain the observed radio flux. Furthermore,  \citet{lora2009} inferred an upper
limit of 2-3$\times$10$^4$ M$_{\odot}$ using N-body simulations of the tidal disruption
of a long lived substructure observed on the north-east side of the UMi major axis. It was also claimed that the compact 
object may not residing in the galactic center but offset from it because of gravitational effects.}

\citet{nucita2013b} analyzed the X-ray data acquired by the Chandra satellite for $\simeq 19.8$ ks in 2011 (Obs. id. 12754) and
found an X-ray source (at J2000 coordinates RA = 15$^{\rm h}$ 09$^{\rm m}$ 14.37$\second$ and Dec = 67$^\circ$ 12$'$ 58.4$''$, with associated error of $\simeq 0.52''$) with an unabsorbed
0.5-7.0 keV flux of F$_{0.5-7.0 keV}^{Unabs}\simeq$4.9$\times 10^{-15}$ erg s$^{-1}$. The X-ray source was spatially coincident
(within $\simeq$1.2$''$) with a source (the same already identified in the radio band by \citealt{maccarone2005}) 
with a flux density of 7.1$\pm$0.4 mJy at 1.4 GHz. 
Such radio source 150914+671258 was detected by NRAO VLA Sky Survey (NVSS) at J2000 coordinates RA = 15$^{\rm h}$ 09$^{\rm m}$ 14.56$\second$ and Dec = 67$^\circ$ 12$'$ 58.9$''$ and the relevant positional error is
$\simeq$2.1$''$ as obtained from the sum in quadrature of the uncertainties associated to the two coordinates. Under the assumption that the observed source is an accreting IMBH, \citet{nucita2013b}
used the fundamental plane relation to evaluate the BH mass, which resulted to be $\left(2.9_{-2.7}^{+33.6}\right)\times$10$^6$M$_{\odot}$.
As stressed by the same authors, the detection algorithm ({\it wavdetect}) was run with a significance threshold of $10^{-6}$ corresponding
to the possibility to have at least one false detection in the CCD where the target is found. Hence, the source was detected by Chandra only
at $\simeq 2.5\sigma$ confidence level.

Here, in order to test the robustness of this result, we analyzed the 2005 {\it XMM}-Newton data and found that, as shown in Fig. \ref{center_UMi},
an X-ray source (Src 3) is clearly detected and superimposed to the location of the galaxy center.
In particular,  Src 3 (with J2000 coordinates RA = 15$^{\rm h}$ 09$^{\rm m}$ 13.1$\second$ and Dec = 67$^\circ$
12$'$ 59.4$''$) has a distance of $\simeq$30$''$ from UMi center (white circle) as reported by \citet{falco1999}
\footnote{\citet{falco1999} assigned a positional uncertaintly of $\simeq$3.5$''$, obtained as a sum in quadrature of the
errors ($\simeq$2.5$''$) associated on RA and Dec, to UMi galaxy center.}, and a galactocentric distance $\simeq$24$''$ if the SIMBAD database coordinates
\footnote{UMi dSph is located, as reported by SIMBAD (yellow circle), at J2000 coordinates
RA = 15$^{\rm h}$ 09$^{\rm m}$ 11.34$\second$ and Dec = 67$^\circ$ 12$'$ 51.7$''$ with a positional uncertaintly
of $\simeq$12.2$''$, due to the sum in quadrature of the RA ($\simeq$2$''$) and Dec ($\simeq$12$''$) errors.}
are considered. In the same Figure, we give the position of the X-ray source detected by Chandra (1'' red circle) and the associated NVSS radio counterpart (2'' magenta circle). Note that the distance between 
the {\it XMM}-Newton and Chandra sources is $\simeq 7.4''$. 

Src 3 is characterized by an absorbed 0.2-12.0 keV flux of 7.3$\pm$2.9)$\times$10$^{-15}$ erg s$^{-1}$ cm$^{-2}$ which
corresponds to an unabsorbed flux  of  F$^{Unabs}_{0.2-12.0~{\rm keV}}$=(7.90$\pm$3.14)$\times$10$^{-15}$ erg s$^{-1}$ cm$^{-2}$. The
unabsorbed  2-10 keV band flux is F$^{Unabs}_{2-10~{\rm keV}}$=(4.05$\pm$1.61)$\times$10$^{-15}$ erg s$^{-1}$ cm$^{-2}$.

Assuming a distance of 73$\pm$10 kpc, as reported by NED, we are left with a luminosity
L$^{Unabs}_{0.2-12.0~{\rm keV}}$=(4.88$\pm$2.37)$\times$10$^{33}$ erg s$^{-1}$ and
L$^{Unabs}_{2-10~{\rm keV}}$=(2.50$\pm$1.21)$\times$10$^{33}$ erg s$^{-1}$.

Following \citet{nucita2013b} and the references therein, we can get
constraints on the possible central BH parameters. In fact, from the correlation
between X-ray Src 3 and the radio 150914+671258 source (at distance of $\simeq$3.6$''$),
we can estimate the BH mass using the fundamental plane relation (\citealt{merloni2003}, \citealt{koerding2006})
\begin{equation}
\label{fundamental}
log{L_R}=\xi_{RX}log(L_X)+\xi_{RM}log(M_{BH})+b_R
\end{equation}
where the mass and the luminosities are in units of solar masses M$_\odot$ and erg s$^{-1}$, respectively, $\xi_{R_X}$=
0.60$^{+0.11}_{-0.11}$, $\xi_{R_M}$=0.78$^{+0.11}_{-0.09}$ and $b_{R}$=7.33$^{+4.05}_{-4.07}$.
Solving for the black hole mass, \citet{merloni2003} obtained
\begin{equation}
\log (M_{BH}) \simeq 16.3 +\log (D) + 1.28\left[(\log (F_R) -0.60 \log (F_X)\right] \pm 1.06~,
\label{eqfin}
\end{equation}
where $D$ is the source distance expressed in Mpc and the last term is a consequence of the scatter in the
fundamental plane relation.

From the 1.4 GHz radio flux density of the source 150914+671258, we obtained the 5 GHz one, assuming
$F(\nu)\propto \nu^{-\alpha_R}$. So, assuming a flat source spectrum ($\alpha_R \simeq 0$), we get
$F(5~{\rm  GHz})\simeq 7.1$ mJy, consequently a radio flux $F_R=(3.5\pm0.2)\times 10^{-16}$ erg cm $^{-2}$ s$^{-1}$
and a radio luminosity of $L_R=(2.19\pm0.62)\times 10^{32}$ erg s$^{-1}$.

Using the previous estimate of the radio (5 GHz) and X-ray (2-10 keV) fluxes in eq. (\ref{eqfin}),
we estimate a mass of $\left(2.76^{+32.00}_{-2.54}\right) \times$10$^6$ M$_{\odot}$ for the putative IMBH.
The rather large uncertainty is mostly due to the intrinsic scatter of the fundamental plane relation. The obtained IMBH mass is
consistent with that estimated by \citet{nucita2013b} when analyzing the 2011 Chandra data.
We also note that this result sligthly depends on the index $\alpha_R$ of the radio
spectral energy distibution. Changing $\alpha_R$, $M_{BH}$ scales with the factor
$(1.4~{\rm GHz}/5~{\rm GHz})^{1.28\alpha_R}$. For example, for $\alpha_R\simeq 0.4$
the BH mass gets reduced by $\simeq 50\%$.

Note also that when scaling the 0.2-12 keV source flux (as detected by the {\it XMM-}Newton) to the range 0.5-7 keV,
one gets a flux consistent (within the errors) to that estimated using the 2011 Chandra data. The corresponding
0.5-7.0 keV luminosity is $L_X=L_{0.5-7.0~{\rm keV}}=(3.21\pm2.17)\times10^{33}$
erg $s^{-1}$. By comparing the bolometric luminosity, calculated as $L_B\simeq 16 L_X$
(\citealt{ho2008}), with the expected Eddington one $L_{Edd}\simeq 1.3 \times 10^{38} \left(M_{BH}/M_{\odot}\right)$
erg s$^{-1}$, one gets $L_B/L_{Edd}\simeq 1.43\times10^{-10}$. This clearly shows that the UMi putative BH is radiatively inefficient.
Indeed, assuming the simplified Bondi accretion scenario, we use eq. (\ref{explum}) with $L_X=L^{Unabs}_{0.2-12.0~{\rm keV}}$
and obtain \begin{equation}
\epsilon \eta \simeq 7.3\times 10^{-11} ~ - ~ 7.3\times 10^{-9}~,
\label{effi}
\end{equation}
that confirms the expected low IMBH accretion efficiency, previously highlighted.

Finally, we evaluate the expected number of background objects to investigate the possibility that the high energy
emission of Src 3 is due to an object standing behind. By using the minimum X-ray detected flux towards UMi,
correctly scaled to the 0.5-2.0 keV band ($F_{0.5-2.0~{\rm keV}}^{Una}\simeq 1.80 \times 10^{-15}$ erg s$^{-1}$ cm$^{-2}$),
we get $N\simeq0.07$ within $25''$ from the center. In spite of the small value of N, the background scenario
cannot be definitively ruled out.

\section{Conclusions}

In this paper, we re-analyzed some deep archival \sat\ data sets in order to full characterize the
high-energy point-like source population of four dSph MW companions. Hence, the ultimate goal, is to classify 
the X-ray sources identified towards our galaxy sample and pinpoint local (or candidate local) sources.

We performed an accurate study obtaining 89 X-ray sources for Draco, 116 for Leo I, 49 for UMa II and 54 for UMi. Albeit these values are statistically consistent 
with those of background AGNs, theoretically reached via the logN-logS relation, we cannot rule out the possibility that 
some sources belong to the same dSphs. This claim is supported by the color-color diagram (based on the ratio between 
$0.2-2.4$ keV X-ray and J band NIR fluxes versus the J-K color). In this way we find that some
X-ray sources, correlating with counterparts in the 2MASS catalogue, have a stellar nature (see Fig. 
\ref{figclass} and Sect. 4.1). 
Further, we perform a statistical source sorting using only the high-energy data. This leads us to
a wider and more complete classification of the X-ray sources in the target field of view.
We reveal two high-energy sources (among which a carbon star) belonging to Draco, one to UMa II and none to Leo I. 
Although the the statistic at our hand is poor, finding a few possible local X-ray sources may represent a problem that need to be addressed.
In fact, dSph galaxies, as globular clusters, host mainly old star populations but, due to the much lower central 
stellar density in dSphs, X-ray sources (either LMXBs or CVs) are expected to be primordial objects and not formed by capture encounters. 
However, these X-ray sources, should already turned off making unlikely to being found in dSphs. Finding X-ray sources in dSphs represent therefore 
a puzzling problem and in any case may allow to test the formation theories of these objects in a contrasting environment with respect to that in globular clusters..

By extrapolating the fundamental $M_{BH} - M_{Bulge}$ relation to the dSph realm, one sees that these galaxies are expected to contain IMBHs 
in their gravitational centers. We thus searched for X-ray sources located within the core radius of the dSphs of our sample. However, 
for Draco, Leo, and UMa we did not identify any high-energy (and radio) sources nearby the galactic centers. Then, we can only put upper limits to the mass and luminosity of the putative central IMBHs.

On the contrary, it is interesting that the UMi center hosts an X-ray source --Src 3-- which correlates in position 
with a radio object. In the IMBH hypothesis, the fundamental plane relation of eq. (5) 
allows us to get an estimate of the mass (a few 
10$^6M_{\odot}$) of the accreting black hole possible hosted in the galaxy. 
We remind that the source was already detected by Chandra in the past but only at the poor 
$\simeq 2.5\sigma$ confidence level, thus leaving the possibility that the Chandra source was a mere fake detection. 

Our analysis, based on {independent} {\it XMM}-Newton data, confirms however Src 3 as a possible X-ray counterpart of an accreting IMBH.

\section*{Acknowledgments}
{We thank M. Guainazzi for stimulating discussions while preparing the manuscript. We also acknowledge the support by the 
INFN project TaSP. We thank the anonymous Referee for the suggestions that greatly improved the paper. One of us (AAN) is grateful to 
Matteo Nucita for reading the draft.}

\end{document}